\documentclass[12pt]{article}
\usepackage{amssymb}
\usepackage{epsfig}
\usepackage{bbm}
\usepackage{bbold}

\usepackage{graphicx}

\def\L{\mathcal L}
\def\e{\varepsilon}

\newcommand{\wt}{\widetilde}

\begin{document}

\def\a{\alpha}
\def\b{\beta}
\def\c{\chi}
\def\d{\delta}
\def\e{\epsilon}
\def\f{\phi}
\def\g{\gamma}
\def\h{\eta}
\def\i{\iota}
\def\j{\psi}
\def\k{\kappa}
\def\l{\lambda}
\def\m{\mu}
\def\n{\nu}
\def\o{\omega}
\def\p{\pi}
\def\q{\theta}
\def\r{\rho}
\def\s{\sigma}
\def\t{\tau}
\def\u{\upsilon}
\def\x{\xi}
\def\z{\zeta}
\def\D{\Delta}
\def\F{\Phi}
\def\G{\Gamma}
\def\J{\Psi}
\def\L{\Lambda}
\def\O{\Omega}
\def\P{\Pi}
\def\Q{\Theta}
\def\S{\Sigma}
\def\U{\Upsilon}
\def\X{\Xi}

\def\ve{\varepsilon}
\def\vf{\varphi}
\def\vr{\varrho}
\def\vs{\varsigma}
\def\vq{\vartheta}

\def\dg{\dagger}                                     
\def\ddg{\ddagger}                                   
\def\wt#1{\widetilde{#1}}                    
\def\mt{\widetilde{m}_1}
\def\mti{\widetilde{m}_i}
\def\rt{\widetilde{r}_1}
\def\mtt{\widetilde{m}_2}
\def\mttt{\widetilde{m}_3}
\def\rtt{\widetilde{r}_2}
\def\mb{\overline{m}}
\def\VEV#1{\left\langle #1\right\rangle}        
\def\be{\begin{equation}}
\def\ee{\end{equation}}
\def\ds{\displaystyle}
\def\ra{\rightarrow}

\def\bea{\begin{eqnarray}}
\def\eea{\end{eqnarray}}
\def\NO{\nonumber}
\def\Bar#1{\overline{#1}}


\def\pl#1#2#3{Phys.~Lett.~{\bf B {#1}} ({#2}) #3}
\def\np#1#2#3{Nucl.~Phys.~{\bf B {#1}} ({#2}) #3}
\def\prl#1#2#3{Phys.~Rev.~Lett.~{\bf #1} ({#2}) #3}
\def\pr#1#2#3{Phys.~Rev.~{\bf D {#1}} ({#2}) #3}
\def\zp#1#2#3{Z.~Phys.~{\bf C {#1}} ({#2}) #3}
\def\cqg#1#2#3{Class.~and Quantum Grav.~{\bf {#1}} ({#2}) #3}
\def\cmp#1#2#3{Commun.~Math.~Phys.~{\bf {#1}} ({#2}) #3}
\def\jmp#1#2#3{J.~Math.~Phys.~{\bf {#1}} ({#2}) #3}
\def\ap#1#2#3{Ann.~of Phys.~{\bf {#1}} ({#2}) #3}
\def\prep#1#2#3{Phys.~Rep.~{\bf {#1}C} ({#2}) #3}
\def\ptp#1#2#3{Progr.~Theor.~Phys.~{\bf {#1}} ({#2}) #3}
\def\ijmp#1#2#3{Int.~J.~Mod.~Phys.~{\bf A {#1}} ({#2}) #3}
\def\mpl#1#2#3{Mod.~Phys.~Lett.~{\bf A {#1}} ({#2}) #3}
\def\nc#1#2#3{Nuovo Cim.~{\bf {#1}} ({#2}) #3}
\def\ibid#1#2#3{{\it ibid.}~{\bf {#1}} ({#2}) #3}

\title{\bf $SO(10)$-inspired solution to the problem of the initial
conditions in leptogenesis}

\author
{\Large Pasquale Di Bari 
\\
{\it\small School of Physics and Astronomy},
{\it\small University of Southampton,}
{\it\small  Southampton, SO17 1BJ, U.K.}
\\
{\Large Luca Marzola} \\
{\it\small 
} 
{\it\small Institute of Physics, 
University of Tartu}, 
{\it\small T\"{a}he 4, EE-51010 Tartu, Estonia}
}

\maketitle \thispagestyle{empty}

\vspace{-8mm}

\begin{abstract}
We show that, within $SO(10)$-inspired leptogenesis, there exists
a solution, with definite constraints on neutrino parameters, 
able simultaneously to reproduce the observed baryon asymmetry and 
to satisfy the conditions for the  independence of the final asymmetry of the initial conditions 
(strong thermal leptogenesis). We find that the wash-out of a pre-existing asymmetry as large 
as ${\cal O}(0.1)$  requires: 
i)  reactor mixing angle  $2^{\circ}\lesssim  \theta_{13}\lesssim 20^{\circ}$,  
in agreement with the experimental result $\theta_{13}=8^{\circ}-10^{\circ}$;
ii) atmospheric mixing angle  $16^{\circ}\lesssim \theta_{23}\lesssim 41^{\circ}$,
compatible only with current lowest experimentally allowed values;
iii) Dirac phase in the range $-\pi/2\lesssim \delta\lesssim \pi/5$, 
with the bulk of the solutions around $\d \simeq - \pi/5$
and such that ${\rm sign}(J_{CP})= - {\rm sign}(\eta_B)$; 
iv)  neutrino masses $m_i$  normally ordered;
v) lightest neutrino mass in the range
$m_1 \simeq (15 -  25)\,$meV, corresponding to 
$\sum_i m_i \simeq (85 - 105) \,{\rm meV}$;
vi)  neutrinoless double beta decay ($0\nu\b\b$)  effective neutrino mass $m_{ee}\simeq 0.8\, m_1$.
All together this set of predictive constraints characterises the solution
quite distinctively, representing a difficultly forgeable, fully testable, signature. 
In particular, the predictions $m_{ee} \simeq 0.8\,m_1 \simeq 15\,{\rm meV}$ can be
tested by cosmological observations and (ultimately) by $0\nu\b\b$ experiments. 
We also discuss different interesting aspects of the solution such as
theoretical uncertainties, stability under variation of the involved parameters, 
forms of the orthogonal and  RH neutrino mixing matrices. 
 \end{abstract}

\newpage

\section{Introduction}

Leptogenesis \cite {fy,reviews} is a cosmological application of the see-saw mechanism \cite{seesaw},
successfully linking two seemingly independent experimental observations: the matter-antimatter
asymmetry of the Universe  and the neutrino (masses and mixing)   
parameters tested in low energy neutrino experiments. 
The matter-antimatter asymmetry can be expressed in terms of 
the baryon-to-photon number ratio, quite precisely and accurately 
determined by CMB observations, 
in particular from Planck (anisotropies plus lensing) data \cite{Planck}
\be
\eta_B^{CMB} = (6.065 \pm 0.090) \times 10^{-10} \, .
\ee
On quantitative grounds, the requirement of successful leptogenesis is nicely supported 
by  neutrino oscillation experiments measuring the atmospheric and the solar neutrino mass scales 
within an optimal (order-of-magnitude) range \cite{bounds}.

If one considers the so called vanilla scenario,
where lepton flavour effects are neglected, a hierarchical RH neutrino spectrum is assumed
and the asymmetry is dominantly produced by the lightest RH neutrinos,
one obtains an upper bound on the neutrino masses  $m_i \lesssim 0.1\,{\rm eV}$ \cite{bound,pedestrians}. 
As a sufficient (but not necessary) condition that guarantees the final asymmetry to be independent of the
initial conditions (strong thermal leptogenesis), a lower bound $m_1 \gtrsim 0.001\,{\rm eV}$ on the lightest neutrino mass is also easily obtained.  This neutrino mass window \cite{window} is quite interesting since  in this way one obtains (at least partially) a testable quantitative link between the matter-anti matter asymmetry and the absolute neutrino
mass scale.

However, any attempt to derive further connections with the low energy neutrino parameters 
encounters serious difficulties,  mainly  for two reasons: 
the first is that, within the minimal picture,
the right-handed (RH) neutrinos responsible for the generation of the asymmetry are too heavy 
to give any observational trace, except for the matter-anti matter asymmetry itself; the second is 
that,  by just combining the requirement of successful leptogenesis  with low energy neutrino 
data, there is not a model independent way to over-constrain the see-saw parameter space 
obtaining testable predictions on future low energy neutrino results. 
In particular, the final asymmetry is completely independent of the parameters 
in the leptonic mixing matrix tested by neutrino oscillation experiments. 

When lepton flavour effects  are taken into account \cite{flavour}, 
the final asymmetry does depend explicitly
on the leptonic mixing matrix. This could raise the hope  that 
leptogenesis can be tested with neutrino oscillations experiments. 
However, the final asymmetry generally still depends 
also on the high energy parameters, associated to the
properties of the heavy RH neutrinos. It turns out that
the observed value of the asymmetry
can be attained  for an arbitrary choice of the 
low energy neutrino parameters. 
As a consequence, inclusion of
flavour effects does not lead to new model independent 
predictions or links with the low energy neutrino parameters. 
This remains true even within restricted scenarios
such as the usual $N_1$-dominated leptogenesis scenario \cite{rius}
or the two RH neutrino scenario \cite{2RHN}. 

Flavour effects have also an impact on the validity of the above mentioned neutrino mass window
and in particular of the lower bound $m_1 \gtrsim 10^{-3}\,{\rm eV}$,
originating from an intriguing conspiracy between 
the measured atmospheric and solar neutrino mass scales and the condition
of successful strong thermal leptogenesis.  This is because, when flavour effects
are considered, it is much easier for a pre-existing asymmetry to escape the RH neutrino
wash-out \cite{problem}.
A solution to the requirement of successful 
strong thermal leptogenesis still exists, but the conditions for its realisation  
become seemingly quite special. 
First of all they imply a tauon $N_2$-dominated scenario, where the final asymmetry is produced 
by the next-to-lightest RH neutrinos in the two flavour regime, implying
$10^{12}\,{\rm GeV}\gtrsim M_2 \gtrsim 10^9\,{\rm GeV}$, dominantly in the tauon flavour and where the 
lightest RH neutrino mass  $M_1 \ll 10^9\,{\rm GeV}$.
In addition, there are a few further conditions on the flavoured decay parameters 
that apparently  make
the whole set  very difficult to be realised in realistic models.
Therefore, the inclusion of flavour effects makes much more difficult to 
satisfy the strong thermal condition. 

On the other hand, there are some phenomenologically significant implications of flavour effects.
For example,  it is interesting that under some conditions on the RH neutrino masses, the 
same source of $C\! P$ violation that could give effects in neutrino oscillations, would also
be sufficient to explain the observed matter-anti matter asymmetry 
within the $N_1$-dominated scenario  \cite{flavorlep, pascoli,diraclep}. 
After the recent discovery of a non-vanishing $\theta_{13}$ in  long baseline
and reactor experiments \cite{theta13} and subsequent global analyses \cite{fogli,gonzalez,valle} 
finding
\be\label{newrangetheta13}
8^{\circ} < \theta_{13} <  10^{\circ}   \hspace{10mm} (\sim 95\% \, {\rm C.L.})  \,  ,
\ee
this scenario would be viable if $|\sin\d| \simeq 1$ and $M_1 \gtrsim 10^{11}\,{\rm GeV}$. 
Though the realisation of successful Dirac phase leptogenesis
is not motivated within a precise theoretical framework, this scenario
could still emerge as an approximated case within some proposed models  
such as, for example, minimum flavour violation \cite{mfv} and
two RH neutrino models \cite{flavorlep,2RH}. 
Therefore, in this respect,  it will be rather interesting to determine 
the value of the Dirac phase during the next years. 

Another important consequence of flavour effects is that  the 
$N_2$-dominated scenario \cite{geometry} applies for a much wider region of the parameter
space. This is because the $N_2$ produced asymmetry  can more easily 
escape the lightest RH neutrino ($N_1$) wash-out  \cite{vives}
and reproduce the observed  asymmetry \cite{bounds}.  An important 
application of this effect is that it rescues
\cite{SO10lep1} the so called 
$SO(10)$-inspired leptogenesis scenario \cite{buchplum,orloff,falcone,branco,smirnov}.
This scenario corresponds to a very well theoretically motivated 
set of ($SO(10)$-inspired) conditions  that over-constrains the see-saw parameter space.
In this way the  final asymmetry becomes much more sensitive
to the low energy neutrino parameters than in the general case. 
Within an unflavoured description, the final asymmetry
is dominated  by the lightest RH neutrino contribution. 
However, in the light of the current neutrino oscillations data,
the RH neutrino mass spectrum turns out to be typically highly hierarchical with 
the lightest RH neutrino mass  $M_1 \ll 10^9 \,{\rm GeV}$ 
\cite{orloff,falcone,branco,smirnov}, well below
the lower bound for successful leptogenesis \cite{dicmb}.  
This result is quite stable under a precise definition of the $SO(10)$-inspired conditions. 
It just holds barring very  fine tuned choices of the parameters 
around `crossing-level' solutions where RH neutrino masses are quasi-degenerate \cite{smirnov}, 
$C\! P$ asymmetries get resonantly enhanced \cite{crv}  
 and successful leptogenesis can be attained \cite{pilaftsis}. 

On the other hand, when flavour effects are taken into account, the asymmetry produced by the $N_2$ decays
can reproduce the observed asymmetry. 
Therefore,  $SO(10)$-inspired  leptogenesis is rescued by 
a thorough account of  lepton and heavy neutrino flavour effects and
it becomes viable \cite{SO10lep1} if
 some interesting constraints on the low energy neutrino parameters
are satisfied \cite{SO10lep2}.  
In particular, a lower bound on the 
lightest neutrino mass, $m_1 \gtrsim 0.001\,{\rm eV}$, holds. 
Moreover inverted ordered neutrino masses are only marginally allowed. 
\footnote{Generalisations of the see-saw mechanism within 
left-right symmetric models with both type I and type II terms  \cite{abada}
or with an inverse see-saw \cite{mohapatra} provide alternative solutions.}

There is, however, also another interesting feature  
of $SO(10)$-inspired leptogenesis \cite{SO10lep2}: it is potentially able to satisfy the
strong thermal condition, since it indeed naturally realises the above mentioned
 tauon $N_2$-dominated scenario.  

In this paper we investigate in detail this potential feature of $SO(10)$-inspired models
to realise successful strong thermal leptogenesis and we indeed show that
there exists a subset of the solutions leading to successful 
$SO(10)$-inspired leptogenesis that also satisfies the strong thermal condition.
We show that this novel solution realising strong thermal condition  within $SO(10)$-inspired leptogenesis, 
implies quite sharp and distinctive constraints on the low energy neutrino parameters, 
in particular on the neutrino masses. Interestingly, these 
non-trivially overlap with current experimental constraints and, as we discuss, 
they can be fully tested by future  experiments. 

The paper is organised as follows.  In section 2 we introduce the notation 
and review the status of low energy neutrino experimental results.
In Section 3 we briefly review the set up for $SO(10)$-inspired leptogenesis, verifying
the results obtained in \cite{SO10lep2} and presenting new improved scatter plots 
that strengthen the conclusions of \cite{SO10lep2} and reveal some new interesting features. 
In Section 4 we briefly review and motivate the conditions for successful strong thermal
leptogenesis. In Section 5, the central section of the paper, 
we combine strong thermal and $SO(10)$-inspired conditions and 
show the existence of a solution implying predictive constraints on neutrino parameters,
briefly discussing the prospects to test them in next years.
In Section 6 we discuss different aspects of this new solution
such as theoretical uncertainties, stability under variation of the involved parameters, 
corresponding forms of the orhtogonal and RH neutrino mixing matrices. 
Finally, in Section 7, we draw the conclusions. 

\section{See-saw mechanism and low energy neutrino data}

Adding three RH neutrinos  to the standard model Lagrangian, one per each generation as predicted by $SO(10)$ models, with Yukawa coupling $h$ and a Majorana mass term $M$, 
a neutrino Dirac mass term $m_D=h\,v$ is generated
by the vacuum expectation value $v$ of the Higgs boson, like for the other massive fermions
and in particular for the charged leptons with Dirac mass matrix $m_{\rm \ell}$. 
In this way, in the basis where charged lepton and right-handed 
neutrino mass matrices are diagonal, their lagrangian mass terms can be written as 
($\a = e, \m, \t$ and $i=1,2,3$)
\be
- {\cal L}_{M} = \,  \overline{\a_L} \, D_{m_{\ell}}\,\a_R + 
                              \overline{\nu_{\a L}}\,m_{D\a i} \, N_{i R} +
                               {1\over 2} \, \overline{N^{c}_{i R}} \, D_{M} \, N_{i R}  + \mbox{\rm h.c.}\, ,
\ee
where $ D_{m_{\ell}} \equiv {\rm diag}(m_e, m_{\mu}, m_{\tau})$ and 
$D_M \equiv {\rm diag}(M_1, M_2, M_3)$, with $M_1 \leq  M_2 \leq M_3$. 

In the see-saw limit, for $M \gg m_D$, the spectrum of neutrino mass eigenstates 
splits into a very heavy set, $N_i \simeq N_{iR} + N_{iR}^c$, with masses almost coinciding with the Majorana masses $M_i$, and into a light set  $\nu_i \simeq \nu_{iL} + \nu_{i L}^c$, with a 
symmetric mass matrix $m_{\nu}$ given by the see-saw formula
\be\label{seesaw}
m_{\nu} = - m_D \, {1\over D_M} \, m_D^T  \,  .
\ee
This is diagonalised by a unitary matrix $U$,
\be\label{leptonic}
U^{\dagger} \,  m_{\nu} \, U^{\star}  =  - D_m  \,  ,
\ee
corresponding to the leptonic mixing matrix, in a way that we can write
\be\label{diagonalseesaw}
 D_m =  U^{\dagger} \, m_D \, {1\over D_M} \, m_D^T  \, U^{\star}     \,   .
\ee
 Neutrino oscillation experiments 
measure two light neutrino mass squared differences, $\Delta m^2_{\rm atm}$
and $\Delta m^2_{\rm sol}$. There are two possibilities:
either light neutrino masses are normally ordered (NO), 
with  $m^{\,2}_3 - m^{\,2}_2 = \Delta m^2_{\rm atm}$ 
and $m^{\,2}_2 - m^{\,2}_1 = \Delta m^2_{\rm sol}$,  
or they are inversely ordered (IO), with
$m^{\,2}_3 - m^{\,2}_2 = \Delta m^2_{\rm sol}$ and 
$m^{\,2}_2 - m^{\,2}_1 = \Delta m^2_{\rm atm}$.
For NO (IO) it is found, for example in \cite{valle} and similarly 
in \cite{fogli,gonzalez},
\footnote{We will neglect throughout the paper the small 
experimental errors on $m_{\rm atm}$ and on $m_{\rm sol}$
since, with very good approximation, all the constraints 
that we will discuss are insensitive to them.}
\be\label{solatmscales}
m_{\rm atm}\equiv \sqrt{m^{\,2}_3 - m_1^{\,2}} = 0.0505\,(0.0493)\,{\rm eV}\,
\hspace{5mm}
\mbox{\rm and} 
\hspace{5mm}
m_{\rm sol}\equiv \sqrt{\Delta m^2_{\rm sol}} = 0.0087\,{\rm eV} \, .
\ee
In this way  there is just one parameter left  
to be measured in order to determine the
so called absolute neutrino mass scale fixing 
the three light neutrino masses. 
This can be conveniently identified 
with the lightest neutrino mass $m_1$. The most stringent upper bound on $m_1$
is derived from cosmological observations. A conservative upper bound on the sum of the neutrino masses has been recently placed by the Planck collaboration \cite{Planck}. Combining Planck and high-${\ell}$ CMB anisotropies, WMAP polarisation  and baryon acoustic oscillation data it is found   
$\sum_i \, m_i \lesssim 0.23\,{\rm eV}\,(95\% {\rm C.L.})$.  
When neutrino oscillation results are combined, 
this translates into an upper bound on the lightest neutrino mass,
\be\label{upperboundm1}
m_1 \lesssim 0.07 \, {\rm eV}  \,  ,
\ee 
showing how cosmological observations  start to corner quasi-degenerate neutrinos. 

In the NO case we adopt for the leptonic mixing matrix the PDG parametrisation
\begin{equation}\label{Umatrix}
U^{\rm (NO)}=\left( \begin{array}{ccc}
c_{12}\,c_{13} & s_{12}\,c_{13} & s_{13}\,e^{-{\rm i}\,\d} \\
-s_{12}\,c_{23}-c_{12}\,s_{23}\,s_{13}\,e^{{\rm i}\,\d} &
c_{12}\,c_{23}-s_{12}\,s_{23}\,s_{13}\,e^{{\rm i}\,\d} & s_{23}\,c_{13} \\
s_{12}\,s_{23}-c_{12}\,c_{23}\,s_{13}\,e^{{\rm i}\,\d}
& -c_{12}\,s_{23}-s_{12}\,c_{23}\,s_{13}\,e^{{\rm i}\,\d}  &
c_{23}\,c_{13}
\end{array}\right)
\cdot {\rm diag}\left(e^{i\,\rho}, 1, e^{i\,\sigma}
\right)\, ,
\end{equation}
where $c_{ij} \equiv \cos\theta_{ij}$ and $s_{ij} \equiv \sin\theta_{ij}$.
Because of the adopted light neutrino mass labelling convention,
in the IO case the leptonic mixing matrix has to be recast  
simply with a proper relabelling of the column index, explicitly
\be
U^{\rm (IO)} =  \left( \begin{array}{ccc}
s_{13}\,e^{-{\rm i}\,\d} & c_{12}\,c_{13} & s_{12}\,c_{13}  \\
s_{23}\,c_{13} & -s_{12}\,c_{23}-c_{12}\,s_{23}\,s_{13}\,e^{{\rm i}\,\d} &
c_{12}\,c_{23}-s_{12}\,s_{23}\,s_{13}\,e^{{\rm i}\,\d} \\
c_{23}\,c_{13} & s_{12}\,s_{23}-c_{12}\,c_{23}\,s_{13}\,e^{{\rm i}\,\d}
& -c_{12}\,s_{23}-s_{12}\,c_{23}\,s_{13}\,e^{{\rm i}\,\d}
\end{array}\right)
\cdot {\rm diag}\left(e^{i\,\sigma}, e^{i\,\rho}, 1  \right) \,  .
\ee
As already discussed, the reactor mixing angle is found in the range eq.~(\ref{newrangetheta13}). 
Current global analyses \cite{valle} find for the solar mixing angle the $2\s$
range $32.6^{\circ} \lesssim \theta_{12} \lesssim 36.3^{\circ}$.

The atmospheric mixing angle $\theta_{23}$, is  now favoured by 
MINOS results to be non-maximal \cite{MINOS}.
This is also confirmed by  global analyses \cite{fogli,gonzalez,valle}, though
with different statistical significance. 
In \cite{fogli}  $\theta_{23}$ is  favoured 
to be in the first octant, finding for NO the   $2\s$  range 
$36.3^{\circ}\lesssim \theta_{23} \lesssim 43.6^{\circ}$.
In \cite{gonzalez} $\theta_{23}$ is also favoured in the first octant for NO
but with a very low statistical significance. In \cite{valle} the (almost octant symmetric)	 
$2\s$ range $38^{\circ}\lesssim \theta_{23} \lesssim 54.3^{\circ}$ is found for NO.
Certainly more data are needed for a robust determination of the octant. 
As we will see in Section 5, our solution will give quite a clear prediction on this point. 

It will also prove useful to introduce the so called orthogonal (or Casas-Ibarra) parametrisation \cite{casas}. The see-saw formula eq.~(\ref{seesaw}) can be recast as an orthogonality condition for a 
matrix $\Omega$. Through $\O$ the neutrino Dirac mass matrix can be expressed as
\be\label{orthogonal}
m_D =  U\,\sqrt{D_m}\, \Omega \, \sqrt{D_M} \,  .
\ee
The $\O$ matrix contains 6 independent high energy parameters encoding the properties of the 3 RH neutrinos  (e.g. the 3 lifetimes and the 3 total $C\!P$ asymmetries) and it is quite useful not only to express the different relevant quantities for the calculation of the asymmetry, and for this reason  
we will employ it as an intermediate step for the calculation of the 
asymmetry, but also to characterise see-saw neutrino models.

\section{$SO(10)$-inspired leptogenesis}

As we discussed in the introduction, without imposing any condition on the nine high energy parameters,
the baryon asymmetry has in general to be calculated taking into account both lepton and heavy
neutrino flavour effects and the calculation should proceed through the solution of a set of density matrix equations  \cite{bcst,flavour,density}.
The condition of successful leptogenesis, $\eta^{\rm lep}_B = \eta_B^{\rm CMB}$, places  an upper bound on the neutrino masses, $m_1 \lesssim 0.12\,{\rm eV}$ \cite{bound,giudice,pedestrians},
holding in the case of $N_1$-dominated leptogenesis and in the one-flavour regime, 
for $M_1 \gtrsim 10^{12}\,{\rm GeV}$.  This is the only existing model independent link between
leptogenesis and low energy neutrino data.

\subsection{General setup}

Now let us see how, by imposing  $SO(10)$-inspired conditions
and barring fine-tuned crossing level solutions \cite{smirnov},
a  RH neutrino mass pattern implying a $N_2$-dominated leptogenesis scenario necessarily emerges, 
where  the calculation of the asymmetry reduces
to a simple analytical expression and  the successful leptogenesis bound implies
constraints on all low energy neutrino parameters \cite{SO10lep2}.
 
The neutrino Dirac mass matrix can be diagonalised by a bi-unitary transformation
\be\label{biunitary}
m_D = V^{\dagger}_L \, D_{m_D} \, U_R   \, ,
\ee
where $D_{m_D} \equiv {\rm diag}(m_{D1}, m_{D2} , m_{D3})$. 
The unitary matrix $V_L$  acts on the
left-handed neutrino fields operating the transformation from the weak basis to the Yukawa basis.
It is the analogous of the CKM matrix in the quark sector, operating the transformation
from the down to the up quark mass basis.  

Inserting the bi-unitary parameterisation for $m_D$ into the diagonalised see-saw formula  
eq.~(\ref{diagonalseesaw}), one can see that $U_R$ diagonalises the matrix
\be
M^{-1} \equiv D_{m_D}^{-1}\,V_L\,U\,D_m\,U^T\,V_L^T\,D_{m_D}^{-1} \,  , 
\ee
explicitly $M^{-1} = U_R\,D_M^{-1}\,U_R^T$.
\footnote{This also implies $D_M = U_R\, M \, U_R^T$, showing that 
$U_R$ operates the transformation of the Majorana mass matrix from the Yukawa basis,
where $m_D$ is diagonal,  to the basis where the Majorana mass matrix
is diagonal.} 
This expression shows that the  RH neutrino mass spectrum, and the matrix $U_R$,
can be expressed in terms of the low energy neutrino parameters, of the three eigenvalues
of $m_D$ and of the six parameters in $V_L$, explicitly $M_i = M_i (m_j, U; V_L, \a_k)$
and $U_R = U_R (m_j, U; V_L, \a_k)$, 
where the three $\a_k$ are the  ratios of the Dirac mass matrix eigenvalues to the
three up quark masses,  explicitly
\be
m_{D1} = \a_1 \, m_u \,  , \,\, m_{D2} = \a_2 \, m_c \, , \,\, m_{D3} = \a_3 \, m_t  \,  .
\ee
 Notice that so far we have not yet restricted the see-saw parameters space, we have just simply
introduced a sort of hybrid parameterisation where, compared to the orthogonal parameterisation 
(cf. eq.~(\ref{orthogonal})), the nine parameters 
$(M_i,\O)$ are replaced by $(\a_k, V_L)$ or compared to the bi-unitary parameterisation
the nine high energy parameters $(M_i,U_R)$ are replaced by $(m_j,U)$, 
i.e. by the nine testable low energy neutrino parameters in $m_{\nu}$.

We now define  $SO(10)$-inspired models those 
respecting the following  set of three (working) assumptions:
\begin{itemize}
\item The matrix $V_L$ 
is restricted within the range $I \leq V_L \leq V_{CKM}$, i.e. the three mixing angles
in $V_L$ are not larger than the corresponding three mixing angles in the CKM matrix. This is the
most important (i.e. restrictive) condition.
\item  We assume $\alpha_i = {\cal O}(1)$.
\item We bar regions in the space of parameters around
crossing level solutions, where at least two RH neutrino masses are non-hierarchical,
more specifically we impose $M_{i+1} \gtrsim 2\,M_i$ ($i=1,2$).
\end{itemize}
The last condition, of a hierarchical RH neutrino spectrum, 
is not restrictive at all. This is because the conditions to realise crossing level solutions
for the RH neutrino mass spectra are very fine tuned \cite{smirnov}, especially when the 
successful leptogenesis bound is imposed. The reason is simple: at the level 
crossings, the $C\!P$
asymmetries are resonantly enhanced and span many orders of magnitude. Consequently, 
the baryon asymmetry is very sensitive to tiny variations of the parameters that have to be
highly fine tuned in order for the successful leptogenesis condition, 
$\eta_B^{\rm lep} = \eta_B^{\rm CMB}$, to be satisfied (as an example
of a scenario realising a crossing level solution see \cite{buccella}).

Under these conditions, and given the current low energy neutrino data, 
the RH neutrino mass spectrum is hierarchical and of the form \cite{branco,SO10lep1}
\be
M_1 : M_2 : M_3  = (\a_1\,m_u)^2 : (\a_2\,m_c)^2 : (\a_3\,m_t)^2 \,  .
\ee
In particular,  from the second working assumption and given the current low energy neutrino data, 
it follows that $M_1 \ll 10^{9}\,$GeV while
$M_2 \gg 10^{9}\,$GeV. It also follows that all the heaviest RH neutrino ($N_3$) 
$C\!P$ asymmetries are  strongly suppressed. In this way the only contribution able to explain the observed asymmetry   is that one from next-to-lightest RH neutrino ($N_2$) decays. Therefore, the only possibility to satisfy the successful leptogenesis bound is within a $N_2$-dominated scenario. Assuming a thermal scenario, this necessarily requires 
that the reheating temperature $T_{\rm RH}\sim M_2$. 
The baryon asymmetry can then be calculated in a double stage, 
taking  into account first the production and wash-out from
the $N_2$'s at $T\sim M_2$ and then the lightest RH neutrino wash-out at $T\sim M_1$.

Let us introduce some standard quantities in leptogenesis. 
The flavoured decay parameters $K_{i\a}$ are defined as
\be
K_{i\a} \equiv {\G_{i\a}+\overline{\G}_{i\a}\over H(T=M_i)}= 
{|m_{D\a i}|^2 \over M_i \, m_{\star}} \,  ,
\ee
where the $\Gamma_{i\a}$'s and the $\bar{\Gamma}_{i \a}$'s can be identified
with the zero temperature limit of the flavoured decay rates into $\a$ leptons, $\Gamma (N_i \ra \phi^\dagger \, l_\alpha)$,
and anti-leptons, $\Gamma (N_i \ra \phi \, \bar{l}_\alpha)$ in a three-flavoured regime, where
lepton quantum states can be treated as an incoherent mixture  of the three flavour components.  
The equilibrium neutrino mass $m_{\star}$ is defined as
\be
m_{\star} \equiv {16\, \pi^{5/2}\,\sqrt{g_*} \over 3\,\sqrt{5}}\,
{v^2 \over M_{Pl}} \simeq 1.08\times 10^{-3}\,{\rm eV}\;.
\ee
The total decay parameters are simply given by $K_i = K_{ie} + K_{i\m} + K_{i\t}$. 
In the orthogonal parametrisation the flavoured and total 
decay parameters can be calculated as
\be
K_{i\a} = \left|\sum_j\,\sqrt{m_j\over m_{\star}}\,U_{\a j}\,\O_{j i}\right|^2 \, ,
\hspace{5mm}
K_i = \sum_i \, {m_j \over m_{\star}} \, |\O_{ji}|^2 \, .
\ee
The efficiency factors at the production, for a vanishing initial $N_2$ abundance, 
are given by the sum of a negative and of a positive contribution,
\be
\k(K_{2\a},K_2)   =\k_{-}^{\rm f}(K_2,K_{2\a})+ \k_{+}^{\rm f}(K_2,K_{2\a}) \, ,
\ee
that are approximated by the following expressions \cite{flavorlep}
\be\label{k-}
\k_{-}^{\rm f}(K_2,K_{2\a})\simeq
-{2\over p_{2\a}^{0}}\ e^{-{3\,\pi \over 8}\,K_{2\a} }
\left(e^{{p_{2\a}^{0}\over 2}\,\overline{N}(K_2)} - 1 \right) 
\ee
and
\be\label{k+}
\k_{+}^{\rm f}(K_2,K_{2\a})\simeq
{2\over z_B(K_{2\a})\,K_{2\a}}
\left(1-e^{-{K_{2\a}\,z_B(K_{2\a})\,\overline{N}(K_2)\over 2}}\right) \, ,
\ee
where
\begin{equation}\label{nka}
\overline{N}(K_2)\equiv {N(K_2)\over\left(1 + \sqrt{N(K_2)}\right)^2} \, ,
\end{equation}
\be
z_{B}(K_{2\a}) \simeq 2+4\,K_{2\a}^{0.13}\,e^{-{2.5\over K_{2\a}}}={\cal O}(1\div 10) \, 
\ee
and $p^0_{2\a} = K_{2\a}/K_2$ is the tree level probability that the 
lepton  quantum state  produced by a $N_2$-decay  
is measured as an $\a$ flavour eigenstate.
The flavoured $C\!P$ asymmetries, 
\be
\ve_{2\a}\equiv -{\G_{2\alpha}-\overline{\G}_{2\alpha}
\over \G_{2}+\overline{\G}_{2}} \,  ,
\ee
can be calculated from \cite{crv}
\be\label{eps2a}
\ve_{2\a} \simeq
\overline{\ve}(M_2) \, \left\{ {\cal I}_{23}^{\a}\,\x(M^2_3/M^2_2)+
\,{\cal J}_{23}^{\a} \, \frac{2}{3(M^2_3/M^2_2-1)}\right\}\, ,
\ee
where we defined \cite{geometry,bounds,2RHN}
\be
\overline{\ve}(M_2) \equiv {3\over 16\,\pi}\,{M_2\,m_{\rm atm} \over v^2} \, , \hspace{3mm} \xi(x)=\frac{2}{3}x\left[(1+x)\ln\left(\frac{1+x}{x}\right)-\frac{2-x}{1-x}\right] \,  ,
\ee
\be
{\cal I}_{23}^{\a} \equiv   {{\rm Im}\left[m_{D\a 2}^{\star}
m_{D\a 3}(m_D^{\dag}\, m_D)_{2 3}\right]\over M_2\,M_3\,\mtt\,m_{\rm atm} }\,   
\hspace{5mm}
\mbox{\rm and}
\hspace{5mm}
{\cal J}_{23}^{\a} \equiv  
{{\rm Im}\left[m_{D\a 2}^{\star}\, m_{D\a 3}(m_D^{\dag}\, m_D)_{3 2}\right] 
\over M_2\,M_3\,\mtt\,m_{\rm atm} } \,   ,
\ee
with $\mtt \equiv (m_D^{\dag}\, m_D)_{2 2}/M_2 = K_2\,m_{\star}$.
The quantities ${\cal I}_{23}^{\a}$ and ${\cal J}_{23}^{\a}$ can  be expressed in the orthogonal 
parameterisation as \cite{bounds,diraclep}
\be
{\cal I}_{23}^{\a} =   {\rm Im} \Big[ \sum\limits_{k,h,l}
{m_{k}\,\sqrt{m_{h}\,m_{l}} \over \mtt \, m_{\rm atm}}\,
\,\O^*_{k2}\,\O_{k3}\,\O^*_{h2}\,\O_{l 3}\,U_{\alpha h}^* \, U_{\alpha l} \Big] \, ,
\ee
\be
{\cal J}_{23}^{\a} =
{\rm Im} \Big[ \sum\limits_{k,h,l} {m_{k}\,\sqrt{m_{h}m_{l}} \over \mtt \, m_{\rm atm}}\,
\, \O^*_{k3}\,\O_{k2}\,\O^*_{h2}\,\O_{l 3} \, U_{\alpha h}^* \, U_{\alpha l}\Big] \, .  
\ee
We can also conveniently define $\ve_{2\tau^{\bot}} \equiv \ve_{2e}+\ve_{2\mu}$ and
$K_{2\tau^{\bot}}\equiv K_{2e}+K_{2\m}$, where $\tau^{\bot}$ indicates a $\tau$ orthogonal
flavour component  that is a coherent superposition of electron and muon components,
in this specific  case those ones of the leptons ${\ell}_2$ produced in the $N_2$ decays.
In this way the final asymmetry in the $N_2$-dominated scenario can be calculated using 
quite simple expressions \cite{vives,bounds,SO10lep2}.  

For $M_2 \ll 10^{12}\,{\rm GeV}$, so that the $N_2$ 
production occurs in the two-flavour regime, the final
asymmetry can be calculated as
\be\label{twofl}
N_{B-L}^{\rm f} \simeq
{K_{2e}\over K_{2\tau^{\bot}}}\,\ve_{2 \tau^{\bot}}\,\kappa(K_{2 \tau^{\bot}})
\, e^{-{3\pi\over 8}\,K_{1 e}}+
{K_{2\mu}\over K_{2 \tau^{\bot}}}\,
\ve_{2 \tau^{\bot}}\,\kappa(K_{2 \tau^{\bot}})\, e^{-{3\pi\over 8}\,K_{1 \mu}}+
\ve_{2 \tau}\,\kappa(K_{2 \tau})\,e^{-{3\pi\over 8}\,K_{1 \tau}} \,  ,
\ee
where we are calculating abundances in a portion of co-moving volume
containing one RH neutrino in ultra-relativistic thermal equilibrium  (so that
$N_{N_i}^{\rm eq}(T\gg M_i) = 1$).
On the other hand for $M_2 \gg 10^{12}\,{\rm GeV}$
 the production occurs in the one-flavour regime and in this case one can use
\be\label{onefl}
N_{B-L}^{\rm f} \simeq \ve_{2}\,\kappa(K_{2})\,\left(
{K_{2e}\over K_2}\, e^{-{3\pi\over 8}\,K_{1 e}}+
{K_{2\mu}\over K_2}\, e^{-{3\pi\over 8}\,K_{1 \mu}}+
{K_{2\tau}\over K_2}\,e^{-{3\pi\over 8}\,K_{1 \tau}} \right) \, .
\ee
These are the expressions for the final asymmetry that we adopt in our 
calculation. In the end, however, the case $M_2 \gg 10^{12}\,{\rm GeV}$,
will prove to be not particularly significant. 
Finally, the baryon-to-photon number ratio from leptogenesis can be calculated simply using
\be\label{etaNBmL}
\eta_B^{\rm lep} = a_{\rm sph}\,{N^{\rm f}_{B-L}\over N_{\g}^{\rm rec}} \simeq 0.96 \times 10^{-2}\,N^{\rm f}_{B-L} \,  ,
\ee
accounting for sphaleron conversion and photon dilution. It is important to notice
that $\eta_B$ does not depend on $\a_1$ and $\a_3$ \cite{SO10lep1}. This reduction of the number
of parameters in the final asymmetry is a key point for  the see-saw parameter
space to be over-constrained by the condition of successful leptogenesis,
thus resulting into constraints on the low energy neutrino data 
that allow the scenario to be testable. 

To our knowledge, there are five, potentially relevant, 
approximations  in this calculation of the final asymmetry:
 \begin{itemize}
 \item In the intermediate regime, for $M_2 \sim 10^{12}\,$GeV, 
 one should calculate the asymmetry solving the density matrix equation. 
We approximate the calculation simply using the eq.~(\ref{twofl}) if $M_2 \leq 10^{12}\,{\rm GeV}$
and the eq.~(\ref{onefl}) if $M_2 > 10^{12}\,{\rm GeV}$.\item We are neglecting phantom terms \cite{flavourcoupling,density}.
\item We are neglecting flavour coupling \cite{flavourcoupling}.
\item We are neglecting the running of neutrino parameters \cite{running} inserting directly, into the expression
          for the final asymmetry, the results from low energy neutrino experiments.
\item We are neglecting momentum dependence.
\end{itemize}
We will shortly discuss the potential impact of these approximations in Section 6,
concluding that actually they work quite  well.

\subsection{Constraints on neutrino parameters from scatter plots}

Let us now present the constraints on neutrino parameters obtained imposing 
the leptogenesis bound, $\eta^{\rm lep}_B=\eta^{CMB}_B$, $SO(10)$-inspired conditions
and assuming vanishing initial asymmetries and $N_2$-abundance.
We have fixed $\alpha_2=5$. This can be considered a realistic close-to-maximum value yielding 
conservative results, considering that $M_2 \propto \a_2^{\,2}$ and that this 
translates into $\eta_B^{\rm lep} \propto \a_2^{\, 2}$ (as far as $M_2 \lesssim 10^{12}\,$Gev).   

The asymmetry $\eta_B^{\rm lep}$ is calculated for differently
randomly (and uniformly) generated points in a region of the parameter space obeying
the $SO(10)$-inspired condition on the unitary matrix $V_L$  (cf. eq.~(\ref{biunitary})).
The unitary matrix $V_L$ is parameterised exactly as the leptonic mixing matrix $U$ 
(cf. eq.~(\ref{Umatrix})) and, therefore, in terms of three mixing angles 
$(\theta_{12}^L, \theta_{23}^L,\theta_{12}^L)$ and three phases, $(\d_L,\r_L,\s_L)$.
The three mixing angles are randomly scanned within the ranges 
$0 \leq \theta_{12}^L  \leq 13^{\circ}$,  $0 \leq \theta_{23}^L  \leq 2.5^{\circ}$, 
and $0 \leq \theta_{13}^L  \leq 0.2^{\circ}$, while
the three phases simply vary within $[0,2\pi]$.

Let us now describe the ranges adopted for the mixing angles.
In order to compare our results with those previously obtained in \cite{SO10lep2}, 
we still adopt the old $\theta_{13}$ ($2\s$) range, 
\be\label{rangetheta13}
0\leq \theta_{13} \leq 11.5^{\circ}  \,  , 
\ee
mainly determined by the CHOOZ upper bound \cite{CHOOZ}. 
However,  in all plots, we also highlight the current experimentally allowed 
much narrower range (cf. eq.~(\ref{newrangetheta13})). 

Also for the solar mixing angle we will continue, in the scatter plots, to adopt the 
same $2\s$ range as  in \cite{SO10lep2}  from \cite{schwetz},
\be\label{rangetheta12}
31.3^{\circ}\leq \theta_{12} \leq 36.3^{\circ}  \,  ,
\ee
just slightly larger than the above mentioned $2\s$ range from current global analyses. 

Finally, for the atmospheric mixing angle we conservatively adopt the range
\be\label{rangetheta23}
35^{\circ} < \theta_{23}  < 52.5^{\circ} \,  .
\ee
Compared to the range used in \cite{SO10lep2} ($38.5^{\circ} < \theta_{23} < 52.5^{\circ}$)
\cite{schwetz} this is  enlarged at low values taking into account, as previously discussed,
that MINOS results \cite{MINOS} and 
one of the global analyses \cite{fogli} find now that values well lower than $38.5^{\circ}$ are allowed.\
In particular, the MINOS collaboration 
find that values as low as $35^{\circ}$ are allowed at about $2\sigma$. 
The Dirac phase and the two Majorana phases are simply varied within $[0,2\pi]$.
Finally, the atmospheric and solar neutrino mass scales are fixed to their 
best fit values (cf. eq.~(\ref{solatmscales})) since the experimental errors are 
sufficiently small that the final asymmetry is not sensitive to them. 

Therefore, the parameter scan is made in a  13-dim parameter space: 
the 6 parameters in $V_L$ plus the $6$ parameters in $U$
plus the lightest neutrino mass $m_1$.   
We are clearly particularly interested in determining testable 
constraints on the 7-dim low energy neutrino  parameter space.  
In Fig.~1 we show, imposing 
\footnote{We consider separately the results for $10\gtrsim M_3/M_2 \gtrsim 2$ and discuss them in Section 6.}
$M_3/M_2 > 10$, the results as projections of the allowed regions 
on the most significant two low energy neutrino parameter planes for NO.  
Since we show projections on planes it is sufficient to 
impose $\eta_B^{\rm lep} > \eta_B^{\rm CMB}$ (in practice we imposed 
$\eta_B > 5.9 \times 10^{-10}$). Two of the panels also  
contain plots of the constraints on derived parameters such as the effective 
$0\nu\b\b$ neutrino mass $m_{ee}=|\sum_i \, m_i \, U^2_{ei}|$ 
and on the $C\!P$ invariant $J_{CP}=c_{12}\,s_{12}\,c_{23}\,s_{23}\,c^2_{13}\,s_{13}\sin\d$. 
In the case of $m_{ee}$ the dashed band is excluded by
the experimental bound $m_{ee} \lesssim 0.75\,$eV ($95\%\,$C.L.) obtained by
the Heidelberg-Moscow and CUORICINO experiments 
(recently tightened by GERDA \cite{neutrinolessnow}).
The allowed found solutions are indicated with yellow points.
\footnote{The red, green and blue points satisfy, in addition to the successful leptogenesis condition,
also the strong thermal condition, as we will discuss in the next sections.}

We do not show results for IO since in Section 5 we will point out
that IO is incompatible with the strong thermal condition, 
our main focus in this paper.
 
Notice that the ranges for the mixing angles shown in the plots are exactly those 
adopted in the scatter plots (cf. eqs. (\ref{rangetheta13}), (\ref{rangetheta12}) and 
(\ref{rangetheta23})). 
We find a perfect agreement with the results of \cite{SO10lep2} (another reason
not to show again the results for IO). However, due to an improved computing procedure, 
we could generate hundred times higher number of points. 
In this way the borders of the allowed regions are very sharply determined, 
as it can be noticed from the figure. 
We fully confirm and strengthen all results found in \cite{SO10lep2}
(we recall that all constraints are obtained for $\a_2 =5$). 
Let us highlight some of the main features of the found solutions. 
\begin{figure}
\begin{center}
\psfig{file=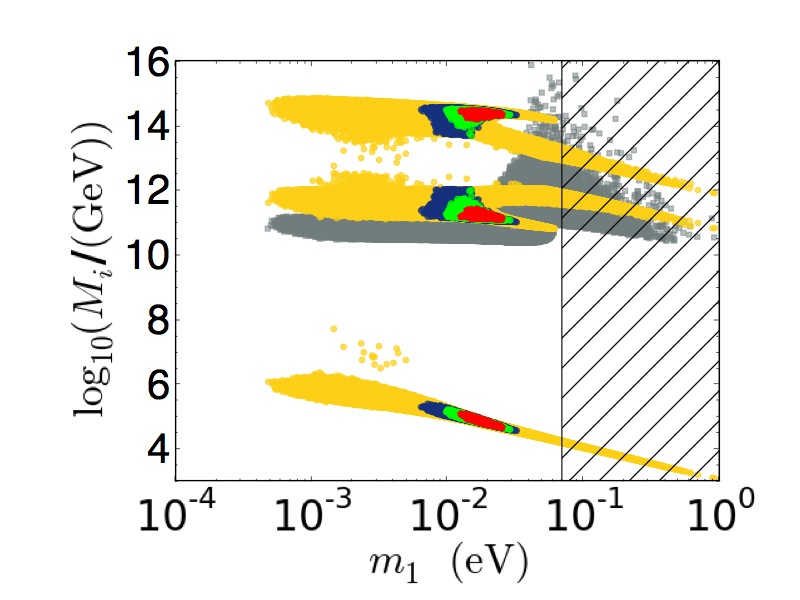,height=50mm,width=56mm}
\hspace{-7mm}
\psfig{file=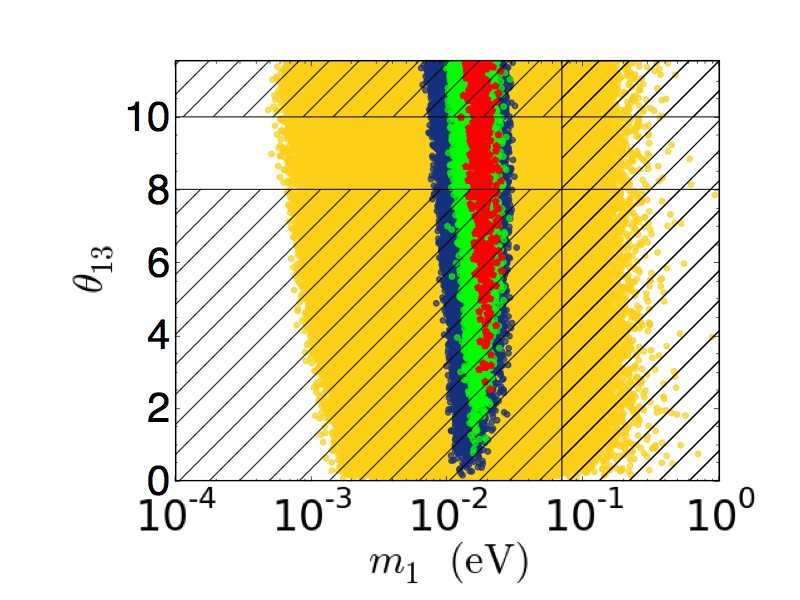,height=50mm,width=56mm}
\hspace{-7mm}
\psfig{file=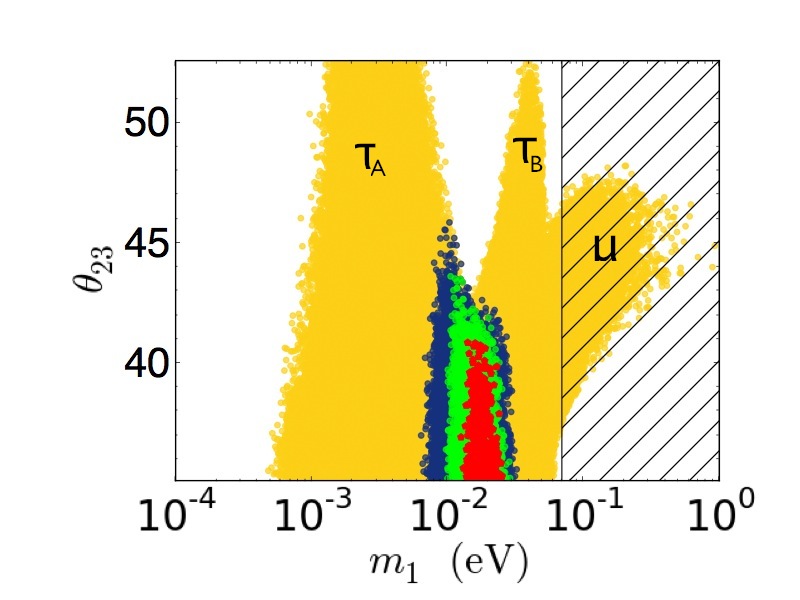,height=50mm,width=56mm} \\
\psfig{file=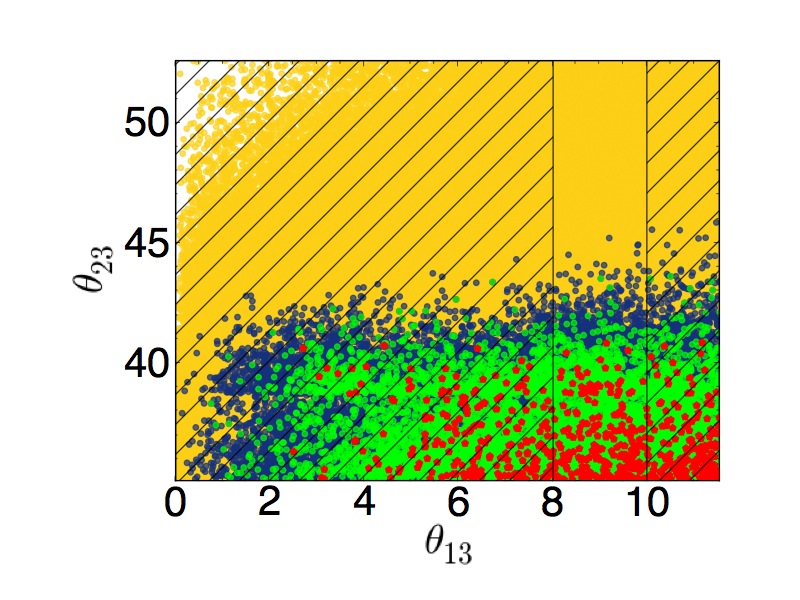,height=50mm,width=56mm}
\hspace{-7mm}
\psfig{file=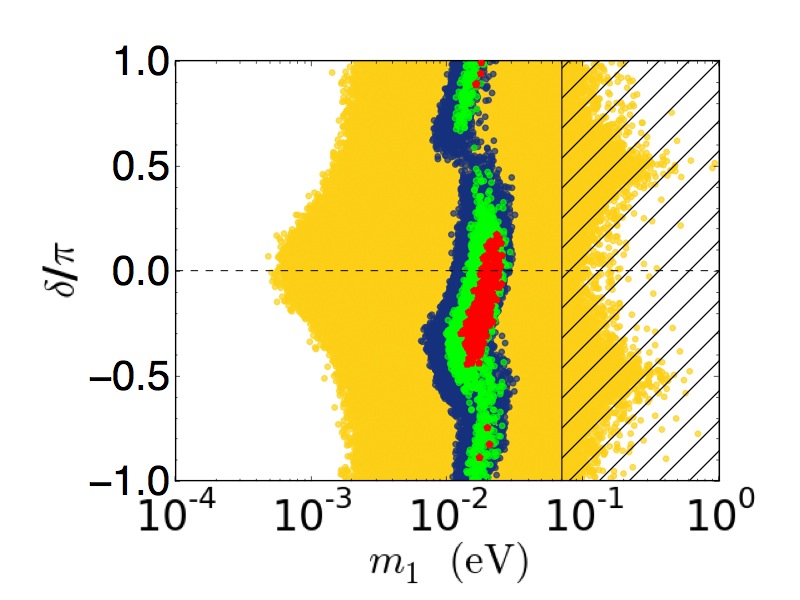,height=50mm,width=56mm}
\hspace{-7mm}
\psfig{file=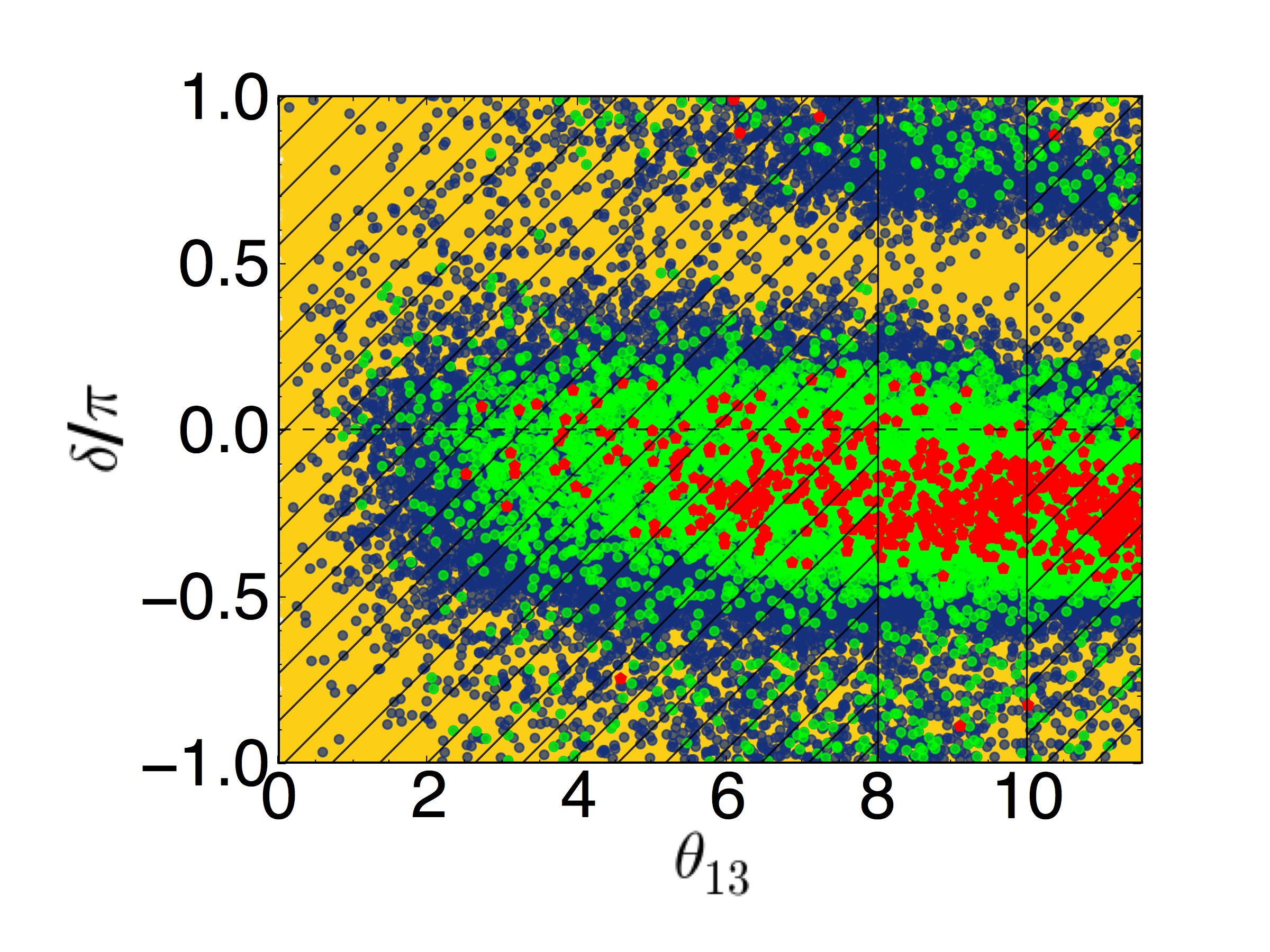,height=50mm,width=56mm} \\
\psfig{file=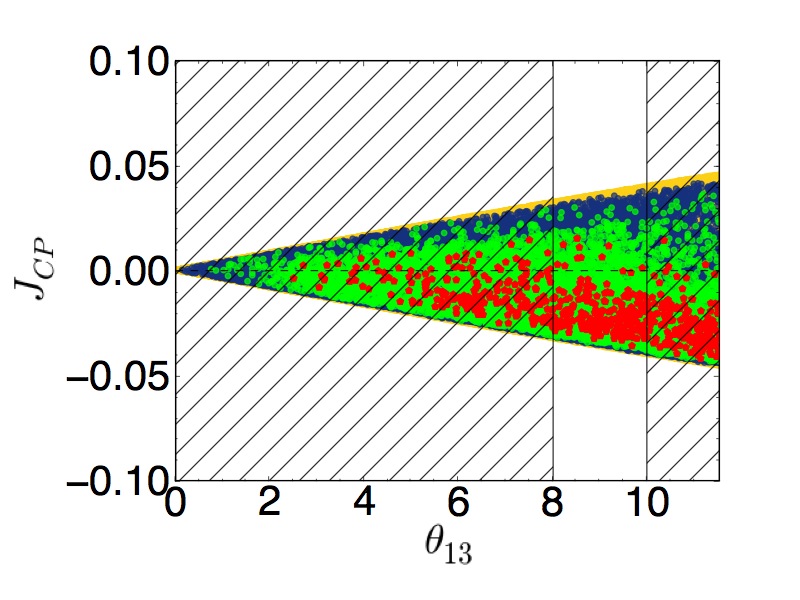,height=50mm,width=56mm}
\hspace{-7mm}
\psfig{file=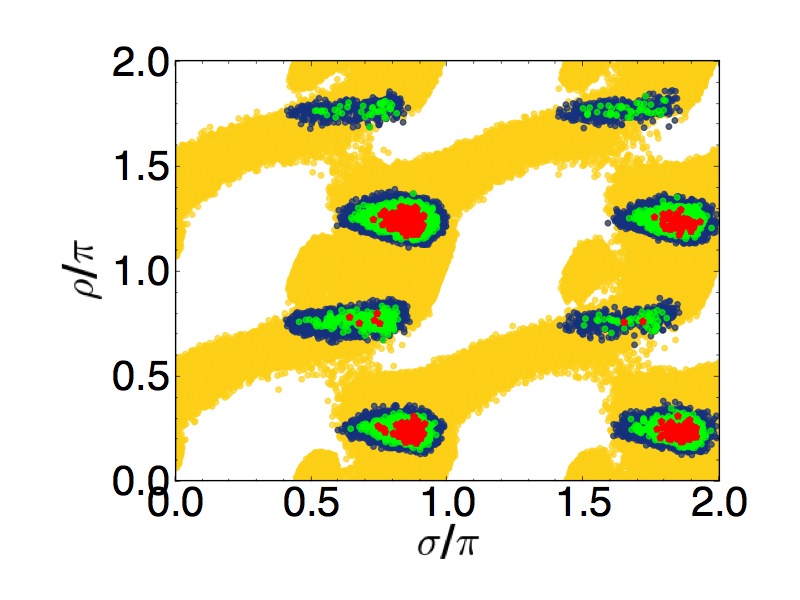,height=50mm,width=56mm}
\hspace{-7mm}
\psfig{file=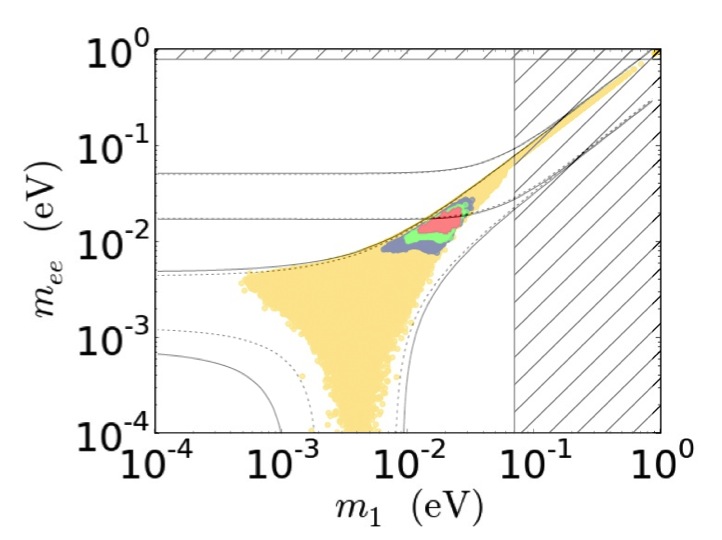,height=50mm,width=56mm}
\end{center}
\vspace{-11mm}
\caption{Scatter plots in the parameter space projected on different planes for NO and $\a_2 = 5$. 
The mixing angles vary within the experimental $2\s$ ranges 
(cf. eqs.~(\ref{rangetheta13}), (\ref{rangetheta12}) and (\ref{rangetheta23})). 
The dashed regions indicate either the values of $m_1$ excluded by 
the CMB upper bound (cf. eq.~(\ref{upperboundm1})), or the values of $m_{ee}$
excluded by $0\nu\b\b$ experiments, or
the values of $\theta_{13}$ excluded by current 
determination (cf. eq.~(\ref{newrangetheta13})). 
All points satisfy the ($\simeq 2\s$) successful leptogenesis bound 
$\eta_B > \eta_B^{\rm CMB} > 5.9 \times 10^{-10}$. 
They are obtained imposing both $SO(10)$-inspired and strong thermal conditions 
for different values of the pre-existing asymmetry. The yellow points correspond to an initial
vanishing asymmetry (the strong thermal condition is ineffective). 
 The blu, green and red points 
are obtained respectively for an initial  value of the  
pre-existing asymmetry $N_{B-L}^{\rm p,i}=10^{-3},10^{-2},10^{-1}$.
In the bottom right panel the dashed (solid) black lines indicate the general (no leptogenesis) allowed bands, both for NO and IO, in
the plane $m_{ee}$ vs. $m_1$
for $\theta_{13}$ in the range eq.~(\ref{newrangetheta13}) (eq.~(\ref{rangetheta13})).}
\label{constrNO}
\end{figure}

\subsubsection{Existence of three types of solutions}

We confirm 
that there are only three types of solutions leading to 
successful $SO(10)$-inspired leptogenesis \cite{SO10lep1,SO10lep2}. We will refer to them
as $\tau_A$, $\tau_B$ and $\mu$-type solutions: the $\tau_A$ and $\tau_B$ types being
characterised by $K_{1\t}\lesssim 1$, implying a tauon-flavour dominant contribution
to the final asymmetry, while the $\mu$-type being characterised by $K_{1\m}\lesssim 1$ and,
therefore, muon dominated. These three types result  respectively into three sets of (partly overlapping) allowed regions, that are now, in our new analysis, quite clearly distinguishable in  two of the plots in Fig.~\ref{constrNO}: in the upper left panel showing 
the constraints in the plane $m_1-M_i$  and in the upper-right panel 
showing the constraints in the $m_1-\theta_{23}$ plane. In this case it should be
noticed how for values $\theta_{23}\gtrsim 45^{\circ}$ the three types 
correspond to well distinguished (non-overlapping) allowed regions.  

In Figure 2 we plot, versus $m_1$, different relevant quantities associated to the three specific sets of parameters specified in the figure caption and realising the three different types: the left panels refer to a $\tau_A$-type solution, the central panels to a $\tau_B$-type solution and the right panels to a $\mu$-type solution. In the bottom panels we plot the contributions to the final asymmetry 
$\eta_B$ from the three different flavours and it can be seen how indeed the $\tau_A$ and the 
$\tau_B$-type solutions are tauon-dominated while the $\mu$-type solution is muon dominated. 
It can be also noticed how the $\tau_A$-type is
characterised by $K_{2\tau}\gg 1$ and $K_{1e}\lesssim 1$ for $m_1 \lesssim 10\,{\rm meV}$,
while $K_{1e} \gg 1$ for $m_1 \gtrsim 10\,{\rm meV}$. On the other hand the $\tau_B$-type is 
characterised by $K_{1e} \gg 1$ for any value of $m_1$ while $1\gtrsim K_{2\t}\gtrsim 20$.
These features will be relevant  when we will impose the strong thermal 
condition in order to understand what kind of subset of the solutions satisfy also
 this additional important property. 
\begin{figure}
\begin{center}
\psfig{file=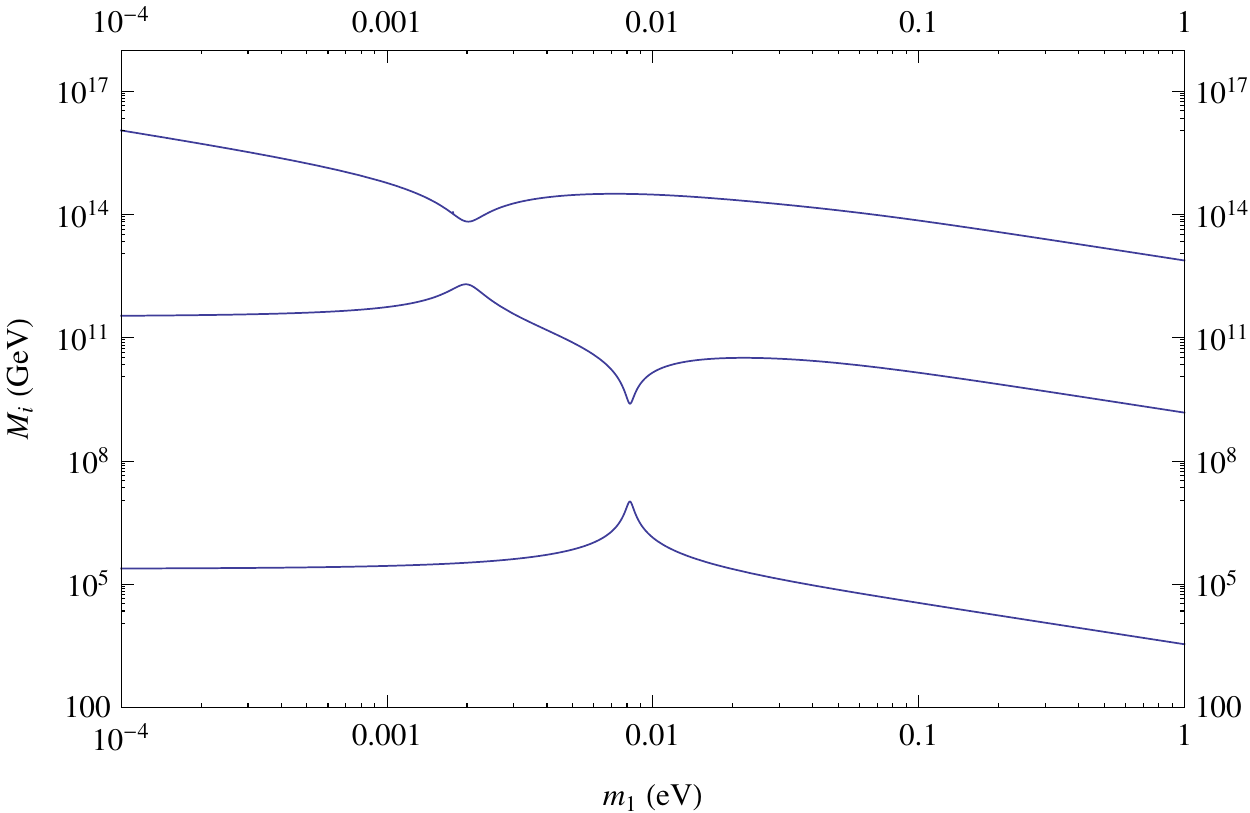,height=38mm,width=45mm}
\hspace{3mm}
\psfig{file=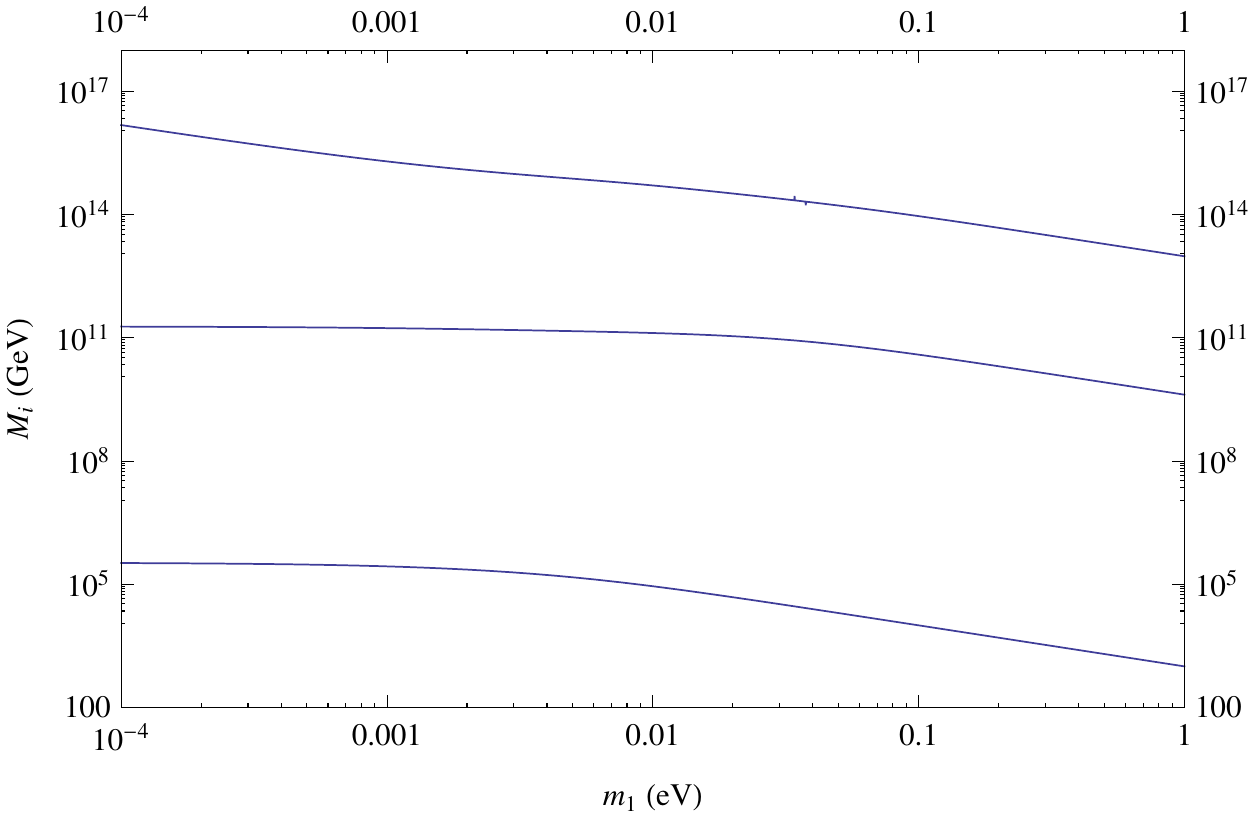,height=38mm,width=45mm}
\hspace{3mm}
\psfig{file=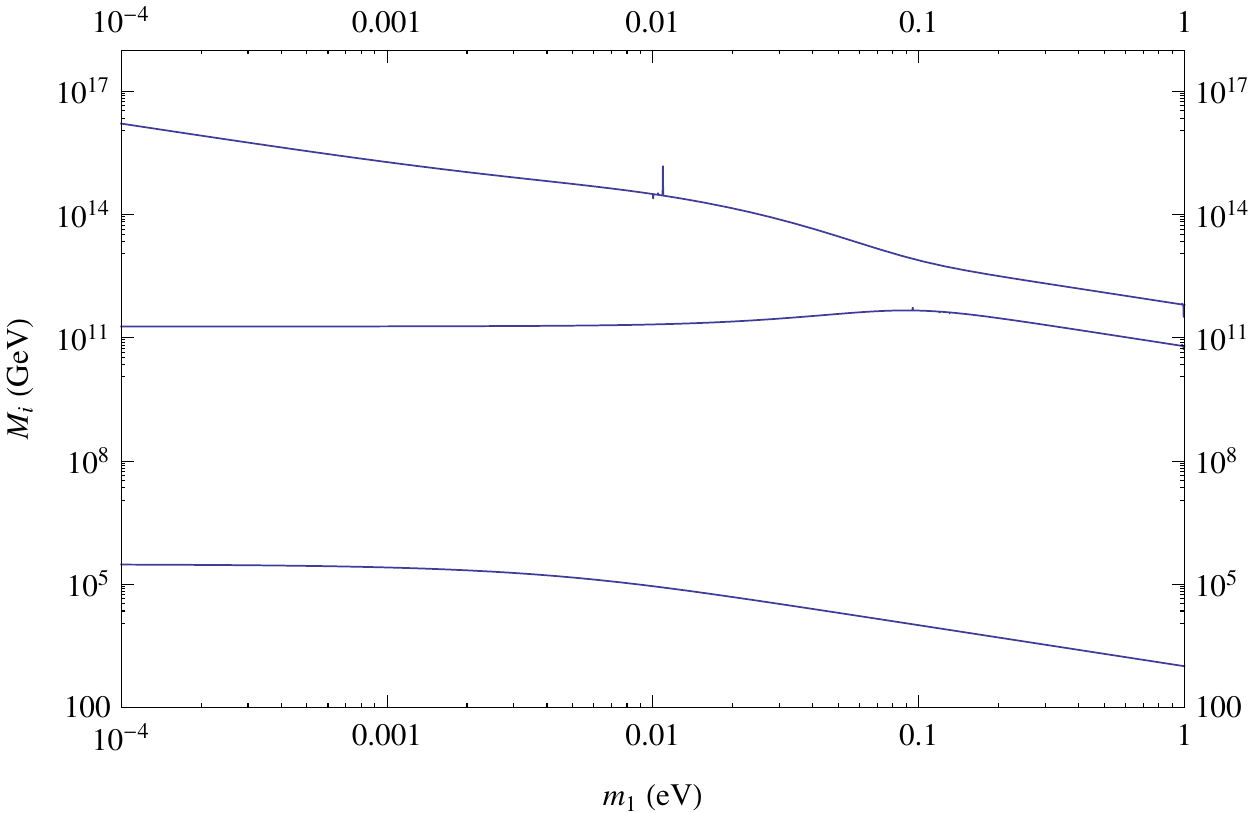,height=38mm,width=45mm} \\
\psfig{file=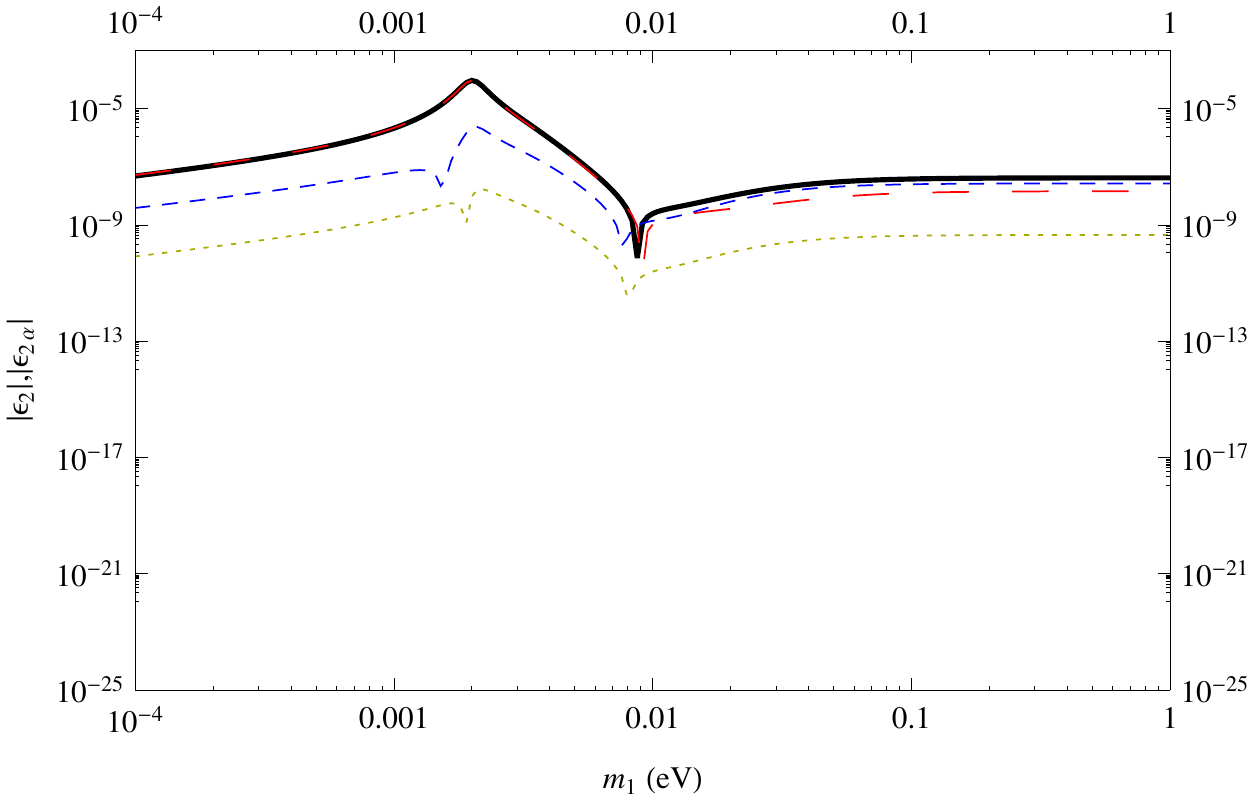,height=38mm,width=45mm}
\hspace{3mm}
\psfig{file=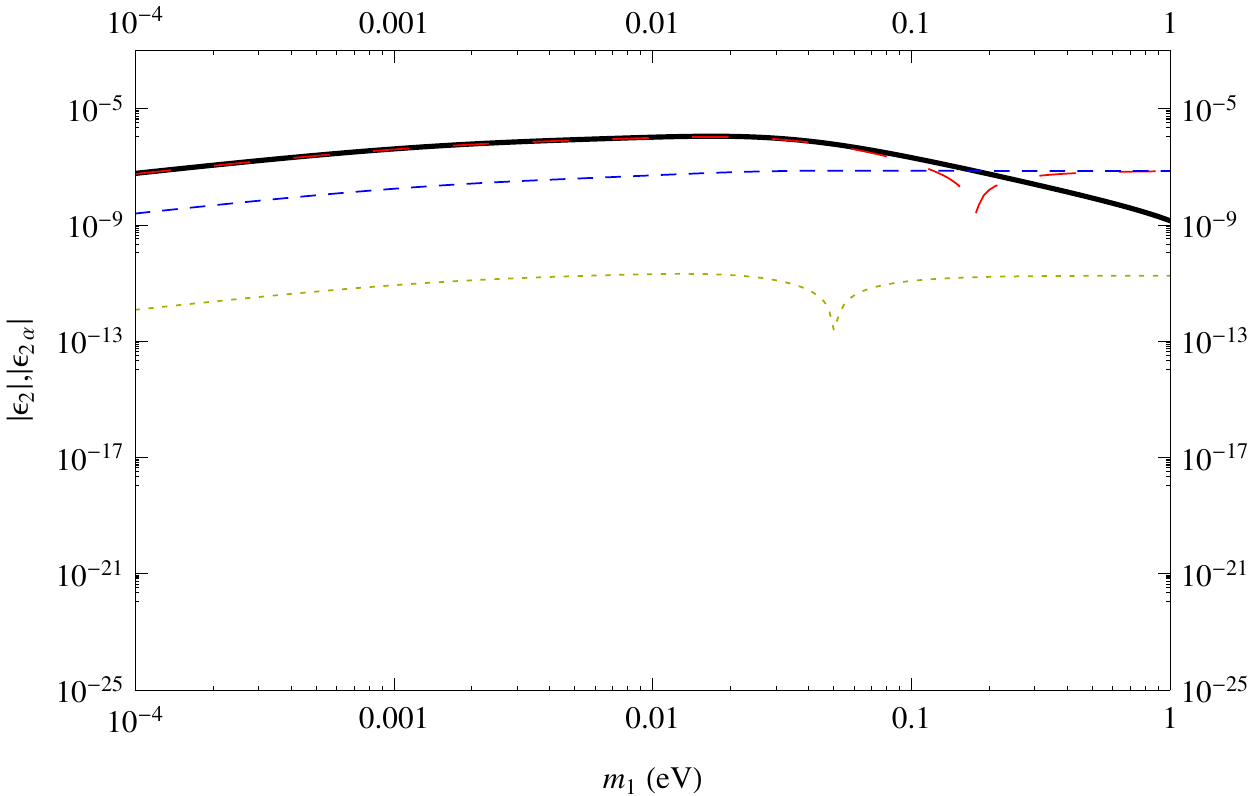,height=38mm,width=45mm}
\hspace{3mm}
\psfig{file=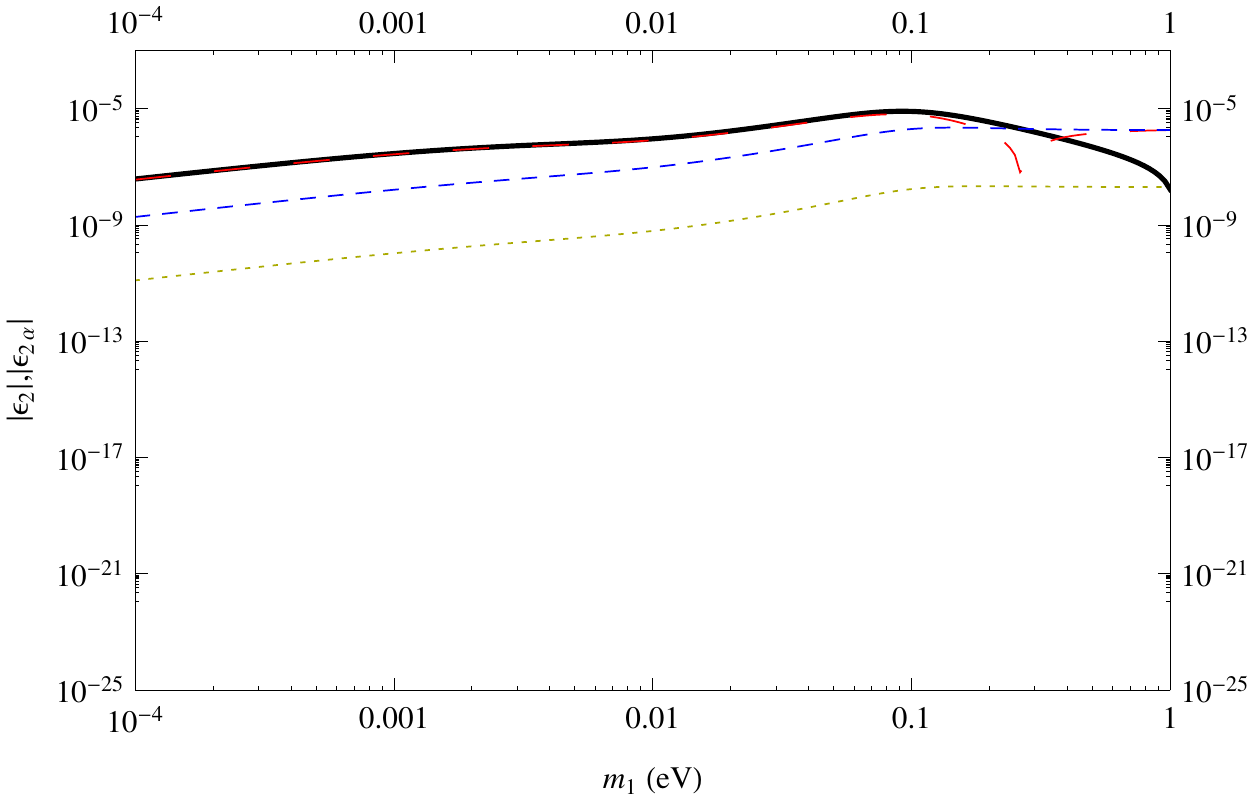,height=38mm,width=45mm} \\
\psfig{file=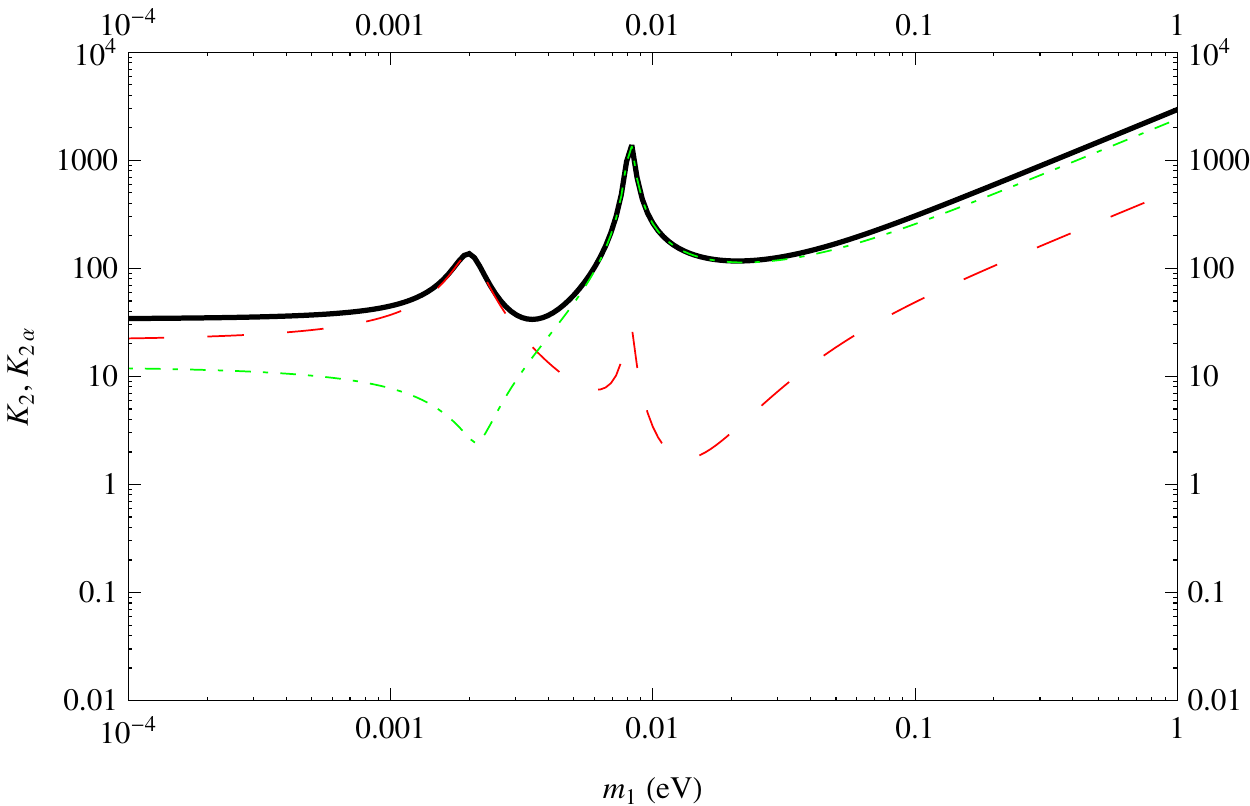,height=38mm,width=45mm}
\hspace{3mm}
\psfig{file=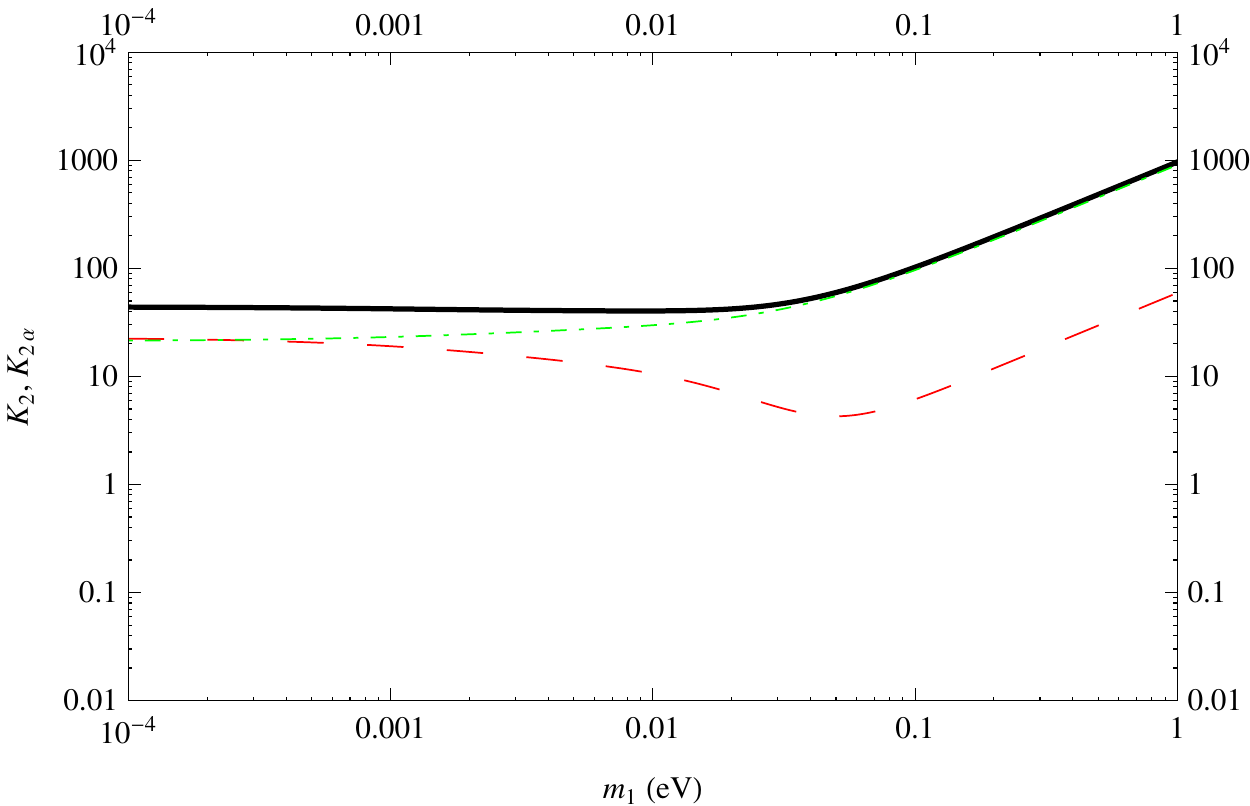,height=38mm,width=45mm}
\hspace{3mm}
\psfig{file=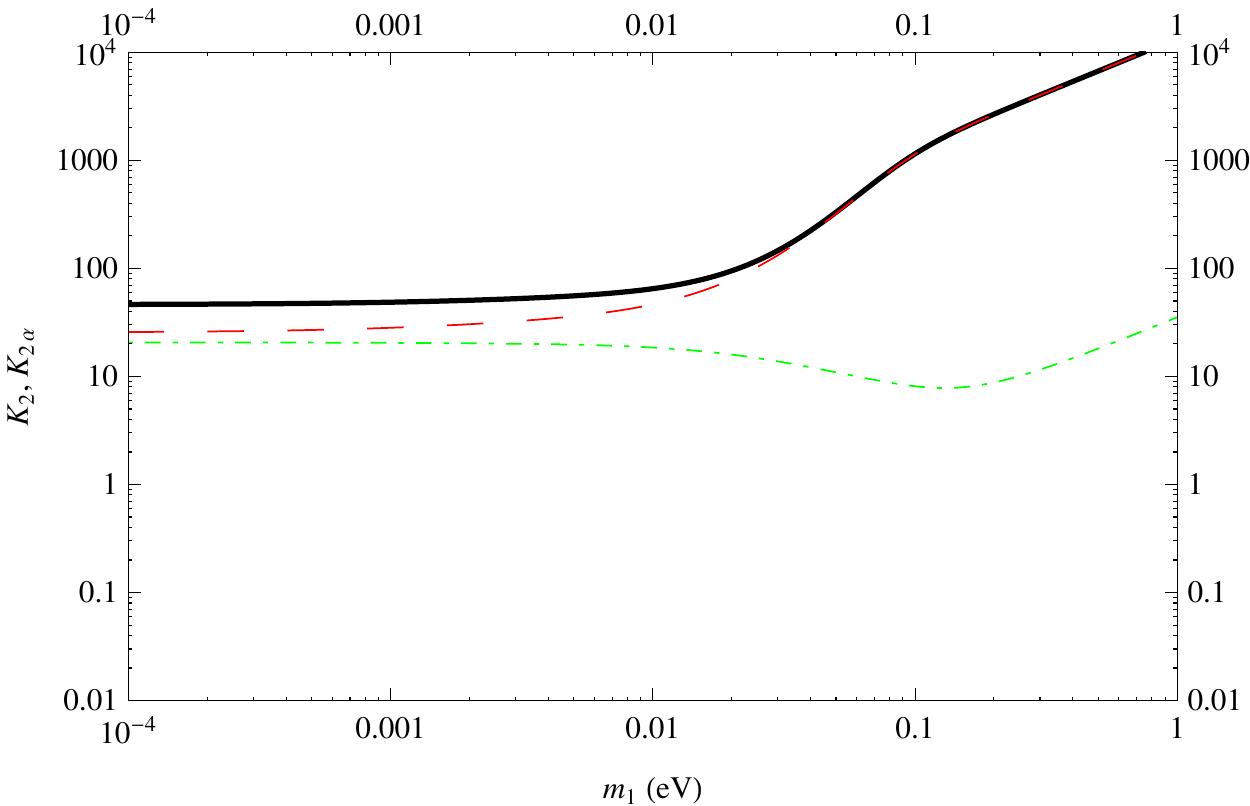,height=38mm,width=45mm} \\
\psfig{file=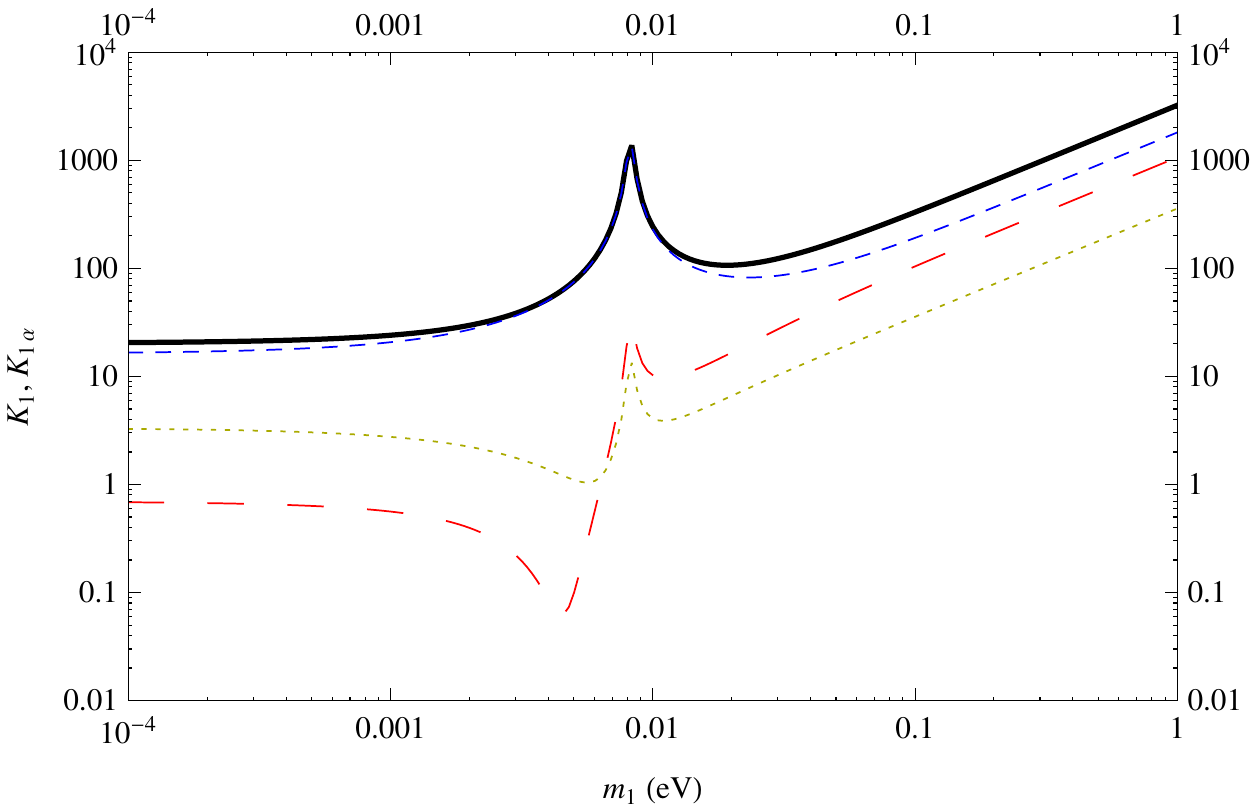,height=38mm,width=45mm}
\hspace{3mm}
\psfig{file=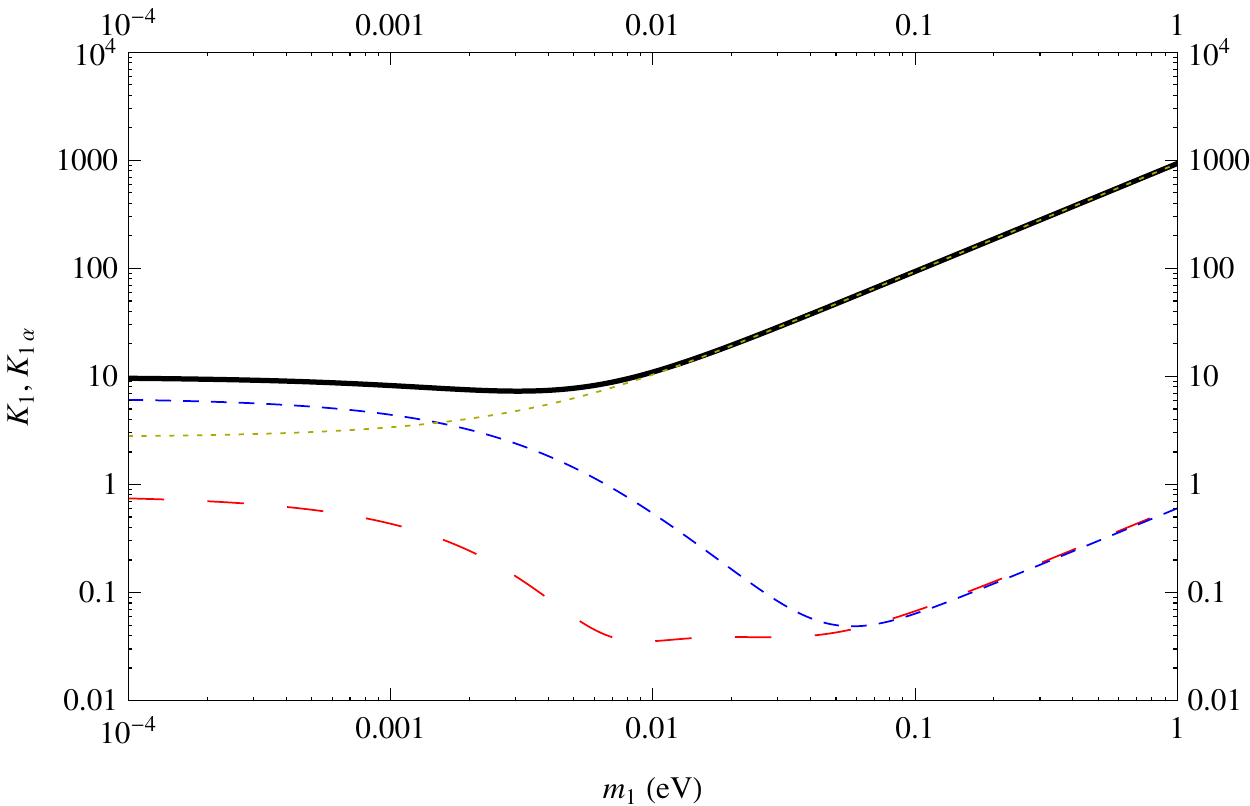,height=38mm,width=45mm}
\hspace{3mm}
\psfig{file=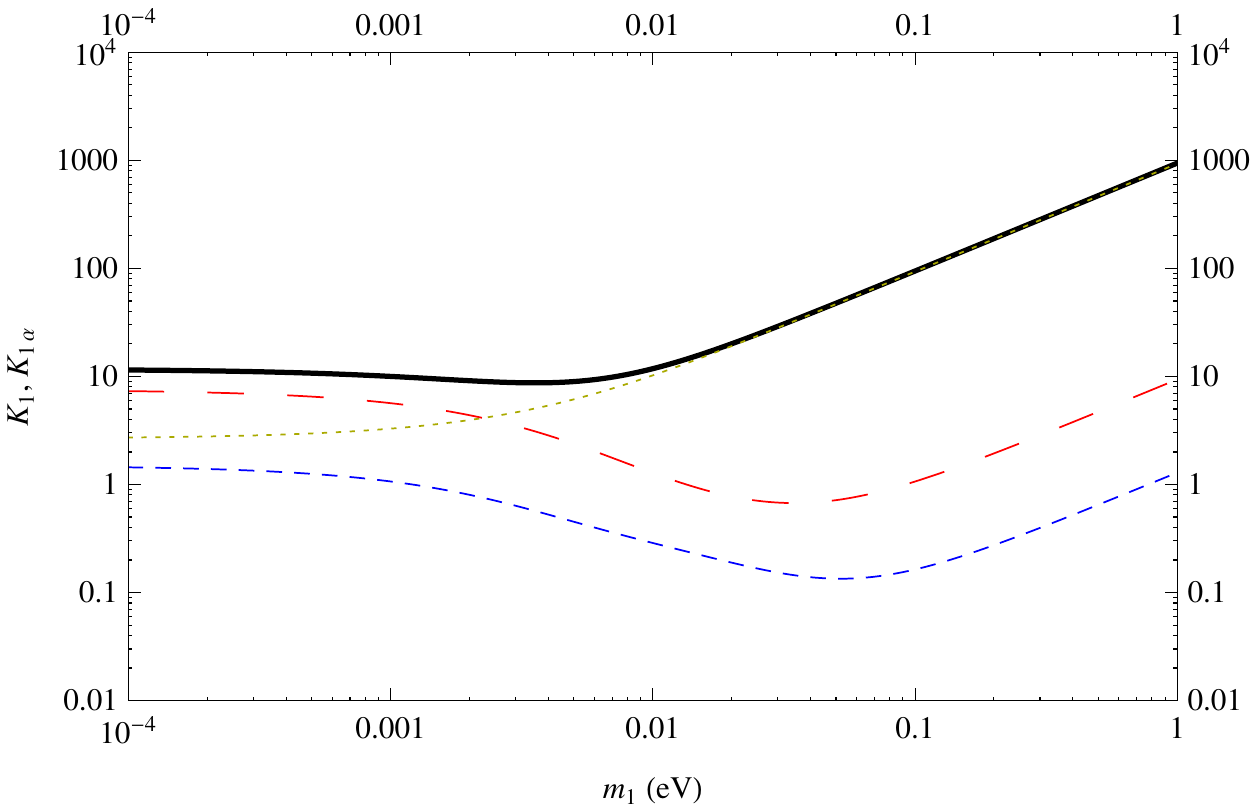,height=38mm,width=45mm} \\
\psfig{file=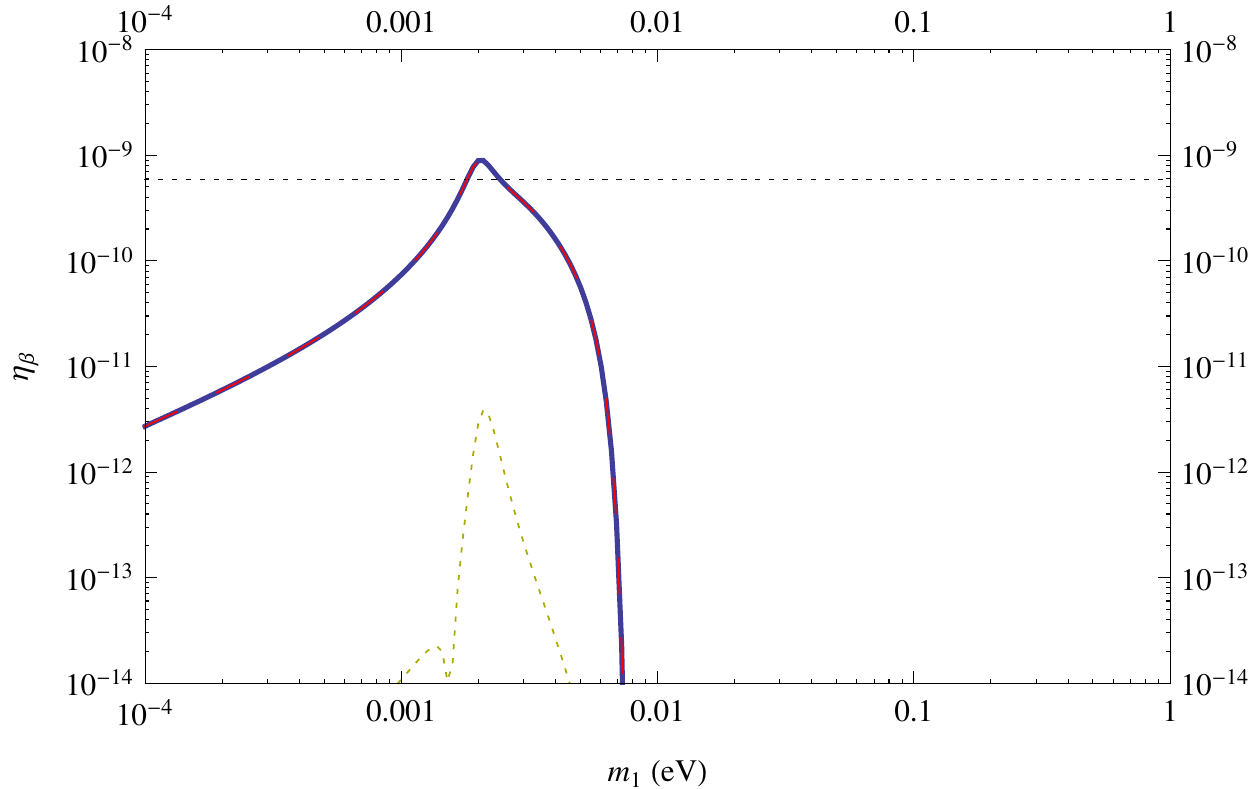,height=38mm,width=45mm}
\hspace{3mm}
\psfig{file=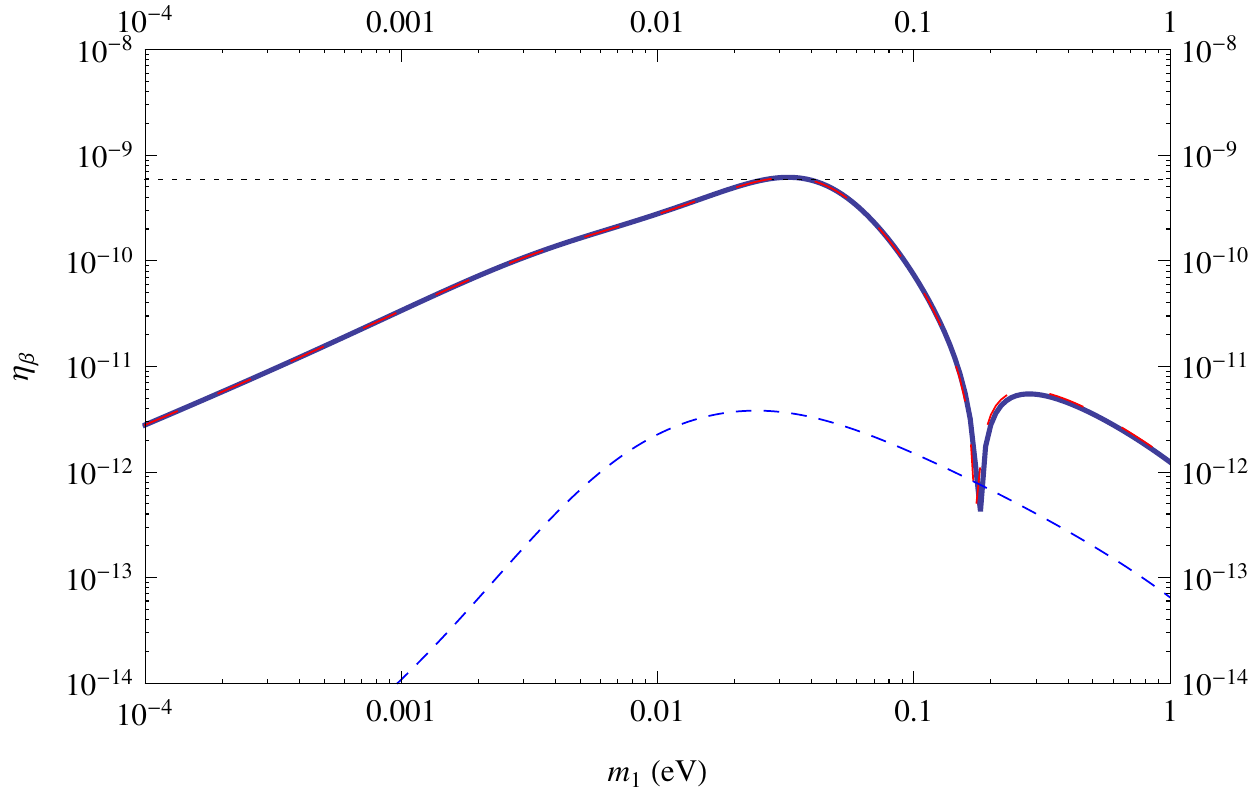,height=38mm,width=45mm}
\hspace{3mm}
\psfig{file=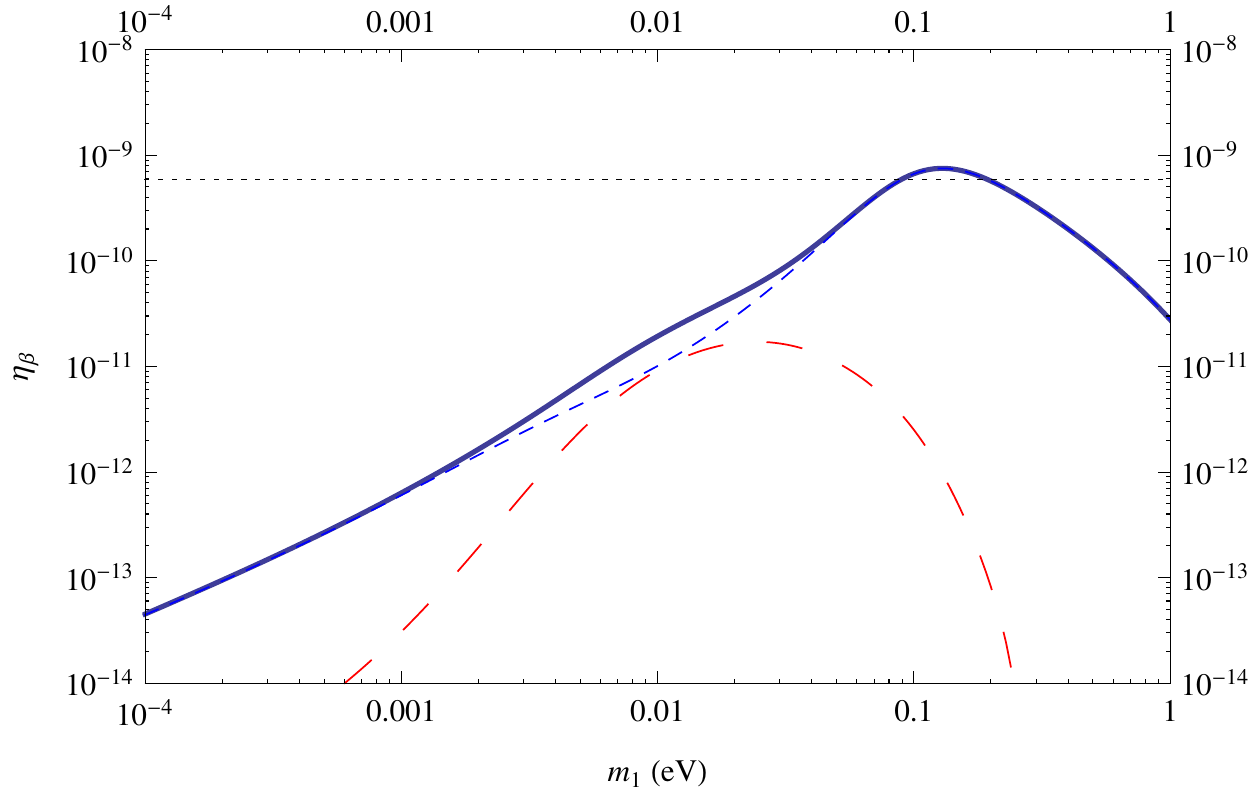,height=38mm,width=45mm}
\end{center}
\vspace{-10mm}
\caption{
Plots of the relevant quantities  for the three following sets of parameters:
$\theta_{13}=(7.9^{\circ},2.8^{\circ},1.4^{\circ})$,
$\theta_{12}=(34^{\circ},34.6^{\circ},36^{\circ})$,
$\theta_{23}=(50^{\circ},48^{\circ},46^{\circ})$,
$\delta=(-0.29,-0.28,0.56)$,
$\rho=(1.4,6.24,3.17)$,
$\s=(3.14,6.02,4.75)$,
$\theta_{13}^L=(0.14^{\circ},0.14^{\circ},0.037^{\circ})$,
$\theta_{12}^L=(6.0^{\circ},0.41^{\circ},5.8^{\circ})$,
$\theta_{23}^L=(2.1^{\circ},2.1^{\circ},1.24^{\circ})$,
$\rho_L=(1.15,0.68,5.1)$, $\sigma_L=(3.7,3.24,2.4)$,
corresponding respectively to $\tau_A$, $\tau_B$ and $\mu$-type solutions.
The long-dashed red lines correspond to $\a=\t$,
the dashed blue lines to $\a=\m$ and the short-dashed dark yellow lines to $\a=e$.
For all three cases $(\a_1,\a_2,\a_3) = (1,5,1)$, though notice that, 
except for the three RH neutrino masses,
all quantities are independent of $\a_1$ and $\a_3$.}
\end{figure}
Let us now discuss the main features of the constraints on the low energy neutrino parameters
in the light of these new results resulting from a much higher amount of solutions
(about two orders of magnitude) compared to the previous ones obtained in \cite{SO10lep2}.

\subsubsection{Lower bound on $m_1$}

First of all we confirm the existence of a lower bound  $m_1 \gtrsim 5 \times 10^{-4}\,{\rm eV}$.  This can be considered quite a conservative and robust lower bound from $SO(10)$-inspired leptogenesis. The origin of this lower bound is due to the fact that  for $m_1 \ll 10^{-3}\,{\rm eV}$ 
one has $M_3 \gg 10^{15}\,$GeV and consequently all the $N_2$ $C\!P$ asymmetries get suppressed 
\cite{SO10lep1,SO10lep2}. 
A new feature, that is interesting to notice in the light of the $\theta_{13}$ measurement,
opening prospects for a measurement of the Dirac phase $\d$, 
is that  the lower bound on $m_1$ depends on $\d$ and in particular the lowest value, 
$m_1 \simeq 5\times 10^{-4}$eV, is saturated for 
$\delta\simeq 0$, while for $|\d|\gtrsim \pi/2$, as very weakly supported by current global analyses,  
one has $m_1 \gtrsim 10^{-3}\,$eV. Therefore, in these models, a determination of $\d$ shows an interesting  interplay with absolute neutrino mass scale experiments. 
 
\subsubsection{Upper bound on $\theta_{23}$ for quasi-degenerate neutrinos} 

Another interesting constraint of this scenario, found in \cite{SO10lep2} and confirmed by our 
analysis, is  the existence of an upper bound on $\theta_{23}$ for sufficiently 
large values of $m_1$, the $\mu$ type region. 
Our new results confirm this constraint as well. This is now determined quite accurately and 
precisely: $\theta_{23}\lesssim 48^{\circ}$ for $m_1 \gtrsim 60\,{\rm meV}$. It should be 
noticed  that the new upper bound from Planck data (cf. eq. (\ref{upperboundm1})) now
basically almost completely rules out this $\mu$ type region at high $m_1$ values. 

\subsubsection{Majorana phases}

As it can be seen in the lower central panel of Fig.~1, the Majorana phases
cannot have arbitrary values but there are some quite large excluded regions. 
Our results for $\a_2 =5$ are fully compatible with the results found in 
\cite{SO10lep2}. In \cite{SO10lep2} results were found for $\alpha_2=4,5$ and showed that the Majorana phases tend to cluster dominantly around disconnected regions for values $\rho\simeq (n+1/2)\,\pi$ and $\sigma \simeq n\,\pi$ and sub-dominantly around regions for $\rho,\sigma \simeq n\,\pi$. Now, since we have found a much greater amount of solutions, the regions are sharply determined
and for $\alpha_2 =5$ the allowed regions are connected. However, the bulk of points still 
falls around  the same values found in \cite{SO10lep2}. 
The differences are then just simply to be ascribed to the 
much higher number of determined points. 

\subsubsection{Dirac phase and $J_{CP}$}

The results for the Dirac phase and for the Jarlskog invariant,
 \be
J_{CP}={\rm Im}[U_{\mu 3}\,U_{e 2}\,U^{\star}_{\m 2}\,U^{\star}_{e 3}] 
=c_{12}\,s_{12}\,c_{23}\,c^2_{13}\,s_{13}\,\sin\d \,  ,
\ee
do not show any constraint and, in particular, no preference for the  sign. 
Compared to the results found in \cite{SO10lep2} we have just found a trivial 
bug in the plot of $J_{CP}$ vs. $\theta_{13}$ shown in \cite{SO10lep2} where $\theta_{13}$ 
was displayed for $\theta_{13} \lesssim 11.5^{\circ}$ in radians instead of
degrees as indicated.  

\subsection{Are the low energy neutrino data pointing in the right direction?}

 A particularly interesting test was performed in \cite{SO10lep2}. 
The allowed regions  for the low energy neutrino parameters 
were also determined without imposing any restriction from low energy neutrino
experiments. In this way one can test how predictive the scenario is and whether
the agreement with current experimental data is particularly significant.
We have repeated this test  and the results  are shown in Fig.~3. 
\begin{figure}
\begin{center}
\psfig{file=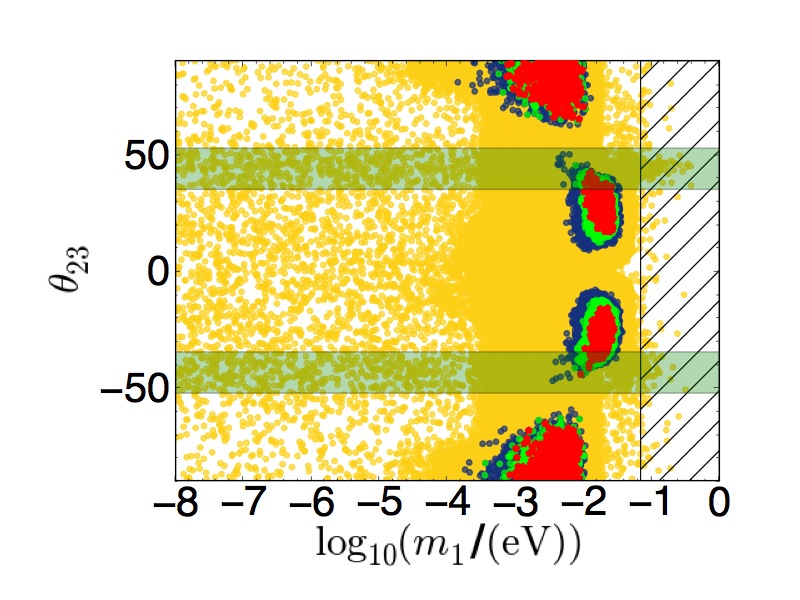,height=50mm,width=55mm}
\hspace{-5mm}
\psfig{file=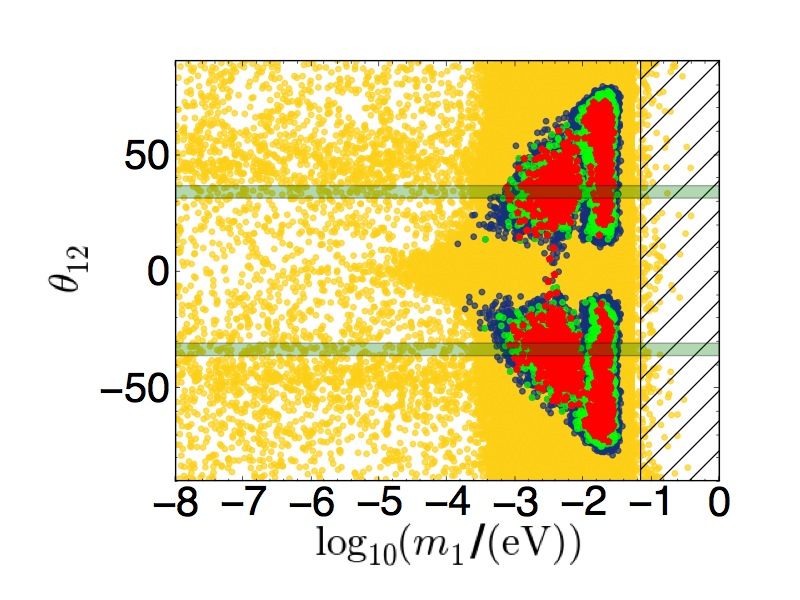,height=50mm,width=55mm}
\hspace{-5mm}
\psfig{file=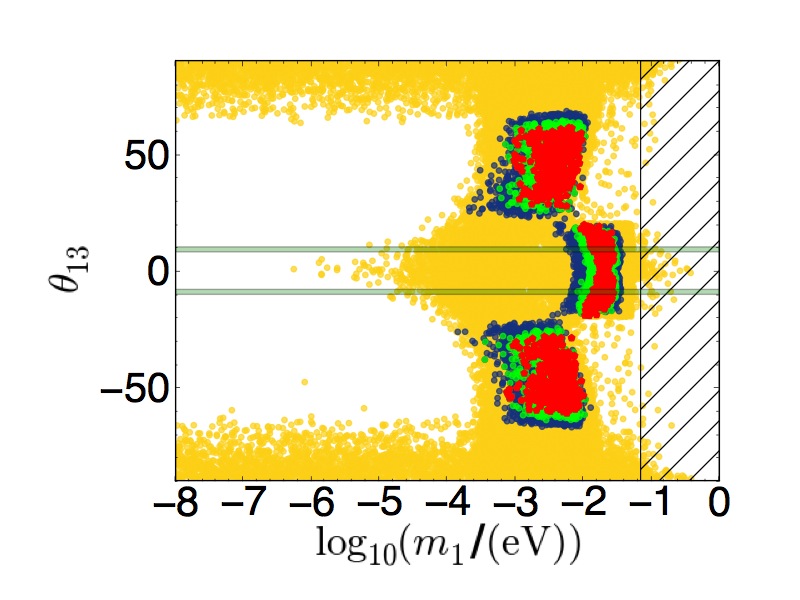,height=50mm,width=55mm} \\
\psfig{file=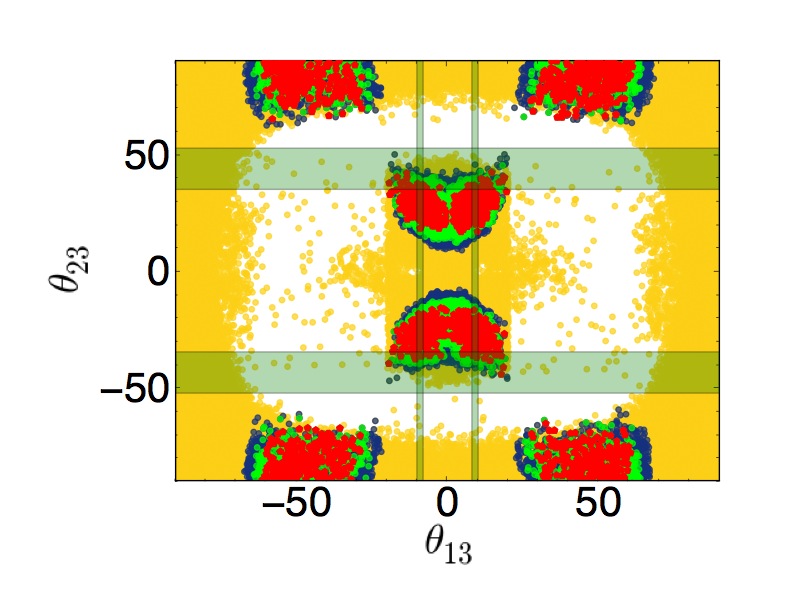,height=50mm,width=55mm}
\hspace{-5mm}
\psfig{file=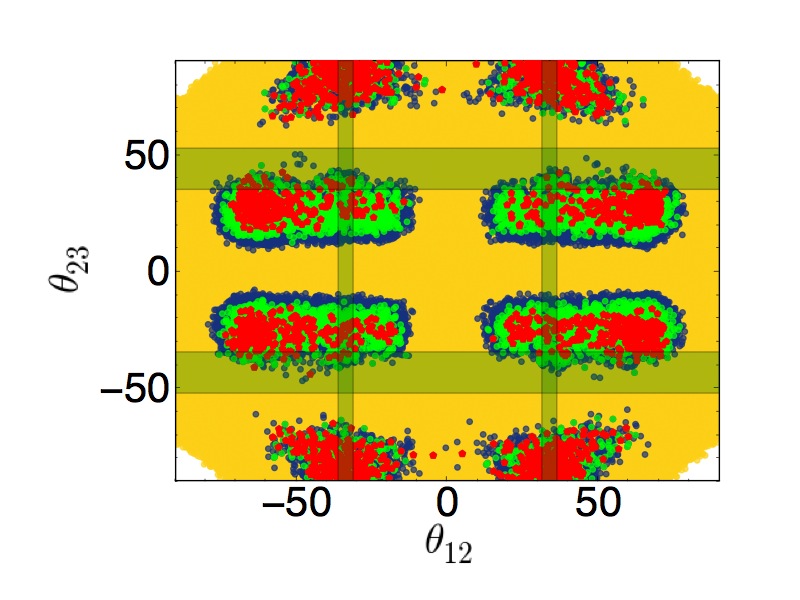,height=50mm,width=55mm}
\hspace{-5mm}
\psfig{file=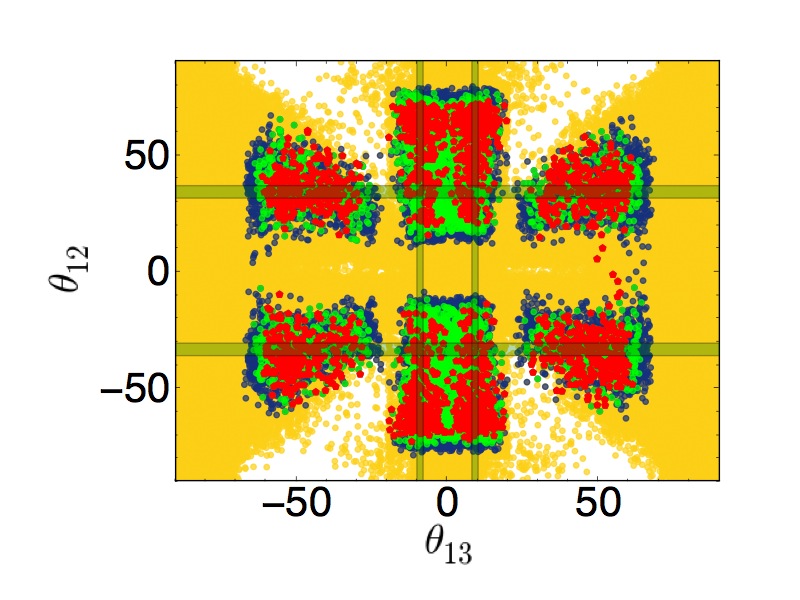,height=50mm,width=55mm} 
\end{center}
\caption{Scatter plots as in Fig.~1 but without imposing 
the experimental information on mixing angles 
from neutrino oscillation experiments. Mixing angles are shown in the range $[-90^{\circ},90^{\circ}]$
since the addition of a Majorana mass term with three RH neutrinos introduces potentially a sign sensitivity
(differently from neutrino oscillations probabilities). In our case since the asymmetry is generated by just one RH neutrino there is no sign sensitivity and the regions at negative values just mirror those
at positive values. 
}
\label{blueplots}
\end{figure}
Also in this case we confirm the results of \cite{SO10lep2}. The huge amount of points now 
clearly determines the existence of  excluded regions.  The fact that the experimental results 
(the green bands) fall in the allowed regions represents a positive test of the model. 
In particular, it is quite interesting to notice (see yellow points in the left bottom panel) 
that the measured value of $\theta_{13}$ implies that the atmospheric
mixing angle range $50^{\circ} \lesssim \theta_{23} \lesssim 70^{\circ}$ is excluded or
that for the measured values of $\theta_{23}$ the range of values 
$20^{\circ} \lesssim \theta_{13} \lesssim 60^{\circ}$ is excluded. 

However, the allowed (yellow) regions  cover a large portion of the parameter space and, therefore, the test is not particularly statistically significant. In other words, neutrino data could have already 
ruled out $SO(10)$-inspired leptogenesis, but the probability that they just by chance fall within the $SO(10)$ allowed regions is too high to draw any  statistically significant conclusion. Indeed, if one looks at the mixing angles, one could say that there was roughly just a $50\%$ probability that the data could exclude $SO(10)$-inspired leptogenesis.  
As we will see, the situation drastically changes when  the strong thermal leptogenesis condition is further imposed.

\section{The strong thermal leptogenesis condition}

We have so far assumed that the observed asymmetry is entirely generated
by leptogenesis. However, there are other possible external mechanisms,
such as gravitational baryogenesis \cite{gravity}   and Affleck-Dine baryogenesis \cite{affleckdine},
able to generate an asymmetry prior the onset of leptogenesis.
In particular, so called grand unified  baryogenesis models \cite{GUTB},
are particularly relevant within our context, since this would be 
quite a natural and extensively studied possibility arising just within 
grand unified $SO(10)$ models  inspiring the scenario we are discussing.
Moreover they are particularly well motivated considering the 
large initial temperatures  required by minimal thermal leptogenesis
(though a non-thermal production would be also plausible). 

These potential sources would compete with leptogenesis and in general, at the large initial reheat temperature required by (minimal) thermal leptogenesis and in particular
by $SO(10)$-inspired leptogenesis,  $T_{\rm RH} \gtrsim 10^{11}\,{\rm GeV}$,
they would typically produce a pre-existing asymmetry well above the observed one,
up to values ${\cal O}(0.1)$. 

Clearly one possibility would be to assume that at the end of the inflationary stage
any asymmetry was completely erased and that no mechanism had
efficiently  produced a pre-existing asymmetry  prior the onset of leptogenesis. 
However, it would be quite attractive, and the constraints on low energy neutrino
parameters much more significant, if the same processes involving RH neutrinos could wash-out 
any pre-existing asymmetry and at the same time produce a final value of the asymmetry 
independent of the initial RH neutrino abundances (strong thermal leptogenesis condition). 
This would be an  analogous situation compared to what happens in Standard Big Bang Nucleosynthesis.

Let us translate this request in quantitative terms.  In the presence of an initial pre-existing
asymmetry, the predicted value of the final $B-L$ asymmetry  would be in general the sum of the  residual value of the pre-existing asymmetry, $N_{B-L}^{\rm p,f}$, plus the genuine leptogenesis contribution from RH neutrino decays, $N_{B-L}^{\rm lep,f}$, or, in terms of the
baryon-to-photon number ratio at the present time,
\be
\eta_{B} = \eta_{B}^{\rm p} +\eta_{B}^{\rm lep}  \,  ,
\ee
where $\eta_{B}^{\rm p}$ and $\eta_{B}^{\rm lep}$ are simply 
given by the eq.~(\ref{etaNBmL}) by replacing $N^{\rm f}_{B-L}$ 
respectively with $N_{B-L}^{\rm p,f}$  and $N_{B-L}^{\rm lep,f}$.
The condition of {\em successful strong thermal leptogenesis}  
can then be expressed as \cite{problem}
\be\label{strongthcondition}
|\eta_{B}^{\rm p}| \ll \eta_{B}^{\rm lep} \simeq \eta_{B}^{\rm CMB} \, .
\ee
Within the simple vanilla leptogenesis scenario, where the asymmetry is
$N_1$-dominated and flavour effects are neglected, the relic value of the
pre-existing $B-L$ asymmetry is simply given by \cite{fy,window}
\be
N_{B-L}^{\rm p,f} = N_{B-L}^{\rm p,i}\, e^{-{3\pi\over 8}\,K_1} \, .
\ee
Considering the relation eq.~(\ref{etaNBmL}) between $N_{B-L}^{\rm f}$ and $\eta_B$,
it is, therefore, simply sufficient to impose $K_1 \gtrsim 15 + \ln N_{B-L}^{\rm p,i}$
to enforce the strong thermal leptogenesis condition. 

  When flavour effects are taken into account, and considering hierarchical RH neutrino mass patterns,
 as we are considering within $SO(10)$-inspired models, strong thermal leptogenesis can be realised only within a tauon-dominated $N_2$-dominated scenario where the dominant contribution to the asymmetry is in the tauon flavour \cite{problem}.  This is because, if $M_2 \lesssim 10^{12}\,$GeV, the tauon components
 of the lepton and anti-lepton quantum states can be measured before the asymmetry is produced
 by the $N_2$-decays. 
 In this way the $\tau$ component of the pre-existing asymmetry can be 
 washed-out by the $N_2$ inverse processes if $K_{2\tau}\gg 1$, and at the same time
 a new tauon component can be afterwards generated by the out-of-equilibrium $N_2$  decays.
 On the other hand, for a generic model, the $e$ and the $\mu$ components can be fully washed out only in the three-flavour regime by the $N_1$ wash-out, i.e. after the $N_2$ leptogenesis, so that they cannot be afterwards regenerated  contributing to $\eta_B^{\rm lep}$.
 
  As we have seen, the tauon dominance condition is naturally satisfied by 
  two of the three types of solutions found in $SO(10)$-inspired leptogenesis.
  \footnote{Notice that this does not happen by chance. Since one assumes the hierarchy of neutrino Yukawa
  couplings like for up quarks (and similarly for the charge leptons) Yukawa couplings,
  the fact that the tauon flavour component is the first to become incoherent at $T \lesssim 10^{12}\,$GeV,
  the reason why one needs a tauon $N_2$-dominated scenario to satisfy the strong thermal condition,
reflects typically into a dominant tauon $C\!P$ asymmetry ($\propto \a_3^2$) and, therefore,
naturally into a tauon $N_2$-dominated scenario within $SO(10)$-inspired models.}
This, therefore,  represents quite a well motivated theoretical framework 
that is a  potential candidate to realise successful strong thermal leptogenesis .
 
 The request of the successful strong thermal condition, however, goes beyond the
 tauon dominance since it also requires quite restrictive additional conditions onto 
 the flavoured decay parameters.
 These additional conditions can be fully understood calculating explicitly the residual value of 
 the pre-existing asymmetry. 
 
 First of all, we can safely assume that the heaviest RH neutrinos are too heavy to be
 thermally produced and, therefore, they do not contribute to the wash-out of the 
 pre-existing asymmetry. This is clearly a conservative assumption since the presence 
 of the heaviest RH neutrino can only introduce an additional wash-out stage of the pre-existing
 asymmetry. However, it should be taken into account that, since $M_3 \gg 10^{12}\,$GeV
 the $N_3$ wash-out acts on a ${\ell}_3$ flavour direction and, therefore, it is in general
 not really helpful in washing out the pre-existing asymmetry, not even along the $\tau$ direction \cite{problem}.
 For this reason an inclusion of such a wash-out would not have in any case any impact on the constraints 
 we will find.

Therefore, if a pre-existing asymmetry is generated at $T \geq T_B^{\rm ext}$
by some external mechanism, at a later stage, for temperatures $T_{B}^{\rm ext}\gg T \gg M_2$, 
this simply remains constant, 
\be
N^{\rm p}_{B-L}(T \gg M_2) = N_{B-L}^{\rm p,i}   \,  .
\ee
For temperatures $10^{12}\,{\rm GeV} \gg T \gg M_2$, because of the the fast tauon lepton interactions, the quantum lepton states become an incoherent admixture of a tauon component
and of a $\tau$ orthogonal  component $\tau^{\bot}$.  
The initial pre-existing asymmetry can then be 
regarded as the sum of two components 
\be
N^{\rm p}_{B-L} = 
N_{\D_\t}^{\rm p,i} + N_{\D_{\tau^{\bot}}}^{\rm p,i} \,    \hspace{15mm}
(10^{12}\,{\rm GeV} \gg T \gg M_2)  \,  ,
\ee
related to the total pre-existing asymmetry simply by
\be
N_{\D_\t}^{\rm p,i} = p^0_{{\rm p}\t}\,  N_{B-L}^{\rm p,i}   \,  ,
\hspace{5mm} 
N_{\D_{\tau^{\bot}}}^{\rm p,i} = (1-p^0_{{\rm p}\t})\, N_{B-L}^{\rm p,i}   \,  ,
\ee 
where $p^0_{p\tau}$ is the tree-level probability of pre-existing leptons
to be in the tauon flavour. In principle, there could be  differences in the
pre-existing  lepton-antilepton flavour compositions and these would 
translate into additional opposite contributions to the flavoured asymmetries, the so called
phantom terms, that, however we can simply neglect in order to simplify the notation.
We will point out in the end  that all results are valid also in the presence of these additional terms.  
  
For temperatures $T\sim M_2$ the $N_2$ processes at the same time will generate a contribution
to $N_{B-L}^{\rm lep}$  and wash-out the pre-existing flavoured asymmetries. 
However, these processes cannot wash-out the  component 
$\t^{\bot}_{2^{\bot}}$ of the pre-existing asymmetry, 
i.e. the projection on the $e-\mu$ plane orthogonal to the heavy neutrino lepton flavour ${\ell}_2$. 
At the end
of this stage, at $T \simeq T_{B2}\simeq M_2/z_{B2}$, the residual values of the pre-existing asymmetries will be then given by three components,
\bea
N_{\D_\t}^{\rm p}(T_{B2})  & = & 
p^0_{{\rm p}\t}\,  e^{-{3\pi\over 8}\,K_{2\t}} \, N_{B-L}^{\rm p,i} \,  ,\\ \nonumber
N_{\D_{\tau^{\bot}_2}}^{\rm p}(T_{B2}) & = & 
(1-p^0_{{\rm p}\t})\,p^0_{p\t^{\bot}_2}\,e^{-{3\pi\over 8}\,(K_{2e}+K_{2\m})}  \, N_{B-L}^{\rm p,i} \,  , \\ \nonumber
N_{\D_{\tau^{\bot}_{2^\bot}}}^{\rm p}(T_{B2}) & = & 
(1-p^0_{{\rm p}\t})\,(1-p^0_{p\t^{\bot}_2}) \, N_{B-L}^{\rm p,i}  \,  .
\eea
At temperatures $T \sim 10^9\,{\rm GeV}$, also muon lepton interactions become effective,
breaking the residual coherence of the $e-\m$ lepton components in  way that
in the range $10^9\,{\rm GeV}\gg T \gg M_1$ the total asymmetry can be regarded
as the sum of three charged lepton flavour components
\be
N_{B-L}^{\rm p}(10^{9}\,{\rm GeV} \gg T \gg M_1) = \sum_{\a=e,\m,\t} \,  N_{\D_\a}^{\rm p}(T_{B2}) , 
\ee
where
\bea\label{NpDal}
N_{\D_\t}^{\rm p}(10^{9}\,{\rm GeV} \gg T \gg M_1) & = & 
p^0_{{\rm p}\t}\,  e^{-{3\pi\over 8}\,K_{2\t}} \, N_{B-L}^{\rm p,i} \,  , \\  \nonumber
N_{\D_\m}^{\rm p}(10^{9}\,{\rm GeV} \gg T \gg M_1) & = & 
(1-p^0_{{\rm p}\t})\,\left[ 
p^0_{\mu\t_2^{\bot}}\, p^0_{p\t^\bot_2}\,
e^{-{3\pi\over 8}\,(K_{2e}+K_{2\m})} + (1-p^0_{\m \t_2^{\bot}})\,(1-p^0_{p\t^\bot_2}) \, \right] 
\, N_{B-L}^{\rm p,i}
\,  ,  \\  \nonumber
N_{\D_e}^{\rm p}(10^{9}\,{\rm GeV} \gg T \gg M_1) & = & 
(1-p^0_{{\rm p}\t})\,\left[ p^0_{e\t_2^{\bot}}\,
 p^0_{p\t^\bot_2}\,e^{-{3\pi\over 8}\,(K_{2e}+K_{2\m})} + 
 (1-p^0_{e \t_2^{\bot}})\,(1-p^0_{p\t^\bot_2}) \, \right] 
\, N_{B-L}^{\rm p,i} \,   
\eea
and where the probabilities $p^0_{\a\t_2^{\bot}}$ are
unambiguously expressed in terms of the decay parameters,
\be
p^0_{e\t_2^{\bot}} = {p^0_{2 e} \over p^0_{2 e} + p^0_{2\mu}} 
= {K_{2 e} \over K_{2e}+K_{2\m}} \,  ,
\ee
analogously for $p^0_{\mu\t_2^{\bot}}$.
These expressions now clearly show that the  tauon component is
the only component of the pre-existing asymmetry that can be completely
washed-out by the $N_2$ wash-out processes.   

Finally, at temperatures $T \sim M_1$, the lightest RH neutrino wash-out processes
act on the flavoured asymmetries in a way that the relic values of the pre-existing
asymmetries flavoured components are given by
\bea\label{finalpas}
N_{\D_\t}^{\rm p,f} & = & 
p^0_{{\rm p}\t}\,  e^{-{3\pi\over 8}\,(K_{1\t}+K_{2\t})} \, N_{B-L}^{\rm p,i} \,  , \\  \nonumber
N_{\D_\m}^{\rm p,f} & = & 
(1-p^0_{{\rm p}\t})\,\,e^{-{3\pi\over 8}\,K_{1\m}} \,\left[ 
p^0_{\mu\t_2^{\bot}}\, p^0_{p\t^\bot_2}\,
e^{-{3\pi\over 8}\,(K_{2e}+K_{2\m})} + (1-p^0_{\m \t_2^{\bot}})\,(1-p^0_{p\t^\bot_2}) \, \right] 
\, N_{B-L}^{\rm p,i}
 ,  \\  \nonumber
N_{\D_e}^{\rm p,f}& = & 
(1-p^0_{{\rm p}\t})\,\,e^{-{3\pi\over 8}\,K_{1e}} \,\left[ p^0_{e\t_2^{\bot}}\,
 p^0_{p\t^\bot_2}\,e^{-{3\pi\over 8}\,(K_{2e}+K_{2\m})} + 
 (1-p^0_{e \t_2^{\bot}})\,(1-p^0_{p\t^\bot_2}) \, \right] 
\, N_{B-L}^{\rm p,i} \,   .
\eea
The most reasonable assumption for the flavour composition of  the pre-existing 
asymmetry is that $p^0_{p\a}\simeq 1/3$, equivalent to assume that the source is flavour 
blind.  In any case, as we will comment, the results are basically insensitive to 
specific choices, unless one select special values corresponding, for example, 
to a  pre-existing asymmetry  entirely in one specific charged lepton flavour.  
In this special case it would be much easier to wash-out the pre-existing asymmetry but on the other hand this would be analogous to assuming a vanishing initial asymmetry, 
while we are interested in finding the general conditions for  the independence of the initial conditions. 
We have, therefore, set $p^0_{p\a}={\cal O}(0.1)$. 

 The expression eq.~(\ref{finalpas}) now explicitly shows that, in order for successful strong thermal leptogenesis to be realised, the final asymmetry has to be necessarily tauon dominated.
 This is because only in the tauon flavour  the wash-out of the pre-existing asymmetry 
 by the $N_2$ inverse processes at $T\sim M_2$ for $K_{2\tau}\gg 1$
 does not prevent that a genuine leptogenesis contribution is afterwards
 generated  by the same $N_2$ decays at $T \simeq T_{B2} \ll M_2$, 
 surviving until the present time for $K_{1\tau}\lesssim 1$.
 On the other hand the electron and muon components of the pre-existing asymmetries 
 can be only fully washed-out by the $N_1$ wash-out processes at $T\sim M_1$
 for $K_{1e},K_{1\m} \gg 1$.
  \footnote{
 There is a caveat: this  conclusion does not hold for fine tuned models where the ${\ell}_2$ tauon orthogonal  component is purely electronic or muonic with huge precision such that
$1-p^0_{\alpha \t_2^{\bot}}\simeq 0$ ($\a= e$ or $\mu$). These models would effectively
correspond  to two-flavour models. In any case this special situation is not
realised in $SO(10)$-inspired models under consideration. 
 } However, this unavoidably implies that together
 also the electron and muon leptogenesis contribution from $N_2$ decays is washed-out, 
 while the $N_1$ decays are ineffective in generating a sizeable asymmetry. 
In this way the final asymmetry has necessarily to be tauon  dominated.
 
 Therefore, the full set of conditions on the flavoured decay parameters can be summarised as
 \cite{problem}
 \be\label{strongthconditions}
 K_{1e}\gg 1  \,  , \hspace{3mm}  K_{1\mu} \gg 1 \, ,  \hspace{3mm} K_{2\tau} \gg 1  \,  ,
  \hspace{3mm}  K_{1\t} \lesssim 1 \, ,
 \ee
with the precise values depending on the precise assumed values of 
$N_{B-L}^{\rm p,i}$.

The same set of conditions is sufficient also if one relaxes the assumption that the pre-existing
leptons and anti-leptons quantum states are not $C\!P$ conjugated of each other. In this case 
the only difference is that in the three-flavour regime one would have additional
contributions to $N_{\D\a}^{\rm p,f}$ with $\a = e,\mu$ in the eq.~(\ref{finalpas})
inside the squared brackets, that are anyway washed out when $K_{1e}, K_{1\mu} \gg 1$.
\footnote{Notice that in the presence of phantom terms in the pre-existing asymmetry 
the caveat pointed out in footnote 4 does not apply.}

In the next Section we will see how this seemingly quite restrictive set of conditions
(cf. eq.~(\ref{strongthconditions})) can be indeed realised within $SO(10)$ inspired leptogenesis, 
translating into quite an interesting set of constraints on the low energy neutrino parameters,  
sharp enough  to be regarded as a quite distinctive signature of the scenario.

Before concluding this section, we would just like to make a brief comment on the possible
existence of a source of baryogenesis posterior to leptogenesis. In this case there is clearly no
condition that can be imposed for its wash-out. 
Simply there should be no experimental evidence for new physics supporting 
an alternative mechanism of baryogenesis.  While a pre-existing asymmetry would be
difficultly testable, a posterior production is more likely to be testable. 
In this case  baryogenesis would occur in a post-inflationary stage during the standard radiation regime. 
Basically the only realistic known  source to be  competitive with leptogenesis would come from some
realisation of electroweak baryogenesis typically requiring
some extension of the Standard Model testable at colliders. 
If some signal of new physics will be found, it would then become important to constraint such a 
possibility for an alternative production of the asymmetry after leptogenesis. 
Since the LHC has not provided evidence for new physics so far,  we do not have to worry 
of any additional condition  to be imposed. 

\section{Strong thermal $SO(10)$-inspired solution}

\subsection{Results on neutrino parameters}

We have imposed the strong thermal condition eq.~(\ref{strongthcondition})
on the solutions with $M_2 < 10^{12}\,$GeV found within the $SO(10)$-inspired scenario discussed in  Section 3
\footnote{More precisely we have imposed that a relic value of the pre-existing asymmetry 
contributes to the final asymmetry less than $10\%$. Notice that we had to select }
finding that this is indeed satisfied by a subset of them.
 This has been done
for three different values of the initial pre-existing $B-L$ asymmetry $N_{B-L}^{\rm p,i}$.

The results can be read off from the same panels of  Fig.~1.
The solutions are indicated with blue, green and red points respectively for
$N_{B-L}^{\rm p,i}=10^{-3}, 10^{-2}$ and $10^{-1}$. One can see that in the different 
neutrino parameter planes, the regions satisfying the strong thermal condition 
are clearly a subset of the allowed regions within $SO(10)$-inspired leptogenesis (the yellow points).
In some cases they introduce such strong and definite constraints on the low energy neutrino
parameters that these  can be regarded as sharp distinctive predictions. 
Let us briefly describe these constraints. 


\subsubsection{IO is excluded}

Even though by imposing $SO(10)$ inspired conditions 
one still finds some marginal allowed regions for IO \cite{SO10lep2}, when the strong thermal
condition is further imposed, no solution is found and for this reason we do not show
any result for IO in the paper.

\subsubsection{Neutrino masses}

The solutions are found for quite a restricted range of
values for the lightest neutrino mass given by $m_1 \simeq (15-25)\,{\rm meV}$ 
($m_1 \simeq (10-30)\,{\rm meV}$) for $N_{B-L}^{\rm p,i}=10^{-1}$ 
($N_{B-L}^{\rm p,i}=10^{-2}$). This range translates into corresponding ranges 
for $m_2$, $m_3$ and $\sum_i  m_i$ given in Table 1. 
\begin{table}[htdp]
\begin{center}
\begin{tabular}{|c|c|c|c|c|c|}
\hline
 $N_{B-L}^{\rm p,i}$ & $m_1$ & $m_2$  & $m_3$ & $\sum_i m_i$ & $m_{ee}$ \\
\hline
$10^{-1}$ & 15--25 &  17--26 & 51--55 & 84--106 & 12--22 \\
\hline
$10^{-2}$ & 10--30 &  13--31 & 50--57 & 73--118 & 9--27\\
\hline
\end{tabular}
\end{center}
\caption{Values of the neutrino masses (in meV) as predicted by strong thermal $SO(10)$-inspired leptogenesis.}
\label{tableneutrinomasses}
\end{table}
The found solution corresponds to NO {\em semi-hierarchical neutrinos}, with the
heaviest neutrino about three times heavier than the two quasi-degenerate lightest ones.
  
\subsubsection{Reactor mixing angle}
 
As one can see from the upper central panel, the bulk of the solutions nicely fall within the 
range now measured by reactor experiments (cf. eq.~(\ref{newrangetheta13})) indicated in the plot. 
 For  $N_{B-L}^{\rm p,i}=10^{-1} (10^{-2})$ 
there is a lower bound $\theta_{13}\gtrsim 2^{\circ} (0.5^{\circ})$.

\subsubsection{Atmospheric mixing angle}

The strong thermal condition cannot be realised for too large values of
the atmospheric mixing angle (upper right panel and central left panel). 
This results into an interesting upper bound $\theta_{23} \lesssim 41^{\circ} (43^{\circ})$
for $N_{B-L}^{\rm p,i}=10^{-1} (10^{-2})$ that provides quite a significant test of the solution, 
since the allowed range is consistent only with current lowest
experimentally allowed values.  

Since the allowed region clearly extends to  values of $\theta_{23}$, lower than
those currently allowed by global analyses, we also determined the lower bound 
of this region repeating the scan for a wider $\theta_{23}$ range compared 
to eq.~(\ref{rangetheta23}), extending to values  as low as zero. The result is 
shown in Fig.~4
 \begin{figure}
\begin{center}
\psfig{file=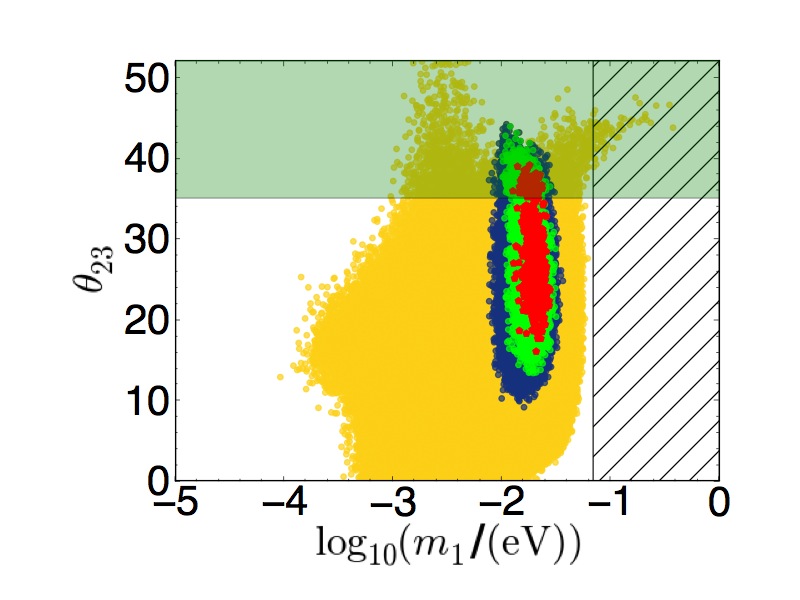,height=60mm,width=70mm}
\hspace{4mm}
\psfig{file=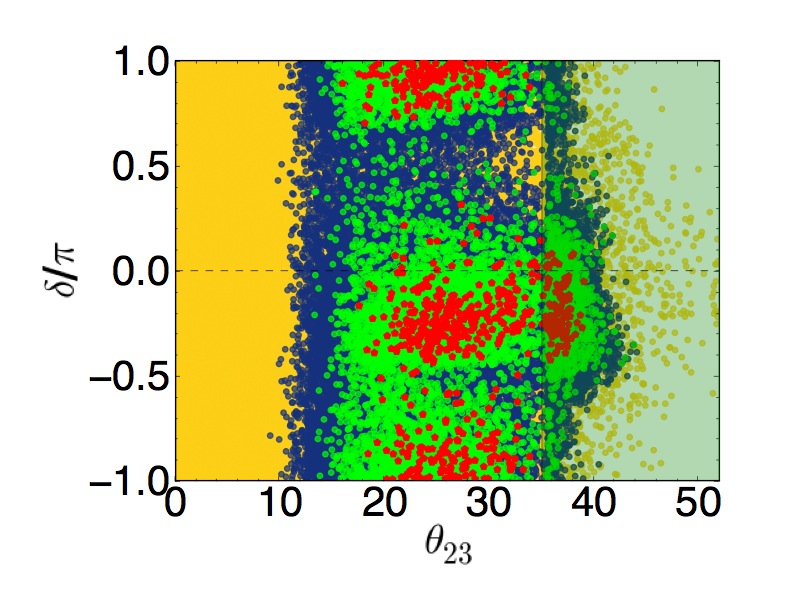,height=60mm,width=70mm}
\end{center}\vspace{-5mm}
\caption{Result of the scatter plot in the plane $\theta_{23}$ vs. $m_1$ (left)
and $\d$ vs. $\t_{23}$ (right) as in Fig.~1 but with $\theta_{23}$ let free 
to variate in a range $0 \leq \theta_{23} \leq 50^{\circ}$.}
\label{atmo}
\end{figure}
and one can see that $\theta_{23}$ can be as low as $\simeq 16^{\circ} (13^{\circ})$ for 
$N_{B-L}^{\rm p,i}=10^{-1} (10^{-2})$.  
\footnote{Notice that in this figure the upper bound is more restrictive than in Fig.~1:
$\theta_{23}\lesssim 40^{\circ}$. This is simply due to the fact that in this figure we
did not generate eneugh points to saturate the bounds. We comment on this aspect
of the constraints at the end of Section 6.}

\subsubsection{Majorana phases}

The allowed regions for the Majorana phases close up around special values.
There are two different kinds of regions: four centred
around $(\sigma,\rho) = (0.8+n,1.25+n)\,\pi$, with $n=0,1$
and four centred around $(\sigma,\rho) = (0.7+n,0.75+n)\,\pi$, with $n=0,1$.
These regions are not perfectly coincident to those obtained without imposing
the strong thermal condition for $\a_2 = 4$ \cite{SO10lep2}. This shows
that they shrink not just around the values that maximise the asymmetry 
irrespectively of the strong thermal condition, but that 
the strong thermal condition influences the values of the Majorana phases.

\subsubsection{$0\nu\b\b$ effective neutrino mass $m_{ee}$}

From the calculation of the effective $0\nu\b\b$ effective neutrino mass,
$m_{ee}=|\sum_i \, m_i \, U^2_{ei}|$, we find that this is quite sharply related 
to the lightest neutrino mass,
and just slightly lower, approximately  $m_{ee} \simeq 0.8 \, m_1$. 
This is clearly an effect of the quite restricted range of allowed values for the Majorana phases. 
We will be back on this point when we will discuss the experimental implications of the solution. 
The allowed range of values for $m_{ee}$ is indicated in the last column of table 1.

\subsubsection{Dirac phase and $C\!P$ violation} 

Very interestingly, having now imposed the strong thermal
condition, the Dirac phase and $J_{CP}$ show a preference for negative
values. In particular, within the measured range for $\theta_{13}$
(cf. eq.~(\ref{newrangetheta13})), the Dirac phase falls dominantly in the 
range $-0.5 \lesssim \d/\pi \lesssim 0.2$ for $N_{B-L}^{\rm p,i}=10^{-1}$. 
Correspondingly one has that the Jarlskog invariant falls in the range  
$-0.04 \lesssim J_{CP} \lesssim 0.02$. 
There is also a sub-dominant region for $|\delta|/\pi \simeq 0.9\,$--$\,1$. 
However, this  marginal region exists only for $\theta_{23} \lesssim 36^{\circ}$.
This can be seen from a plot $\delta$ vs. $\theta_{23}$  
that we are not showing in  Fig.~1 but we are showing it in Fig.~4 (right panel) for 
an extended range of $\theta_{23}$ but, as discussed, for a reduced data set.
As one can see, values $\delta \simeq \pi$ are found even only for 
$\theta_{23} \lesssim 35^{\circ}$.  This is interesting interplay between
$\delta$ and $\theta_{23}$.

\subsubsection{Summary}

We summarise in Table 2 the main features of the solution sorted 
according to a possible chronological  order of their experimental test. 
The first line  is the lower bound on the reactor neutrino mixing angle that has 
been already successfully tested.
\footnote{Preliminary results on the lower bound on $\theta_{13}$
and on the upper bound on $\theta_{23}$ 
were presented in \cite{preliminary}.}
\begin{table}[htdp]
\begin{center}
\begin{tabular}{|c|c|c|}
\hline
$N^{\rm p,i}_{B-L}$ & $10^{-1}$ & $10^{-2}$ 
\\ \hline
$\theta_{13}$ & $\gtrsim 2^{\circ}$ & $\gtrsim 0.5^{\circ}$ \\
\hline
$\theta_{23}$ &  $\lesssim 41^{\circ}$ & $\lesssim 43^{\circ}$ \\
\hline
ORDERING & NORMAL & NORMAL \\ \hline 
$\delta$ & $- \pi/2 \div \pi/5$  & $\notin [0.4\,\pi, 0.7\,\pi]$ \\
               & $\simeq \pi$ (marginal, only for $\theta_{23} \lesssim 36^{\circ}$) &\\
\hline
$m_1$ & $ (15 \div 25) \, $ meV & $(10 \div 30)\,{\rm meV}$ \\ \hline 
$m_{ee}$ & $\simeq 0.8 \, m_1 \simeq (12 \div 20)\,{\rm meV}$ & $(8 \div 24)\,{\rm meV}$  \\
\hline
\end{tabular}
\end{center}
\caption{Summary of the set of conditions on low energy neutrino data from
$SO(10)$-inspired strong thermal leptogenesis imposing the wash-out of a 
pre-existing asymmetry as large as $10^{-1}$ and $10^{-2}$.}
\label{tablesummary}
\end{table}%
 
\begin{figure}
\begin{center}
\hspace{-4mm}
\psfig{file=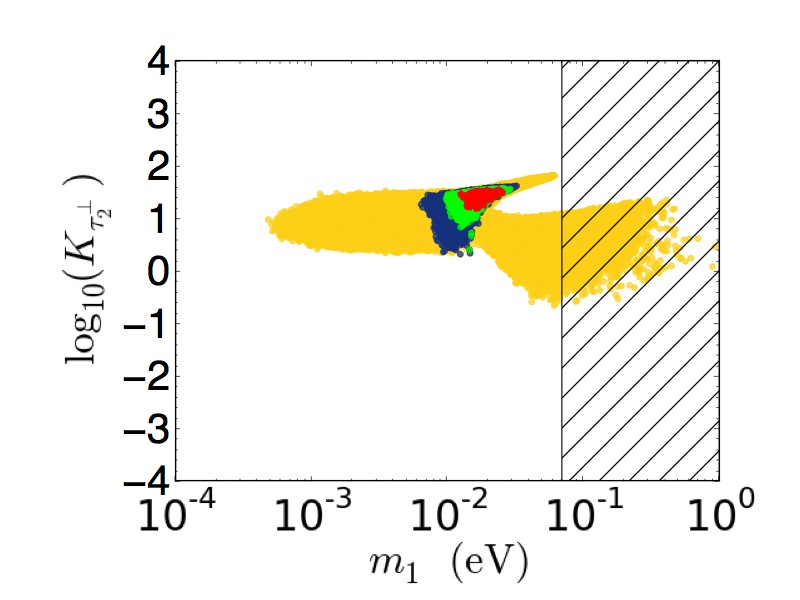,height=48mm,width=54mm} 
\hspace{-4mm}
\psfig{file=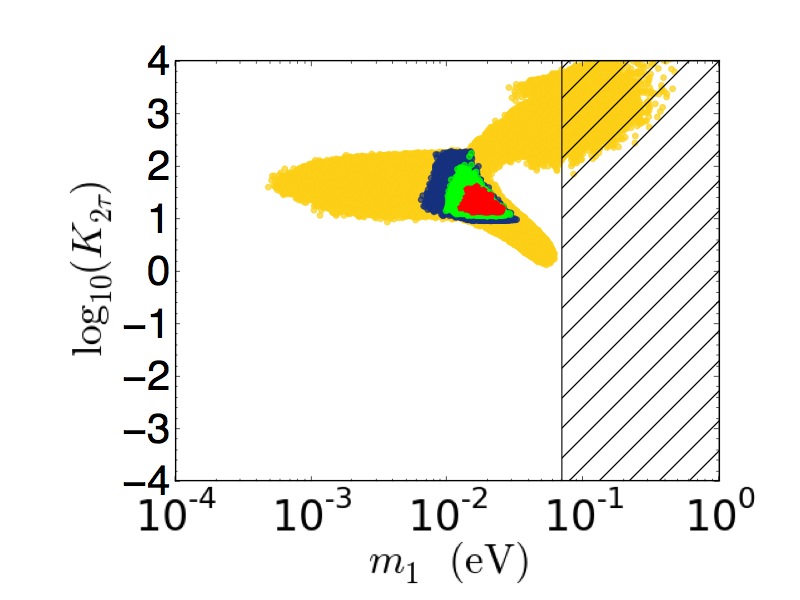,height=48mm,width=54mm}
\\
\psfig{file=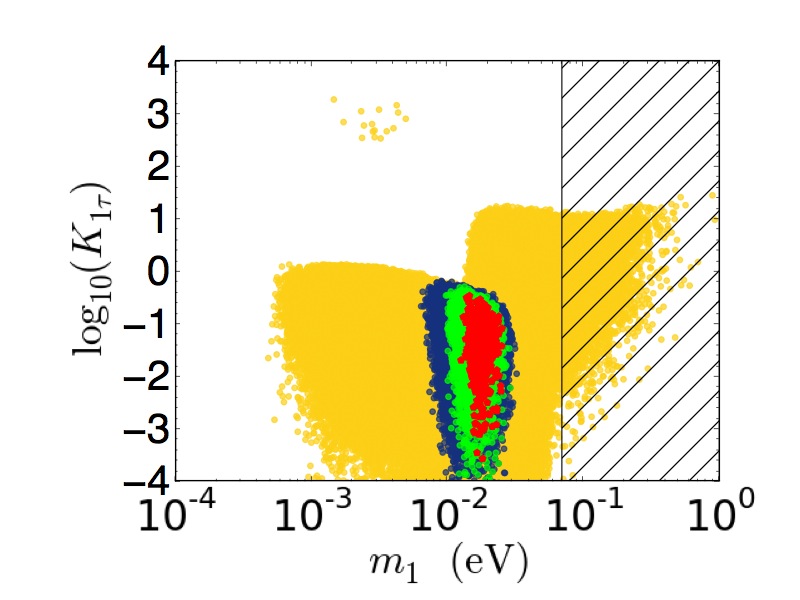,height=48mm,width=54mm}
\hspace{-4mm}
\psfig{file=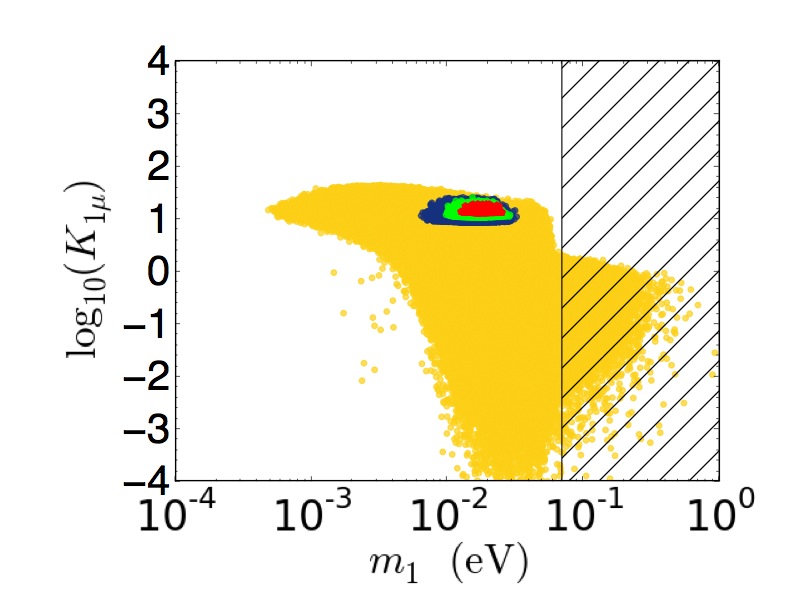,height=48mm,width=54mm}
\hspace{-4mm}
\psfig{file=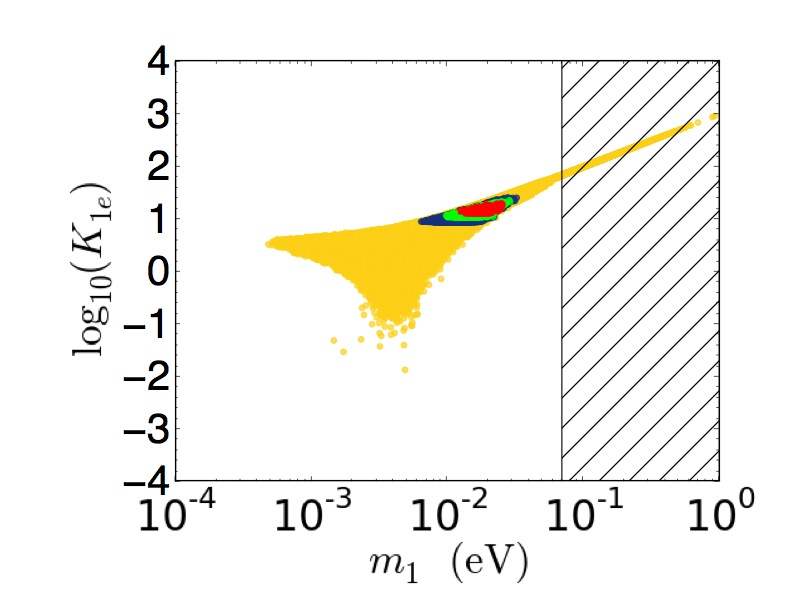,height=48mm,width=54mm} \\
\psfig{file=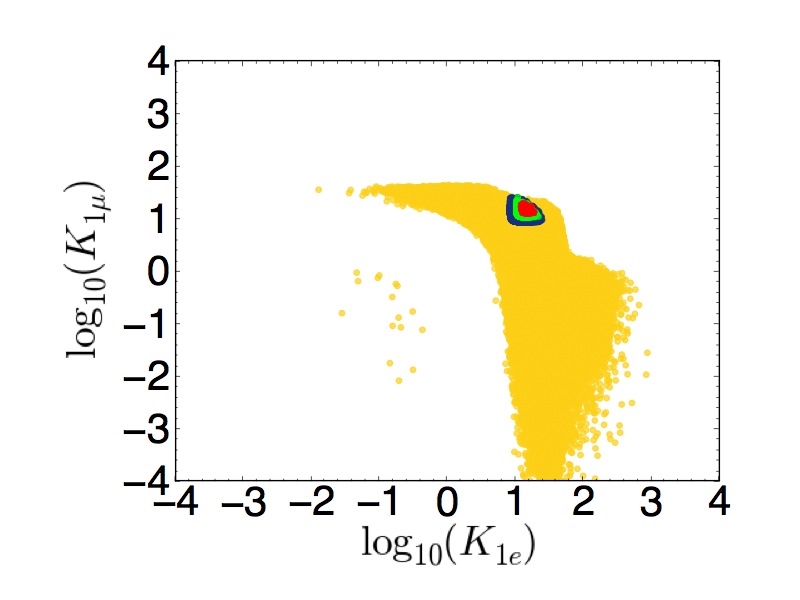,height=48mm,width=54mm}
\hspace{-4mm}
\psfig{file=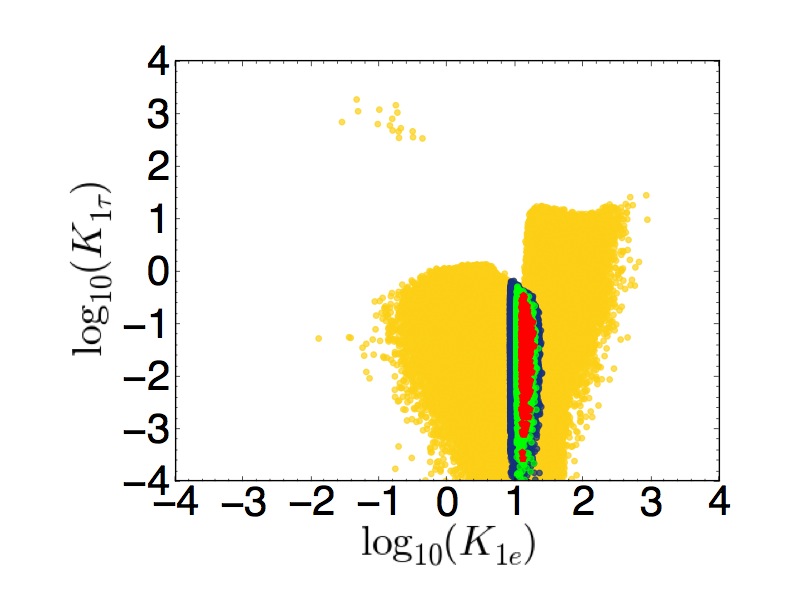,height=48mm,width=54mm} 
\hspace{-4mm}
\psfig{file=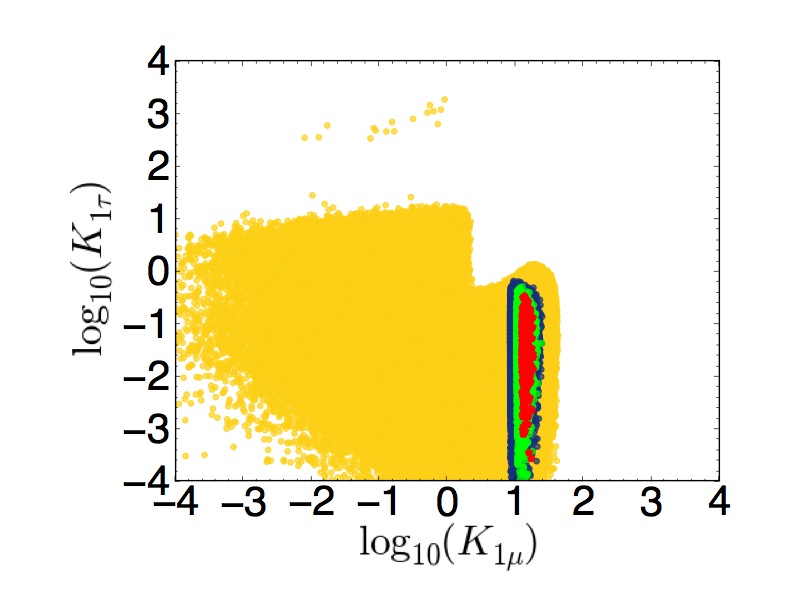,height=48mm,width=54mm} 
\end{center}
\vspace{-5mm}
\caption{Results of the scatter plots for the decay flavoured parameters plotted
versus either $m_1$ (first two rows) or versus themselves (only for the $K_{1\a}$'s).}
\label{Kialpha}
\end{figure}
 
\subsection{Constraints on  flavour decay parameters} 
 
The natural parameter of leptogenesis are the nine flavoured asymmetries
$\varepsilon_{i\a}$ and the nine flavour decay parameters $K_{i\a}$.
As we have seen these can be re-expressed in terms of the nine parameters
in the low energy neutrino matrix and of the nine theoretical parameters, 
six to describe the $V_L$ and the three eigenvalues of the neutrino Dirac mass matrix. 
In order to have a useful insight on the constraints on the low energy neutrino parameters
discussed in the previous subsection, we show in Fig.~5 plots of  
the flavour decay parameters $K_{1\a}$. In this way we can see what are
the values of the relevant flavour
decay parameters that realise the strong thermal condition. 
These plots confirm that the solution we have found 
realises the conditions eq.~(\ref{strongthconditions}). 

Let us discuss them in more detail. In the lower panels of Fig.~5 we have plotted the flavour decay parameters $K_{1\a}$ versus each other. These panels clearly confirm that the conditions  eq.~(\ref{strongthconditions}) 
are indeed fulfilled. It is in particular interesting to notice how the
two conditions $K_{1e}, K_{1\mu} \gg 1$ are satisfied only for the particular subset
of the region realising $SO(10)$-inspired leptogenesis.

Looking at the  panels where the $K_{1\alpha}$'s are plotted versus $m_1$,
one can see that the condition $K_{1e} \gg 1$  can only be satisfied  for 
$m_1 \gg 10^{-3}\,$eV, while the condition $K_{1\mu}\gg 1$ can only be satisfied for
$m_1 \lesssim 0.1\,{\rm eV}$. In addition the plot $K_{2\tau}$ vs. $m_1$, in the bottom
right panel, shows that
$K_{2\tau} \gg 1$ implies $m_1 \lesssim 30\,{\rm meV}$, further 
restricting the upper bound on $m_1$. In this respect, notice that in that panel 
one has not to consider the region extending at $K_{2\tau} \gtrsim 100$ and $m_1$
larger than $0.1\,$ eV, since this corresponds to the muon type solutions. 
Therefore, the quite narrow range of values of $m_1$
realising successful strong $SO(10)$ inspired leptogenesis is a consequence 
of the dependence of the relevant $K_{i\a}$ on $m_1$ in combination with the
strong thermal conditions. 
 
\subsection{Link between the sign of the asymmetry and the sign of $C\!P$ violation}
 
 The results for the Dirac phase $\d$ and for $J_{CP}$, showing an asymmetry between
 positive and negative sign values with negative values clearly favoured,  are quite interesting and motivate an understanding of their origin. The only physical quantity that can favour one sign compared to the other is the same positive sign of the observed matter-antimatter asymmetry. 
 Therefore, we performed a simple check,  working out again the constraints on 
 $\d$ and on $J_{CP}$ but this time imposing $\eta_B^{\rm lep} = - \eta_B^{\rm CMB}$ (in practice we imposed $\eta_B^{\rm lep} < - 5.9 \times 10^{-10}$). The results are shown in the right panels of Figure 5
 and compared with those of Fig.~1 displayed again in the left panels. 
 One can see that despite the much lower amount of points in the data set, 
 they clearly show that now the favoured ranges of $\d$ and $J$ 
 switch to positive values. 
 Therefore, we can conclude that the solution favours values of $\d$ and $J_{CP}$
 with opposite sign compared to the values of the matter-antimatter asymmetry. 
 \begin{figure}
\begin{center}
\psfig{file=d_vs_t13.jpg,height=58mm,width=64mm} 
\hspace{-4mm}
\psfig{file=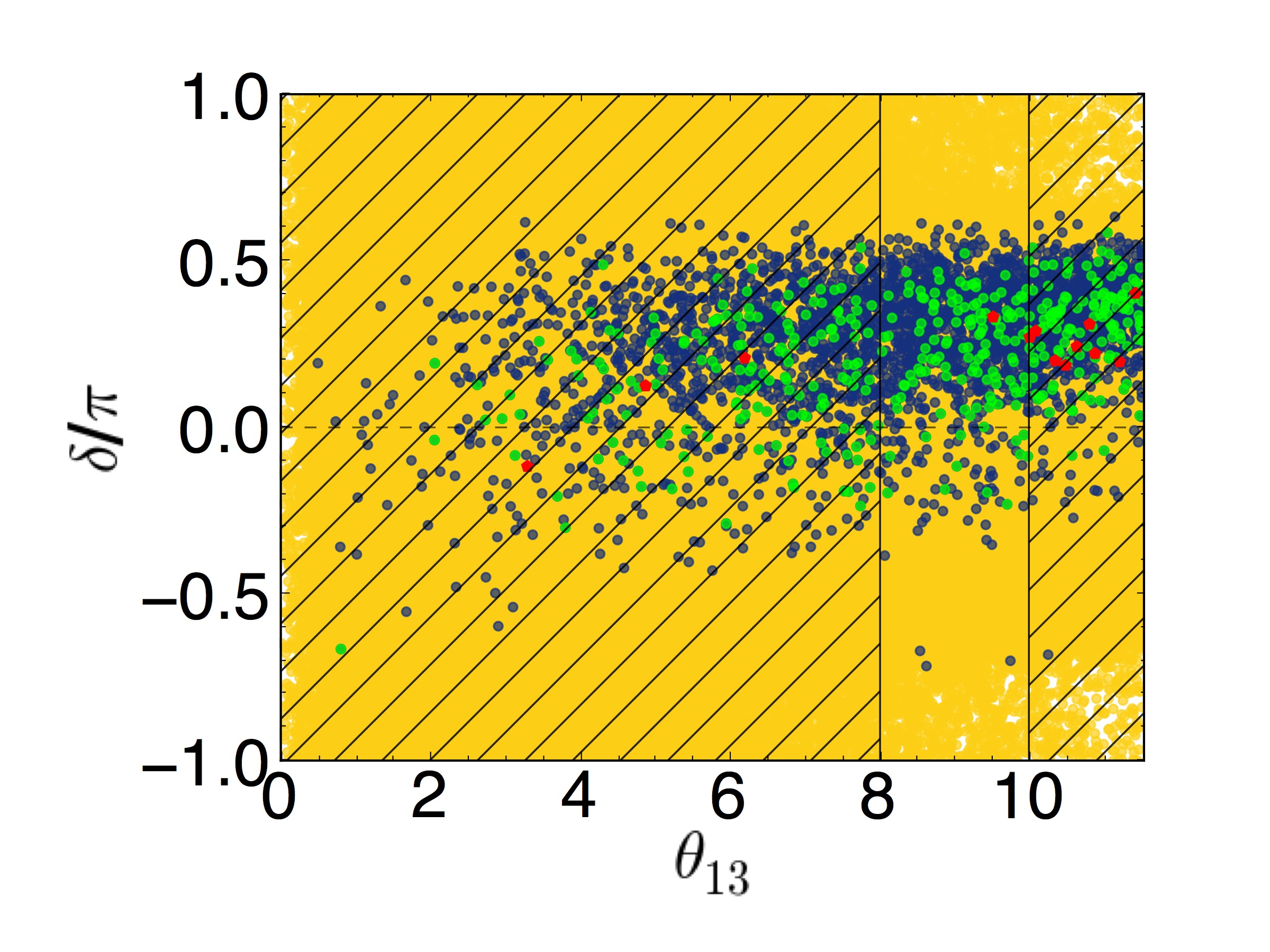,height=58mm,width=64mm} 
\\
\psfig{file=J_vs_t13.jpg,height=58mm,width=64mm} 
\hspace{-4mm}
\psfig{file=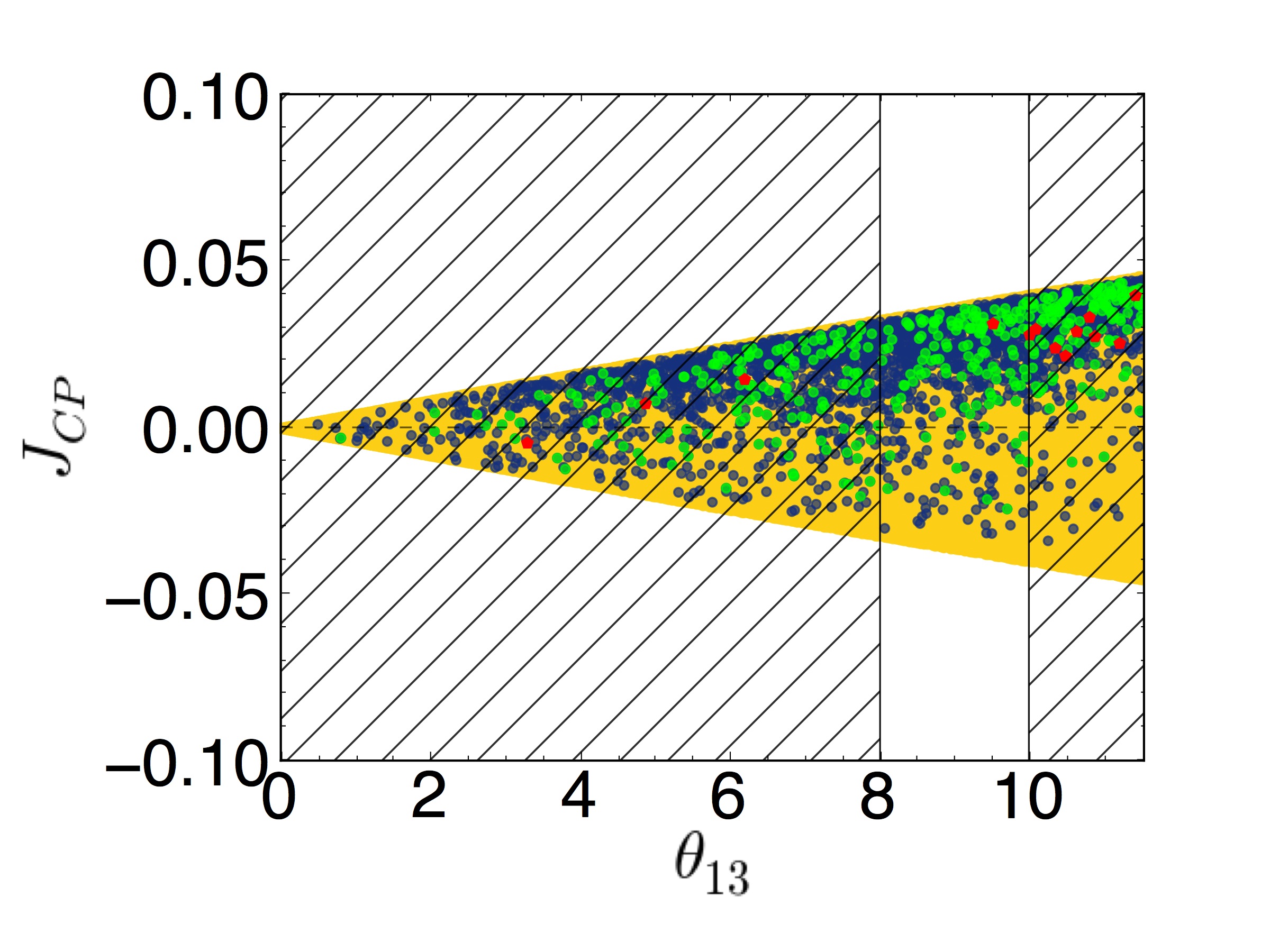,height=58mm,width=64mm} 
\\
\psfig{file=d_vs_m1.jpg,height=58mm,width=64mm} 
\hspace{-4mm}
\psfig{file=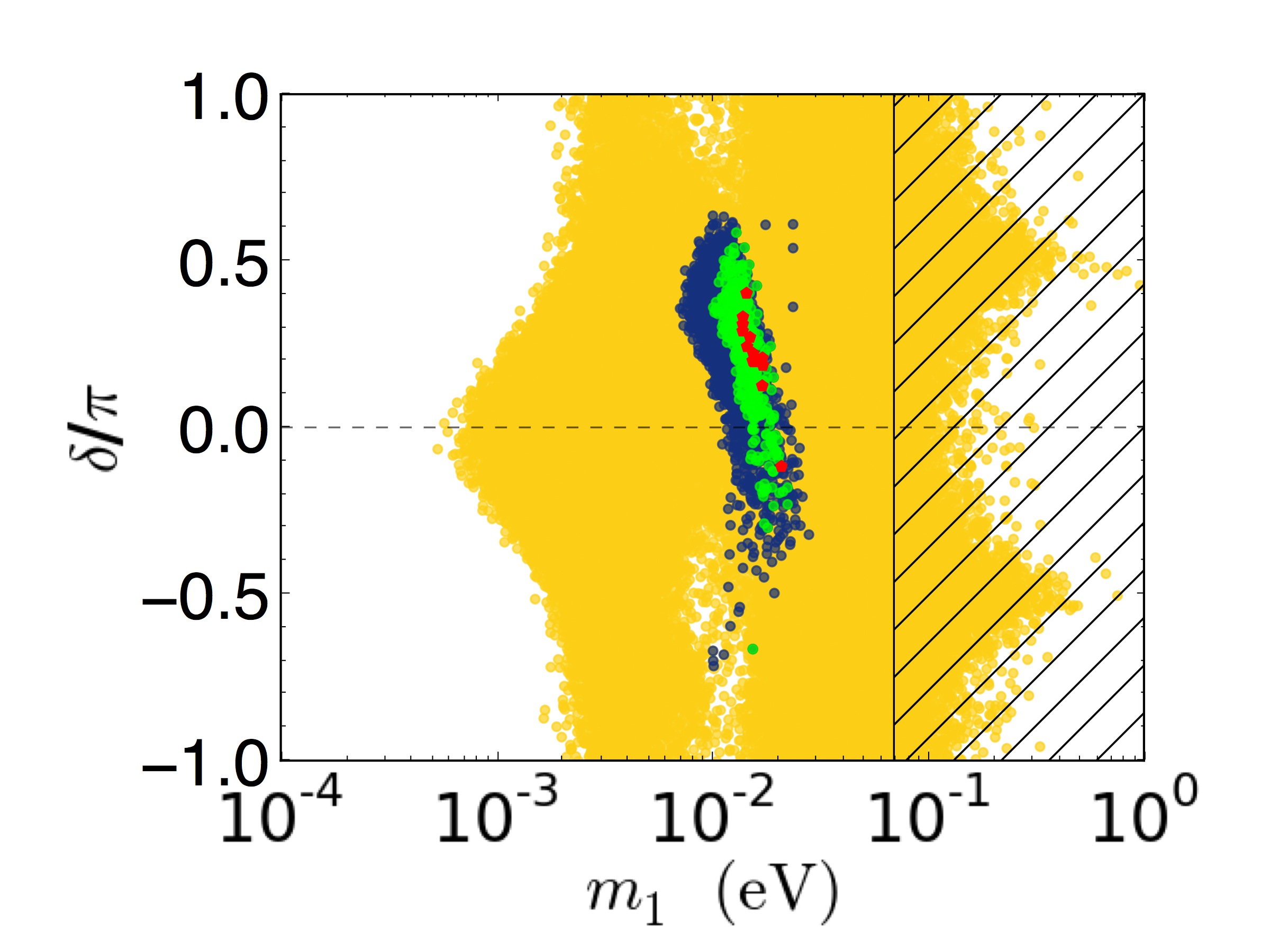,height=58mm,width=64mm} 
\end{center}
\caption{Results showing how $C\!P$ violation in neutrino mixing in this
scenario is linked to the sign of the matter-antimatter asymmetry. The left panels
are obtained imposing $\eta_B^{\rm lep} = \eta_B^{\rm CMB}$, the right panels 
imposing $\eta_B^{\rm lep} = -\eta_B^{\rm CMB}$.}
\label{blueplots}
\end{figure}

\subsection{Are low energy neutrino data pointing in the right direction?}

 We repeated the same test as in the case of $SO(10)$-inspired leptogenesis
 finding the constraints on low energy neutrino parameters without any restriction from
 low energy neutrino experiments.  The results are shown in the same Fig.~3. The
 allowed regions (red, green and blu points) are again subsets of those obtained without imposing the strong thermal condition  (yellow points).
 This time the allowed regions represent a much smaller fraction compared to the whole parameter space
and, therefore, an agreement with experimental data would be
much more statistically significant. 
  
 Let us briefly discuss these results focusing first on the upper panels showing the allowed regions for the mixing angles versus $m_1$. First of all one can again notice  that for negative values of the mixing angles one obtains mirrored regions. Let us then concentrate on  
positive values of the mixing angles.  

One can see that there are two well distinguished allowed regions:
a much larger one for $10^{-4}\,{\rm eV} \lesssim m_1 \lesssim 10^{-2} \,{\rm eV}$ and a  
smaller one for $10^{-2}\,{\rm eV} \lesssim m_1 \lesssim 3 \times 10^{-2} \,{\rm eV}$
 (for $N_{B-L}^{\rm p,i}=10^{-1}$).
 
 Both regions are compatible with the measured value of the solar mixing angle but
 whilst the first one, at small $m_1$, would require unacceptably large values of
 $\theta_{23}$ and $\theta_{13}$, the second one, for large $m_1$, is perfectly compatible
 with the measured value of $\theta_{13}$ but  only  with the lowest 
 experimentally allowed range of values of  
 $\theta_{23}$, i.e. for $\theta_{23}\lesssim 41^{\circ}$.

It is interesting that, just within the three mixing angles parameter space, 
the fraction occupied by the allowed regions is lower than $\sim 10\%$. It should be also added that
IO is excluded even in this case. If one also takes into account the allowed values for $m_1$, we can say that  the
chance to hit randomly both the allowed regions, for a logarithmic scan of $m_1$ between
$10^{-5}$ and $0.1\,$ eV, can be quantified to be about $1\%$.  
If one considers that the Majorana phases further restrict the values of $m_{ee}$
compared to the general case (see in the bottom left panel in Fig.~1),
one arrives to a probability lower than $0.2\%$.  Finally, taking into account the half
chances for the mass ordering, one arrives to the conclusion that  the solution occupies roughly a portion that represents roughly $0.1\%$ of the total accessible volume in parameter space.
This gives an approximated estimation of the statistical 
significance that  a positive test of the solution would have, i.e. it gives an estimation of the 
probability that the allowed region corresponding to the solution 
can be centred by the experimental data just accidentally. 
  
However, this value of the `success rate' is dominated by the large excluded region. 
For the found solution at $m_1 \simeq 20\,$meV the success rate would be much smaller, 
$\sim 10^{-7}$.  Imposing the current experimental ranges for the mixing angles 
((cf. eqs.~(\ref{rangetheta13}), (\ref{rangetheta12}) and (\ref{rangetheta23}))) this rate does not increase 
simply because, despite the fact that $\theta_{12}$ and $\theta_{13}$ fall in the allowed regions,
the range for $\theta_{23}$ eq.~(\ref{rangetheta23}) is only marginally compatible. 
However, if future experimental data will find 
values $\theta_{23} \lesssim 39^{\circ}$, the success rate will interestingly increase
by an order-of-magnitude. In this case one could say that indeed
low energy neutrino data start to show some convergence around the solution. From this point 
of view a more precise experimental determination of the atmospheric mixing angle
represents, in short terms, a crucial test of the solution.  

\subsection{Testing the solution}

A very attractive feature of the solution is that the constraints on neutrino parameters that we have just discussed, summarised in Table 2, can be fully tested. In the case of mixing parameters,
 even by low energy neutrino experiments that are either already taking data  
or scheduled.  In this respect the large value found for $\theta_{13}$, 
is not only in agreement with the solution, but it is also a key ingredient that will make possible 
to determine the atmospheric mixing angle octant,  the  neutrino mass ordering and (of course) 
the Dirac phase  during next years.  

The atmospheric mixing angle is already now favoured to be non-maximal, as discussed in Section 2.
 It is also encouraging that in \cite{fogli} the best fit value is found to be 
$\theta_{23}\simeq 38^{\circ}$, quite well inside the allowed region (cf. Fig.~ 4).
By combining T2K and NO$\nu$A data,
such low values will be either determined  within a  $\sim 3\s$ C.L. range of 
$2^{\circ}$, excluding the second octant, 
or otherwise be excluded at $\sim 3\,\s$ \cite{lindner,agarwalla}. 
At the same time the fact that the solution favours the `experimentally favourable combination' 
of NO and negative values of $\d \sim -\pi/5$, makes also possible 
a $\sim 3\s$ determination of the ordering and of the sign of $\d$ 
by a combination of T2K and NO$\nu$A results \cite{lindner}. 
\footnote{A higher statistically significant determination of the ordering 
can be obtained with PINGU. This would be between
$3\,\s$ and $10\,\s$ after 5 years of operation depending on the
reconstruction accuracies \cite{razzaque}. A combination 
with Daya Bay II would contribute to further improve the statistical significance \cite{blennowschwetz}.} 

Cosmological observations are potentially able to determine
a lightest neutrino mass in the range $m_1 = (15-25)\,$ meV,
corresponding to $\sum_i m_i \sim (84-106)\,$meV,
improving the current upper bound eq.~(\ref{upperboundm1}).
In this respect it is interesting that a combination of the Planck 
results on Sunyaev-Zeldovich  cluster counts with Planck CMB results and BAO
hints at non-vanishing neutrino masses   
$\sum_i m_i = (0.22 \pm 0.09) \, {\rm eV}$ \cite{plancksz}.

Notice that values of  $\sum_i m_i \sim 0.1\,$eV also correspond 
to inverted hierarchy  (i.e. IO for $m_1 \rightarrow 0$). From this point of
view it is important that the mass ordering can be independently determined 
with neutrino oscillation experiments, able to disentangle our semi-hierarchical
NO solution from an inverted hierarchical solution. 

The allowed range for the $00\nu\b$ decay effective 
neutrino mass (cf. Table 2) is certainly the most challenging constraint to
be tested. In the bottom right panel of Fig.~1 we have also over-imposed
the general allowed regions in the plane $m_{ee}$ vs. $m_1$
from current experiments, both for NO and for IO. 
As one can see, the allowed region corresponding to the solution falls into
a range of $m_{ee}$ that is also corresponding to the values expected for
inverted hierarchy. These
values are not accessible to current ongoing experiments, nor
even to planned experiments such as SUPERNEMO, NEXT, Lucifer, MJD that will
at most able to exclude values of $m_{ee}$ above $50\,$meV (for a recent
discussion see \cite{neutrinoless}). However, there is a great international effort 
for the study of new experiments able to test values 
in the range 10--20 \, meV, since these would exclude inverted hierarchy.
Again, it is then important that the mass ordering can independently be determined
by neutrino oscillation experiments able to distinguish
our NO semi-hierarchical solution from  inverted hierarchy.

If both $m_1$ and $m_{ee}$ will be measured with sufficient precision, a comparison will provide
an additional test of our solution that predicts $m_{ee}\simeq 0.8\,m_1$,
due to the particular values of the Majorana phases.
\footnote{This is easy to see analytically replacing $(\s,\r)\simeq (0.8,1.25)\,\pi$ 
in the expression for $m_{ee}$.}
In the bottom right panel of Fig.~1, 
the region between the black lines is the allowed region from low energy neutrino
experiments (no leptogenesis) for NO (lower region) and IO (higher region).
For NO the ratio $m_{ee}/m_1$ can, in general, be in the range $m_{ee}/m_1 \simeq 0.3-1$
(corresponding analytically to  $\cos 2\theta_{12} \, \cos^2 \theta_{13} - 
\sin^2 \theta_{13} \leq m_{ee}/m_1 \leq 1$ \cite{rodejohann}). 
Therefore, a result $m_{ee} \simeq 0.8\, m_1 \simeq 15\,$meV, 
would further very strongly support the solution. 
Notice that a determination of both $m_1$ and $m_{ee}$  would still not be able
to fully determine  the two Majorana phases. This would provide ideally
an even stronger test of the solution making quite precise predictions on both. 
However, even if we still miss  a way to fully determine the Majorana phases, 
in case of a multiple agreement of all low energy neutrino experiments 
with the presented constraints, the probability  that this is just accidental 
would be really low (as discussed $\sim 0.1\%$ if one considers both regions
together, or even just $\sim 10^{-7}$ if one only considers the discussed solution compatible
with current neutrino oscillation data), 
quite an interesting feature of the solution prospectively. 

\section{An exploded view of the solution}

In this section we discuss some important aspects of the solution. 

\subsection{Constraints on the parameters in the RH neutrino mixing matrix $V_L$}

We have so far focused on the constraints on the low energy neutrino parameters that can be tested
in experiments. However, the solution is also determined by the 6 parameters in the matrix $V_L$. 
Indeed the first of the working assumption defining $SO(10)$-inspired models, $I \leq V_L \leq V_{\rm CKM}$ 
does not completely fix $V_L$ but allows some variation within a restricted range.  In Fig.~7 we show the constraints
on the six parameters in the $V_L$. 
\begin{figure}
\begin{center}
\psfig{file=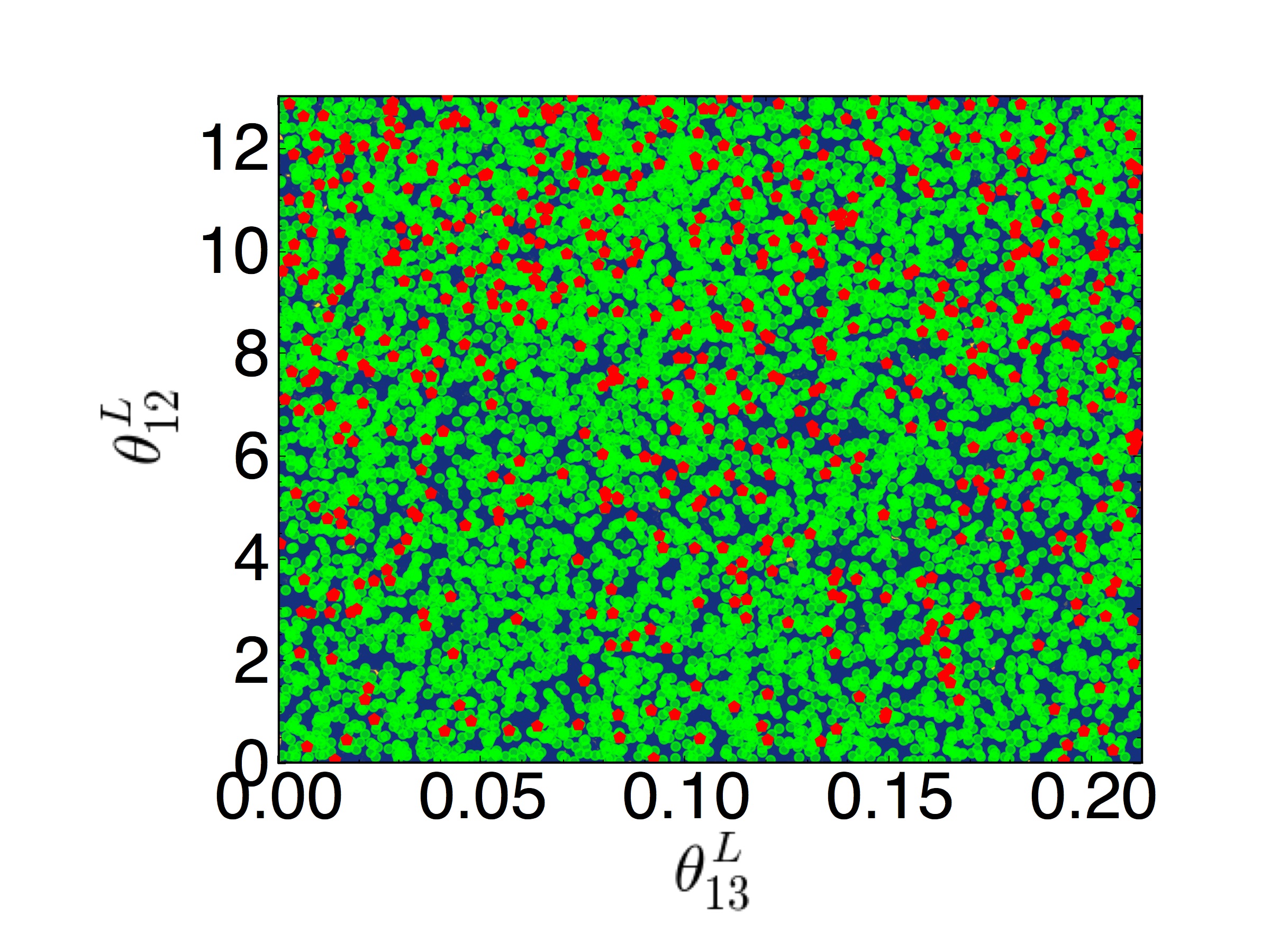,height=48mm,width=54mm}
\hspace{-4mm}
\psfig{file=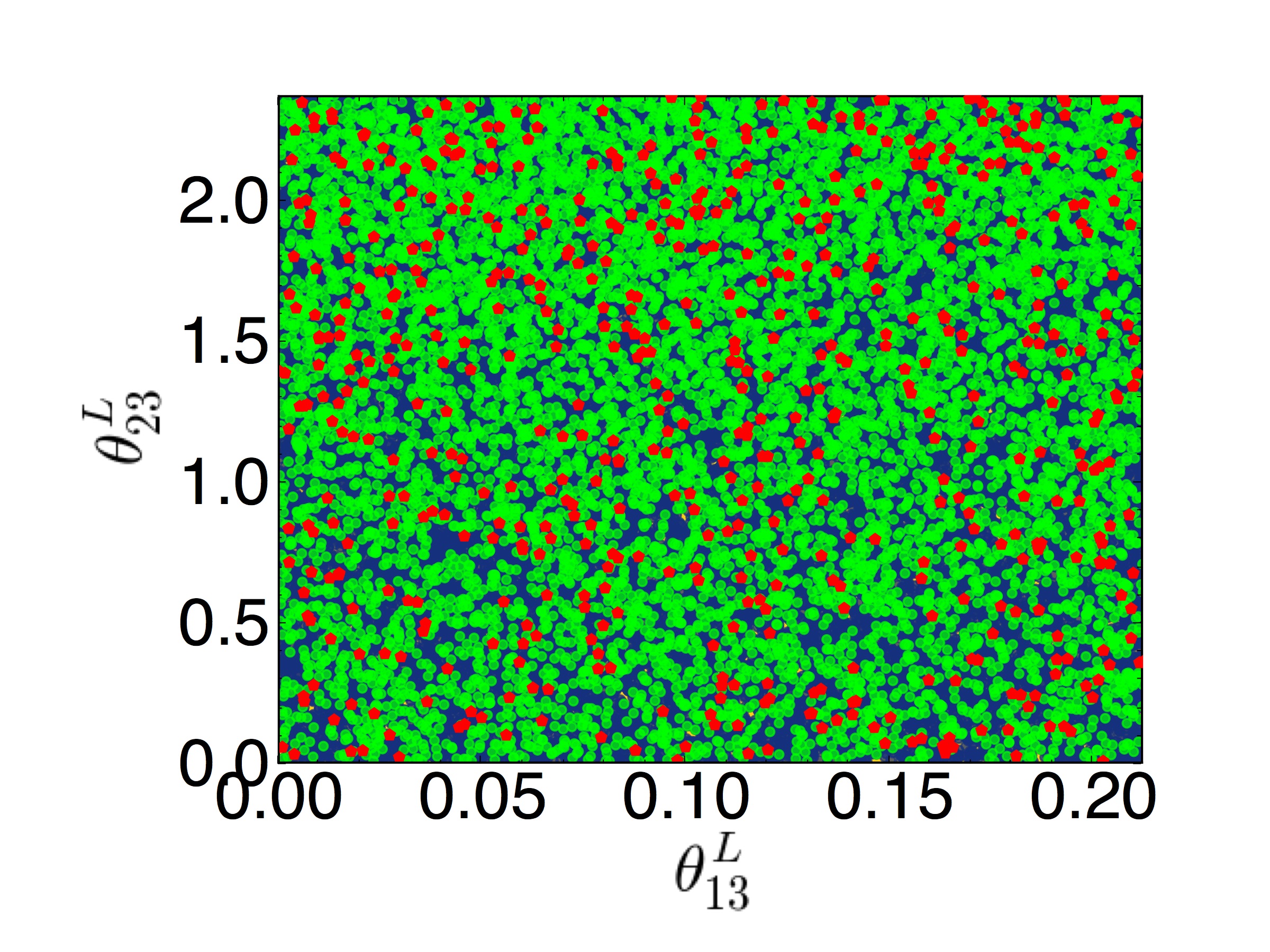,height=48mm,width=54mm}
\hspace{-4mm}
\psfig{file=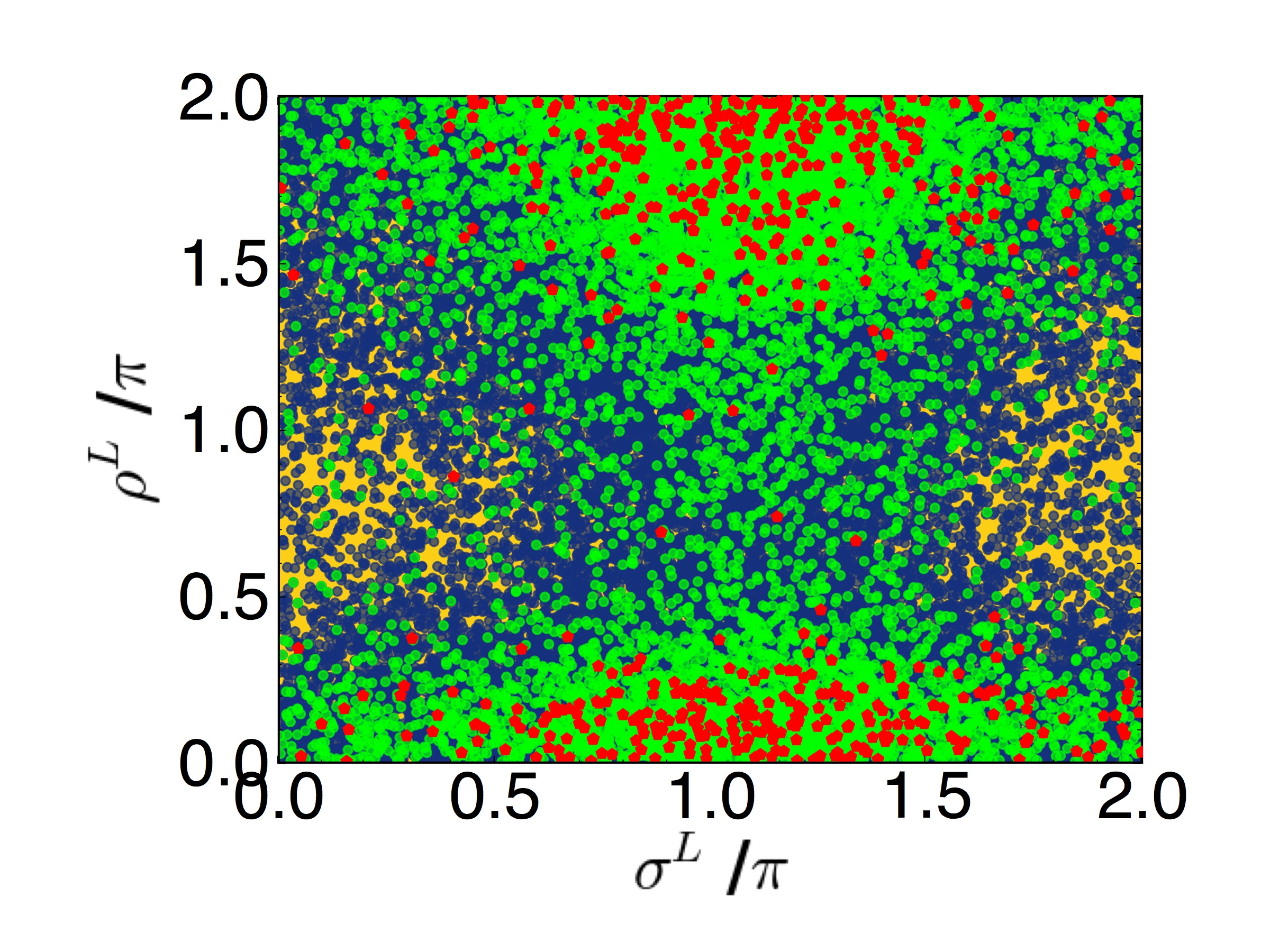,height=48mm,width=54mm} 
\end{center}
\caption{Results of the same scatter plots as in Fig.~1 for the six parameters in the unitary matrix $V_L$.}
\label{constrNOVL}
\end{figure} 
As one can see from the central panel, the points distribute quite uniformly 
for $\theta_{23}^L$ and $\theta_{13}^L$, those strongly restricted by the $SO(10)$-inspired condition, 
while there is a slight preference for high values of $\theta_{12}^L$, 
maybe an indication for a slight preference of $V_L = V_{\rm CKM}$ compared to $V_L = I$,
though solutions for $V_L = I$ are anyway possible. 

The points also seem to distribute uniformly in $\delta_L$ that, therefore, 
is not constrained  even when the strong thermal condition is added.
On the other hand, as it can be seen in the right panel of Fig.~7, the solution 
favours values of the Majorana-like phases in a region around $(\sigma^L, \rho^L) \simeq (\pi,0) $.  

\subsection{A benchmark point}

In this subsection we show in Fig.~8 the same plots shown in Fig.~2 in the case of
a benchmark point in the space of parameters that does respect the strong thermal
leptogensis. This has been simply chosen as a point that is located in a central position within the
`red' allowed regions ($N_{B-L}^{\rm p,i} = 0.1$).  

The results are quite interesting because they show directly how, for $m_1 \simeq 20\,{\rm meV}$
and $\theta_{23} \lesssim 41^{\circ}$, the $K_{i\alpha}$  are indeed able to fulfil all the conditions
eq.~(\ref{strongthconditions}). 
\begin{figure}
\vspace*{-15mm}
\begin{center}
\hspace{-4mm}
\psfig{file=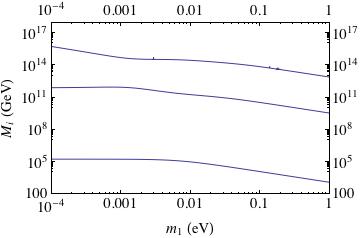,height=46mm,width=50mm}
\hspace{-4mm}
\psfig{file=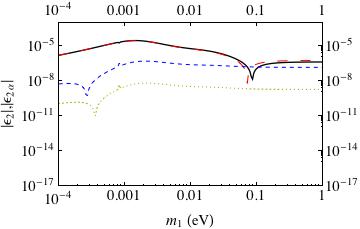,height=46mm,width=50mm} 
\hspace{-4mm}
\psfig{file=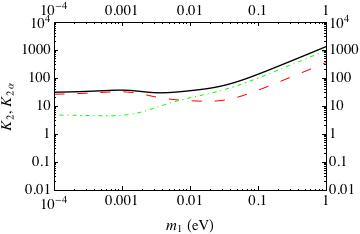,height=46mm,width=50mm} \\
\psfig{file=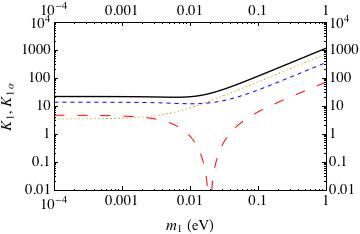,height=46mm,width=50mm} 
\hspace{-4mm}
\psfig{file=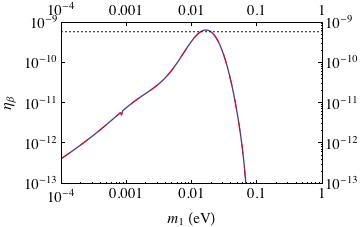,height=46mm,width=50mm} 
\end{center}
\vspace{-10mm}
\caption{Plots of the relevant quantities (same set of $(\a_1,\a_2,\a_3)$ values, line and colour conventions as in Fig.~2) for a particular choice 
of the values of the parameters able to realise successful strong $SO(10)$-inspired leptogenesis
for $N_{B-L}^{\rm p,i}=0.1$. The set of values of the parameters is given by:
$\theta_{13}=9^{\circ}$, $\theta_{12}=34^{\circ}$, $\theta_{23}=38^{\circ}$,
$\delta=-0.17\,\pi$, $\rho=0.20\,\pi$, $\s=0.87 \pi$,
$\theta_{13}^L=0.032^{\circ}$, $\theta_{12}^L=6.3^{\circ}$, $\theta_{23}^L=2.2^{\circ}$,
$\delta_L = 2.35$, $\rho_L=0.33$, $\sigma_L=3.76$.}
\label{benchmark}
\end{figure}
This benchmark point shows how the strong thermal $SO(10)$-inspired solution 
is a  proper combination of the  type $\tau_A$ and $\tau_B$ solutions
discussed in 3.2.1 and  shown in Fig.~2 for a specific set of values of the parameters. 
As for the type $\tau_A$ solution, there is a sharp dip for $K_{1\tau}$ around
a particular value of $m_1$ such that $K_{1\t} \lesssim 1$ and in coincidence $K_{1\m}\gg 1$. 
However, while for the type $\tau_A$ solution one could not respect the condition 
$K_{1e} \gg 1$ at the dip, now one can see that this is realised as in type $\tau_B$, so that now 
one has simultaneously $K_{1e} \gg 1$ and  $K_{1\mu} \gg 1$. Another important hybrid feature is that
now, as for type $\tau_B$, one has $K_{2\tau}\sim 10$ instead of $K_{2\t}\sim 1000$
as for the type $\tau_A$. This minimises the wash-out of the asymmetry produced by $N_2$ decays
still allowing a wash-out of the pre-existing tauon asymmetry and makes possible to 
fulfil  jointly successful leptogenesis  and strong thermal condition.  

These are the main features of the solution that can be realised, in particular, only if $m_1\simeq 20\,$meV
and $\theta_{23} \lesssim 41^{\circ}$.  Notice that a value  $N_{B-L}^{\rm p,i} = {\cal O}(0.1)$ represents basically the maximum value of the pre-existing asymmetry that can be washed-out respecting simultaneously the successful
leptogenesis bound. The main reason is that $K_{2\t}$ has to be necessarily much larger than $1$ and this necessarily
introduces a wash-out at the production. For values $N_{B-L}^{\rm p,i} \gg 0.1$ the value of $K_{2\tau}$
for the wash-out of the pre-existing asymmetry becomes so large that the asymmetry produced by $N_2$
decays is too strongly washed-out to reproduce the observed asymmetry. 

\subsection{Stability of the solution}

The solution has been determined fixing some parameters and it is then 
important to discuss whether a variation of these parameters can significantly change the
constraints. We have already shown and discussed the dependence on the initial value of the
pre-existing asymmetry $N_{B-L}^{\rm p,i}$.

The plots in Fig.~1 have been obtained for $(\a_1,\a_2,\a_3)=(1,5,1)$. 
We verified that indeed there is no dependence of the constraints
on $\a_1$ and $\a_3$ as discussed. A value $\a_2 =5$ should be considered close to the maximum within
$SO(10)$-inspired models. We have also determined which would be the minimum value of $\a_2$ for the
existence of a solution satisfying the strong thermal condition, finding
$\a_2 \simeq 4$ for $N_{B-L}^{\rm p,i}=0.1$. 

We have checked the sensitivity of the solution to  the maximum values of the angles $\theta_{ij}^{L}$ in $V_L$
allowed by the $SO(10)$-inspired conditions.
We have tripled the values of $\theta_{13}^L$ and $\theta_{23}^L$ finding that the only constraint that changes significantly is the upper bound on $\theta_{23}$ that is relaxed to $\theta_{23} \lesssim 43^{\circ}$
for $N_{B-L}^{\rm p,i}=0.1$. We did not try to modify the Cabibbo-like angle $\theta_{12}^L$ since this is
already significantly large. 

The solution corresponds to a RH neutrino mass spectrum that is genuinely hierarchical
since $M_2 \simeq 3\times 10^{11}\,{\rm GeV}$ while $M_3 \simeq \a_3^{\, 2} \, 10^{14}\,{\rm GeV} > 10^{12}\,{\rm GeV}$ for $\a_3\gtrsim 0.1$.  For example, we have relaxed
the condition $M_3/M_2 > 10$ imposed in the results of Fig.~1, redetermining separately 
the constraints for $10 \geq M_3/M_2 \geq 2$ finding no new solutions
for successful strong thermal leptogenesis.
\footnote{We adopted the hierarchical limit for the calculation of the asymmetry. In this limit
the wash-out of $N_3$ on the asymmetry produced by the $N_2$ is neglected. This would start
to produce some effect only for $(M_3-M_2)/M_2 \lesssim 1$ \cite{beyond}.} 
We  have found a new different marginal solution only at 
$m_1 \simeq 0.3\,$eV, incompatible with the cosmological bounds. This means  that our analysis does not exclude  the existence of solutions with quasi-degenerate RH neutrino masses. However, it should be stressed that 
the strong thermal condition that we have imposed on the relic value of the pre-existing asymmetry (cf. (eq.~\ref{finalpas})) is valid only in the hierarchical case. An analysis of the wash-out of a pre-existing asymmetry 
for quasi-degenerate RH neutrino masses is still missing. 
In any case it is important to notice that these possible new solutions would in case
correspond to different constraints on the low energy neutrino parameters 
and would be, therefore, experimentally distinguishable from our solution. 

\subsection{Theoretical uncertainties}

Let us now briefly comment on the approximations that we made in the calculation
of the asymmetry and on the kind of corrections one could expect removing them.

We are not solving density matrix equations. When we do not impose the strong thermal condition
(the yellow points), these can be important when
$M_2$ falls around $10^{12}\,{\rm GeV}$. For $M_2$  above $10^{12}\,$GeV
we have calculated the asymmetry at the production in the unflavoured case (cf. eq.~(\ref{onefl})).
Around $M_2\simeq 10^{12}\,$GeV there is, therefore, a discontinuity.
However, since flavour effects, for $M_2 < 10^{12}\,$GeV, tend to enhance the asymmetry reducing the
wash-out, most of the solutions lie below $10^{12}\,$ GeV. 
When we impose the strong thermal condition for non-vanishing 
$N_{B-L}^{\rm p,i}$, we have simply excluded points for $M_2 \gtrsim 10^{12}\,$GeV,
since the strong thermal condition can be satisfied only if the $N_2$ wash-out occurs in the two-flavour regime. 
This is the reason why the allowed regions satisfying the strong thermal conditions  
 sharply cut above $M_2 = 10^{12}\,GeV$. A calculation from a solution of the density matrix equation would 
 then just simply smoothly describe the transition but 
this would not significantly affect the results found on the constraints on
the low energy neutrino parameters since the bulk of points are found for
$M_2 \simeq 3\times 10^{11}\,$ GeV. 

We are neglecting phantom terms \cite{flavourcoupling}. These can affect the $SO(10)$-inspired
solutions (the yellow points) but not certainly the strong thermal $SO(10)$-inspired solution since
phantom terms can only be present in the electron and muon asymmetries that are fully washed-out,
while the final asymmetry is strictly tauon dominated. 

We are neglecting the running of neutrino parameters from the  high energy scales to low energies. 
However, since our solution is semi-hierarchical and NO, 
the running is negligible for all practical purposes \cite{running}. 
For the atmospheric mixing angle, the only parameter for which it could be potentially relevant to calculate the running,  since we have found compatibility only with the lowest allowed experimental values, the running is 
at most about $0.01^{\circ}$ from $M_2 \sim 10^{11}\,$GeV to low energies,
a variation that clearly is completely negligible for any practical purpose. 

A potential important correction is flavour coupling \cite{flavourcoupling}. This can have two effects: it can 
alter the asymmetry at the production, usually this goes in the direction of increasing the asymmetry, and it could 
make the conditions for strong thermal leptogenesis tighter. The two effects would tend even to cancel with each other.  Notice, however, that the flavour coupling would disappear in the limit $K_{1\tau}/(K_{1e}+K_{1\mu}) \rightarrow 0$, since this would correspond to a case where the tauon flavour decouples. However, this is exactly the case realised in the solution and, therefore, again, we do not expect great effects from flavour coupling at the lightest RH neutrino wash-out. The only effect might be a small enhancement  of the asymmetry at the production that could slightly relax the constraints. The  atmospheric neutrino mixing angle upper bound is the most sensitive constraint to corrections and, therefore, this might motivate an account of flavour coupling. 

An account of momentum dependence also does not produce significant corrections 
since this can alter the lightest RH neutrino wash-out only for $K_{1\alpha} \gg 1$ \cite{vives2}, 
while in our case the production occurs in the tauon flavour and $K_{1\t}\lesssim 1$.

In conclusion we cannot envisage sources of significant corrections, though an account of 
flavour coupling might be justified by a precise determination of the upper bound  on the atmospheric mixing angle
in connections with future experimental results. 

\subsection{The orthogonal and the RH neutrino mixing matrices}

It is also interesting to discuss the form of the 
RH neutrino mixing matrix $U_R$ and of the orthogonal matrix $\O$ corresponding to
the solution. For the set of values of the parameters corresponding to the benchmark
point discussed in  6.2, in particular $(\a_1, \a_2, \a_3)= (1,5,1)$, and for $m_1 = 20\,$meV,
the resulting RH neutrino mixing matrix $U_R$ is given by
\be
U_R \simeq \left( 
\begin{array}{ccc}
 e^{-i\,0.8\,\pi} & 5\times 10^{-4} \, e^{i\,0.6\,\pi}  &  8 \times 10^{-6}  \\
5\times 10^{-4}\, e^{i\,0.6\,\pi}  &  e^{-i\,\pi} & 2 \times 10^{-2} \, e^{i\,0.3\,\pi}  \\
1.5  \times 10^{-6} \, e^{-i\,0.2\,\pi} & 2\times 10^{-2}\, e^{-i\,0.2\,\pi} & e^{i\,0.15\,\pi} 
\end{array}
\right) \,  .
\ee
Contrarily to the final asymmetry and to the $K_{i\a}$, the 
$U_R$ depends not only on $\a_2$ but also on $\a_1$ and $\a_3$
in a way that off-diagonal terms tend to be damped for a higher hierarchy.
For example if we choose $(\a_1,\a_2,\a_3)=(5,5,5)$, so that
$M_3/M_2$ increases while $M_2/M_1$ decreases, we obtain
\be
U_R \simeq \left( 
\begin{array}{ccc}
 e^{-i\,0.8\,\pi} & 3\times 10^{-3} \, e^{i\,0.6\,\pi}  &  8 \times 10^{-6}  \\
3\times 10^{-3}\, e^{i\,0.6\,\pi}  &  e^{-i\,\pi} & 3 \times 10^{-3} \, e^{i\,0.3\,\pi}  \\
1.5  \times 10^{-6} \, e^{-i\,0.2\,\pi} & 3 \times 10^{-3} \, e^{-i\,0.2\,\pi} & e^{i\,0.15\,\pi} 
\end{array}
\right) \,  ,
\ee
showing that the $23$ off-diagonal entries decreased while the $12$ increased and the $13$ stayed constant.
A more drastic enhancement of the, for example, $23$ off-diagonal terms is obtained lowering $\a_3$ 
to $\a_3 = 0.05$ obtaining ($\a_1=1$, $\a_2=5$)
\be
U_R \simeq \left( 
\begin{array}{ccc}
 e^{-i\,0.8\,\pi} & 5\times 10^{-4} \, e^{-i\,0.4\,\pi}  &  1.5 \times 10^{-4} \, e^{-i\,0.1\,\pi} \\
5\times 10^{-4}\, e^{i\,0.6\,\pi}  &  0.95  & 0.3  \, e^{i\,0.3\,\pi}  \\
3 \times 10^{-5} \, e^{-i\,0.2\,\pi} & 0.3 \, e^{i\,0.9\,\pi} & 0.95\, e^{i\,0.15\,\pi} 
\end{array}
\right) \,  .
\ee
These three examples show how there is a flexibility in the choice of $\a_1$ and $\a_3$ 
that can potentially be useful in order to minimise the fine tuning to obtain a softly semi-hierarchical 
light neutrino mass spectrum ($m_3/m_2 \simeq 3$, $m_2 \simeq m_1$) from highly hierarchical
neutrino Yukawa couplings \cite{casaibarraalbu}.

On the other hand the orthogonal matrix $\O$ is very slightly dependent on $(\a_1,\a_2,\a_3)$.
This can be calculated  from \cite{SO10lep1}
\be
\O = D_m^{-{1\over 2}} \, U^{\dagger} \, V_L^{\dagger} \, D_{m_D} \, U_R \, D_M^{-{1\over 2}}   \,  .
\ee
For the benchmark choice of parameters one finds
 \be\label{orthogonal}
\Omega \simeq \left( 
\begin{array}{ccc}
0.8 \, e^{-i\,0.9\,\pi} & 0.7 \, e^{i \,0.1\,\pi}  &  0.4 \, e^{- i \,0.7\,\pi} \\
0.7\, e^{i\,0.9\,\pi}  & 0.7 \, e^{-i\,0.7\,\pi} & 0.8   \\
0.4\, e^{- i\,0.2\,\pi} & 0.7  & 0.7 \, e^{ i \,0.1\,\pi} 
\end{array}
\right) \,  .
\ee
The very slight dependence of the orthogonal matrix on $\alpha_1$ and $\a_3$ is consistent with the
independence of the $K_{i\a}$ of $\a_1$ and $\a_3$ and, consequently, 
combined with the independence of $\ve_{2\t}$ of $\a_1$ and $\a_3$ \cite{SO10lep2},
of the the constraints   on the low energy neutrino parameters we obtained as well.  

This kind of orthogonal matrix (cf. eq.~(\ref{orthogonal})) shows that there are no fine-tuned cancellations
in the see-saw formula. However, each light neutrino mass $m_i$ receives contribution from all three
terms $\propto 1/M_{j}$, not just from one as in the case of 
an orthogonal matrix close to the identity or to one of the other five
forms obtained from the identity permuting rows and columns \cite{geometry},
so called form dominance models \cite{formdominance}. 

\subsection{Remarks on future developments}

We comment on two aspects that will be discussed in detail in future works \cite{preparation}. 
First, analytic insight into the results obtained from the scatter
plots would be certainly desirable requiring a dedicated analysis.  

Second, our analysis does not specify the confidence level of the constraints on the different
low energy neutrino parameters. This requires a full determination of the probability distribution functions
for the different parameters. This will also be discussed in a future work.  
Here we just want to notice that these distribution functions are highly non-trivial to determine since
even though one starts, as input, 
from simple gaussian experimental ranges, the complicated dependence of the asymmetry 
on the parameters makes in a way that the successful leptogenesis bound and the strong thermal condition
produce, as output,  quite complicated distributions functions to be determined with a statistical procedure.
\footnote{For some preliminary results see \cite{preliminary}.} 
For example the difference in the upper bounds on $\theta_{23}$ 
in Fig.~1 and in Fig.~4  is an indication of a different C.L., since they are determined 
with two data sets with a substantial different number of points. Consider the case of
an initial pre-existing abundance $N_{B-L}^{\rm p,i}=0.1$.  
In the first case a number of about two thousands red points was obtained 
to saturate the bound $\theta_{23} \lesssim 41^{\circ}$, given that the initial ranges for the uniform scan 
on the low energy neutrino parameters are fixed. 
The C.L. on this bound corresponds roughly to $\simeq 99.95\%$. 
In the second case the number of red points is much lower, about three hundreds
and the upper bound, $\theta_{23}\lesssim 40^{\circ}$, 
does not saturate the maximum value. The C.L. in this case corresponds roughly
to $99.7\%$. This implies that a future experimental result $\theta_{23}\gtrsim 40^{\circ}$ would 
strongly disfavour the solution.   A precise determination of $\theta_{23}$ will be, therefore, 
a crucial test for the solution.

\section{Final discussion}

We presented a novel solution to the problem of the initial conditions in leptogenesis within 
$SO(10)$-inspired models. It is particularly interesting that this yields 
definite constraints on all low energy neutrino parameters, 
sharp enough to have, all together,  quite a strong predictive power. 
It is encouraging that the solution requires a non-vanishing  value of the reactor mixing angle 
in agreement with the measured range. 
This should be considered in addition to the well known leptogenesis conspiracy for which the solar and atmospheric neutrino mass scales are just about ten times higher than the equilibrium neutrino mass $m_{\star}\sim 10^{-3}\,{\rm eV}$, in a way that the decay parameters tend to be $K_{i\a}={\cal O}(1-10)$,
a key feature for the realisation of the strong thermal condition 
in the flavoured $N_2$-dominated scenario.
However, the full set of constraints on neutrino parameters from the strong thermal $SO(10)$-inspired solution   is still far to be fully tested. As we discussed, it is distinctive enough 
that it can be regarded as a signature of the solution, hard to be mimicked or to  agree just accidentally
with the experimental data.  It clearly predicts NO neutrino masses, atmospheric
mixing angle well in the first octant and it strongly favours a negative Dirac phase, around $\d \sim - \pi/5$. 
Indeed these features should all be tested during next years by neutrino oscillation experiments.
At the same time the absolute neutrino mass scale predictions, $m_{ee}\simeq 0.8\,m_1 \simeq 15\,{\rm meV}$,
also  provide  quite a definite feature that should be tested with cosmological observations 
and  (ultimately with) $00\nu\b$ decay experiments. 
It will be exciting to see  whether future experimental data will further
support the presented solution or rule it out. In any case this
provides an example  of a motivated  falsifiable minimal high 
energy scale leptogenesis scenario.

\vspace{-1mm}
\subsection*{Acknowledgments}

We wish to thank A.~Abada, E.~Bertuzzo, S.~Blanchet, M.~Re Fiorentin, S.~King, S.F.~King, 
S.~Huber, S.~Lavignac, A.~Mirizzi,  M.~Schmidt and T.~Yanagida for useful comments and discussions.
We acknowledge financial support  from the NExT/SEPnet Institute.
PDB also  acknowledges financial support  from the STFC Rolling Grant ST/G000557/1 and from the  
EU FP7  ITN INVISIBLES  (Marie Curie Actions, PITN- GA-2011- 289442).
LM also acknowledges financial support from the European Social Fund (grant MJD387).


\vspace{-3mm}
\begin{thebibliography}{99}

\bibitem{fy}  
M.~Fukugita and T.~Yanagida,
  Phys.\ Lett.\ B {\bf 174}, 45 (1986); 

\bibitem{reviews}  
For recent reviews on leptogenesis see
S.~Blanchet and P.~Di Bari,
New J.\ Phys.\  {\bf 14} (2012) 125012;
T.~Hambye, New J.\ Phys.\  {\bf 14} (2012) 125014.
  
\bibitem{seesaw} P.~Minkowski,
Phys.\ Lett.\ B {\bf 67},  421 (1977);
M. Gell-Mann, P. Ramond and
R. Slansky,  {\em Proceedings of the Supergravity Stony Brook Workshop}, New
York 1979,  eds. P. Van Nieuwenhuizen and D. Freedman; T. Yanagida,  {\em
Proceedings of the Workshop on Unified Theories and Baryon Number in the
Universe},  Tsukuba, Japan 1979, eds. A. Sawada and A. Sugamoto;
R. N. Mohapatra, G. Senjanovic, Phys. Rev. Lett. {\bf 44},  912 (1980).

\bibitem{Planck}
  P.~A.~R.~Ade {\it et al.}  [Planck Collaboration],
  arXiv:1303.5076 [astro-ph.CO].

\bibitem{bounds}
S.~Blanchet and P.~Di Bari,
  Nucl.\ Phys.\ B {\bf 807} (2009) 155.



\bibitem{bound}
W.~Buchmuller, P.~Di Bari and M.~Plumacher,
  Phys.\ Lett.\ B {\bf 547} (2002) 128.
   
 \bibitem{pedestrians}
 W.~Buchmuller, P.~Di Bari, M.~Plumacher,
  Annals Phys.\  {\bf 315 } (2005)  305-351.

\bibitem{window}
W.~Buchmuller, P.~Di Bari and M.~Plumacher,
  Nucl.\ Phys.\ B {\bf 665} (2003) 445.


\bibitem{flavour}
 A.~Abada, S.~Davidson, F.~-X.~Josse-Michaux, M.~Losada and A.~Riotto,
  JCAP {\bf 0604} (2006) 004;
E.~Nardi, Y.~Nir, E.~Roulet and J.~Racker,
  JHEP {\bf 0601} (2006) 164.
 
 \bibitem{rius}
 S.~Davidson, J.~Garayoa, F.~Palorini and N.~Rius,
  Phys.\ Rev.\ Lett.\  {\bf 99} (2007) 161801.
 
 \bibitem{2RHN}
S.~Antusch, P.~Di Bari, D.~A.~Jones and S.~F.~King,
  {Phys.\ Rev.\ } D {\bf 86} (2012) 023516.


 
\bibitem{problem}
  E.~Bertuzzo, P.~Di Bari, L.~Marzola,
  Nucl.\ Phys.\  {\bf B849 } (2011)  521-548.

\bibitem{flavorlep}
 S.~Blanchet, P.~Di Bari,
  JCAP {\bf 0703 } (2007)  018.

\bibitem{pascoli}
 S.~Pascoli, S.~T.~Petcov and A.~Riotto,
  Phys.\ Rev.\ D {\bf 75} (2007) 083511.

\bibitem{diraclep}
A.~Anisimov, S.~Blanchet and P.~Di Bari,
  JCAP {\bf 0804} (2008) 033.
  
 \bibitem{theta13}
 K.~Abe {\it et al.}  [T2K Collaboration],
  Phys.\ Rev.\ Lett.\  {\bf 107} (2011) 041801;
  P.~Adamson {\it et al.}  [MINOS Collaboration],
  arXiv:1108.0015 [hep-ex];
Y.~Abe {\it et al.}  [DOUBLE-CHOOZ Collaboration],
  arXiv:1112.6353 [hep-ex].
F.~P.~An {\it et al.}  [DAYA-BAY Collaboration],
  arXiv:1203.1669 [hep-ex];
S.~-B.~K.~f.~R.~collaboration,
  arXiv:1204.0626.

\bibitem{fogli}
G.~L.~Fogli, E.~Lisi, A.~Marrone, D.~Montanino, A.~Palazzo and A.~M.~Rotunno,
  Phys.\ Rev.\ D {\bf 86} (2012) 013012.
  \bibitem{gonzalez}
M.~C.~Gonzalez-Garcia, M.~Maltoni, J.~Salvado and T.~Schwetz,
  arXiv:1209.3023 [hep-ph].

\bibitem{valle}
D.~V.~Forero, M.~Tortola and J.~W.~F.~Valle,
  Phys.\ Rev.\ D {\bf 86} (2012) 073012.



\bibitem{mfv}
S.~Uhlig, 
JHEP {\bf 0711} (2007) 066.

\bibitem{2RH}
S.~F.~King,
  Nucl.\ Phys.\  B {\bf 576} (2000) 85;
 P.~H.~Frampton, S.~L.~Glashow and T.~Yanagida,
  Phys.\ Lett.\ B {\bf 548} (2002) 119;
 A.~Ibarra and G.~G. Ross, \emph{Phys. Lett. } B {\bf 591} (2004) 285; 
A.~Abada, S.~Davidson, A.~Ibarra, F.~-X.~Josse-Michaux, M.~Losada and A.~Riotto,
 JHEP {\bf 0609} (2006) 010.
 
 \bibitem{geometry}
P.~Di Bari,
 Nucl.\ Phys.\  {\bf B727 } (2005)  318-354.
 
 
 \bibitem{vives}
 O.~Vives,
  Phys.\ Rev.\  {\bf D73 } (2006)  073006.
  
  
\bibitem{SO10lep1}
P.~Di Bari and A.~Riotto,
  Phys.\ Lett.\ B {\bf 671} (2009) 462.


 
\bibitem{buchplum}
W.~Buchmuller and M.~Plumacher,
  Phys.\ Lett.\ B {\bf 389} (1996) 73.

\bibitem{orloff}
E.~Nezri and J.~Orloff,
  JHEP {\bf 0304} (2003) 020
  [hep-ph/0004227].

\bibitem{falcone}
F.~Buccella, D.~Falcone and F.~Tramontano,
  Phys.\ Lett.\ B {\bf 524} (2002) 241.

\bibitem{branco}
G.~C.~Branco, R.~Gonzalez Felipe, F.~R.~Joaquim and M.~N.~Rebelo,
  Nucl.\ Phys.\ B {\bf 640} (2002) 202.
 
\bibitem{smirnov}
E.~K.~Akhmedov, M.~Frigerio and A.~Y.~Smirnov, JHEP {\bf 0309}, 021 (2003).

\bibitem{dicmb}
 S.~Davidson and A.~Ibarra,
  Phys.\ Lett.\ B {\bf 535} (2002) 25;
W.~Buchmuller, P.~Di Bari and M.~Plumacher,
  Nucl.\ Phys.\ B {\bf 643} (2002) 367 [Erratum-ibid.\ B {\bf 793} (2008) 362].

\bibitem{crv}  
L.~Covi, E.~Roulet and F.~Vissani,
  Phys.\ Lett.\ B {\bf 384} (1996) 169.
 
 \bibitem{pilaftsis}
A.~Pilaftsis and T.~E.~J.~Underwood,
  Nucl.\ Phys.\ B {\bf 692} (2004) 303.
 
 \bibitem{SO10lep2}
P.~Di Bari, A.~Riotto,
  JCAP {\bf 1104 } (2011)  037.

\bibitem{abada}
A.~Abada, P.~Hosteins, F.~-X.~Josse-Michaux and S.~Lavignac,
  Nucl.\ Phys.\ B {\bf 809} (2009) 183.

\bibitem{mohapatra}  
S.~Blanchet, P.~S.~B.~Dev and R.~N.~Mohapatra,
  Phys.\ Rev.\ D {\bf 82} (2010) 115025.
 
\bibitem{MINOS}
 R.~Nichol [MINOS Collaboration],
  Nucl.\ Phys.\ B, Proc.\ Suppl.\  {\bf 235-236} (2013) 105.

\bibitem{casas}
J.~A.~Casas and A.~Ibarra,
  Nucl.\ Phys.\ B {\bf 618} (2001) 171.

\bibitem{buccella}
F.~Buccella, D.~Falcone, C.~S.~Fong, E.~Nardi and G.~Ricciardi,
  Phys.\ Rev.\ D {\bf 86} (2012) 035012.

\bibitem{bcst}
R.~Barbieri, P.~Creminelli, A.~Strumia, N.~Tetradis,
  Nucl.\ Phys.\  {\bf B575 } (2000)  61-77.

\bibitem{giudice}
G.~F.~Giudice, A.~Notari, M.~Raidal, A.~Riotto and A.~Strumia,
  Nucl.\ Phys.\ B {\bf 685} (2004) 89.

\bibitem{density}
 S.~Blanchet, P.~Di Bari, D.~A.~Jones and L.~Marzola,
  JCAP {\bf 1301} (2013) 041.

\bibitem{flavourcoupling}
 S.~Antusch, P.~Di Bari, D.~A.~Jones and S.~F.~King,
  Nucl.\ Phys.\ B {\bf 856} (2012) 180.
  
\bibitem{running}
K.~S.~Babu, C.~N.~Leung, J.~T.~Pantaleone,
  Phys.\ Lett.\ B {\bf 319} (1993) 191;
S.~Antusch, J.~Kersten, M.~Lindner, M.~Ratz,
  Nucl.\ Phys.\ B {\bf 674} (2003) 401;
S.~Antusch, J.~Kersten, M.~Lindner, M.~Ratz, M.~A.~Schmidt,
  JHEP {\bf 0503} (2005) 024.


\bibitem{CHOOZ}
M.~Apollonio {\it et al.}  [CHOOZ Collaboration],
  Phys.\ Lett.\ B {\bf 466} (1999) 415.


\bibitem{schwetz}
 T.~Schwetz, M.~A.~Tortola and J.~W.~F.~Valle,
  New J.\ Phys.\  {\bf 10} (2008) 113011.

\bibitem{neutrinolessnow}
 E.~Andreotti {\it et al.},
 \emph{Astropart.\ Phys.\ }  {\bf 34} (2011) 822;
  M.~Agostini {\it et al.}  [GERDA Collaboration],
  arXiv:1307.4720 [nucl-ex].

\bibitem{gravity}
 R.~Kallosh, A.~D.~Linde, D.~A.~Linde and L.~Susskind,
  Phys.\ Rev.\  D {\bf 52} (1995) 912
  [arXiv:hep-th/9502069];
 H.~Davoudiasl, R.~Kitano, G.~D.~Kribs, H.~Murayama and P.~J.~Steinhardt,
  Phys.\ Rev.\ Lett.\  {\bf 93} (2004) 201301.



\bibitem{affleckdine}
I.~Affleck and M.~Dine,
  Nucl.\ Phys.\  B {\bf 249} (1985) 361.


\bibitem{GUTB}
 M.~Yoshimura,
  Phys.\ Rev.\ Lett.\  {\bf 41} (1978) 281
  [Erratum-ibid.\  {\bf 42} (1979) 746];
 S.~Dimopoulos and L.~Susskind,
  Phys.\ Rev.\  D {\bf 18} (1978) 4500;
  D.~Toussaint, S.~B.~Treiman, F.~Wilczek and A.~Zee,
  Phys.\ Rev.\  D {\bf 19} (1979) 1036;
E.~W.~Kolb and S.~Wolfram,
  Nucl.\ Phys.\  B {\bf 172} (1980) 224
  [Erratum-ibid.\  B {\bf 195} (1982) 542].
 E.~W.~Kolb, A.~D.~Linde and A.~Riotto,
  Phys.\ Rev.\ Lett.\  {\bf 77} (1996) 4290.

\bibitem{preliminary}
Talks by P.~Di Bari and L.~Marzola at the DESY Theory Workshop 2011, 
27-30 September 2011, Hamburg.

\bibitem{lindner}
P.~Huber, M.~Lindner, T.~Schwetz and W.~Winter,
  JHEP {\bf 0911} (2009) 044.




\bibitem{agarwalla}
S.~K.~Agarwalla, S.~Prakash and S.~U.~Sankar,
  arXiv:1301.2574 [hep-ph].

\bibitem{razzaque}
E.~K.~Akhmedov, S.~Razzaque and A.~Y.~Smirnov,
  JHEP {\bf 02} (2013) 082.

\bibitem{blennowschwetz}
 M.~Blennow and T.~Schwetz,
  arXiv:1306.3988 [hep-ph].

\bibitem{plancksz}
  P.~A.~R.~Ade {\it et al.}  [Planck Collaboration],
  arXiv:1303.5080 [astro-ph.CO].

\bibitem{neutrinoless}
 B.~Schwingenheuer,
  Annalen Phys.\  {\bf 525} (2013) 269;
O.~Cremonesi,
  Nucl.\ Phys.\ Proc.\ Suppl.\  {\bf 237-238} (2013) 7.

\bibitem{rodejohann}
W.~Rodejohann,
  Int.\ J.\ Mod.\ Phys.\ E {\bf 20} (2011) 1833.

\bibitem{beyond}
S.~Blanchet and P.~Di Bari,
  JCAP {\bf 0606} (2006) 023.

\bibitem{vives2}
 J.~Garayoa, S.~Pastor, T.~Pinto, N.~Rius and O.~Vives,
  JCAP {\bf 0909} (2009) 035.


\bibitem{casaibarraalbu}
J.~A.~Casas, A.~Ibarra and F.~Jimenez-Alburquerque,
  JHEP {\bf 0704} (2007) 064.

\bibitem{formdominance}
S.~F.~King, Rept.\ Prog.\ Phys.\  {\bf 67} (2004) 107;
M.~C.~Chen and S.~F.~King,
  JHEP {\bf 0906} (2009) 072.

\bibitem{preparation}
In preparation.

\end{thebibliography}
\end{document}